\documentclass[conference]{IEEEtran}
\IEEEoverridecommandlockouts
\usepackage{cite}
\usepackage{amsmath,amssymb,amsfonts}
\usepackage{algorithmic}
\usepackage{colortbl}
\usepackage{graphicx}
\usepackage{textcomp}
\usepackage{xcolor}
\usepackage{tabularx}
\usepackage{multirow}
\usepackage{array}
\usepackage{threeparttable}
\usepackage{ltablex}
\usepackage{makecell}
\usepackage{amssymb} 
\usepackage{pifont} 
\usepackage{placeins}
\usepackage{afterpage}
\usepackage{float}
\usepackage{longtable}
\usepackage{hyperref}
\usepackage{bm}
\usepackage{booktabs}

\usepackage[utf8]{inputenc}

\def\BibTeX{{\rm B\kern-.05em{\sc i\kern-.025em b}\kern-.08em
    T\kern-.1667em\lower.7ex\hbox{E}\kern-.125emX}}

\newcommand{\xmark}{\ding{55}} 
\newcolumntype{Y}{>{\centering\arraybackslash}p{7.5cm}}
\newcolumntype{C}[1]{>{\centering\let\newline\\\arraybackslash\hspace{0pt}}m{#1}}
\newcolumntype{L}[1]{>{\raggedright\let\newline\\\arraybackslash\hspace{0pt}}m{#1}}

\begin{document}
\bstctlcite{IEEEexample:BSTcontrol}

\title{End-to-End Deep Learning in Wireless \color{black}Communication \color{black}Systems: A Tutorial Review\\
\thanks{This work was supported by the \color{black}US \color{black} National Science Foundation under Grant~\#~2114779.}

\thanks{Abdelrahman Elfikky is with the Department of Computer Science, University of Arkansas at Little Rock, Little Rock, AR, USA (email: aelfikky1@ualr.edu).}

\thanks{Zouheir Rezki, and Jorge Cortez are with the Department of Electrical and Computer Engineering, University of California, Santa Cruz, California, USA (email: zrezki@ucsc.edu, jcorte15@ucsc.edu).}

\thanks{Youssef Boumhaout is with the African Leadership Academy, Johannesburg, South Africa (email: youssef.boumhaout@gmail.com).}

\thanks{Anne Xia is with the Department of Computer Science, Cornell University, New York, USA (email: ax53@cornell.edu).}

\thanks{Abdulkadir Celik is with the School of Electronics and Computer Science, University of Southampton, SO17 1BJ, United Kingdom. (email: a.celik@soton.ac.uk).}


\thanks{Georges Kaddoum is with the \'{E}cole de technologie sup\'{e}rieure (ETS), Universit\'{e} du Qu\'{e}bec, Montr\'{e}al, QC, Canada (email: georges.kaddoum@etsmtl.ca).}
}

\author{Abdelrahman~Elfikky, \textit{Member, IEEE}, Zouheir~Rezki, \textit{Senior Member, IEEE}, Jorge~Cortez, \\ Youssef Boumhaout, Anne~Xia, Abdulkadir Celik, \textit{Senior Member, IEEE}, and Georges~Kaddoum, \textit{Senior Member, IEEE}}

\maketitle


\begin{abstract}
	The physical layer (PHY) in wireless communication systems has traditionally relied on model-based methods \color{black}that are often optimized individually as independent blocks to perform \color{black} tasks \color{black}such as \color{black} modulation, coding, and channel estimation. However, these approaches face challenges \color{black} when it comes to \color{black} capturing real-world nonlinearities, hardware imperfections, and increasing complexity in modern networks. This paper surveys advancements in applying deep learning (DL) for end-to-end PHY optimization \color{black}by incorporating \color{black} the autoencoder (AE) model as a powerful end-to-end DL framework \color{black}to enable \color{black} joint transmitter and receiver \color{black}optimization and address \color{black} challenges like dynamic channel conditions and scalability. We review cutting-edge DL models; their applications in PHY tasks such as modulation, error correction, and channel estimation; and their deployment in real-world scenarios, including point-to-point communication, multiple access, and interference channels. \color{black}This work highlights \color{black} the benefits of learning-based approaches over traditional methods, \color{black}offering \color{black} a comprehensive resource for researchers and engineers \color{black}looking \color{black} to innovate in next-generation wireless systems. Key insights and future directions are discussed to bridge the gap between theory and practical implementation.
\end{abstract}


\section{Introduction}

\color{black}
\IEEEPARstart{M}{odern} life relies on seamless wireless connectivity, with the rapid growth of the Internet of Things (IoT) driving the need for more intelligent and adaptive networks. To meet these demands, the International Telecommunication Union’s IMT-2030 framework for 6G expands beyond 5G’s capabilities by introducing AI-native systems as a core pillar \cite{IMT2030REF1}. This marks a shift toward self-optimizing, autonomous networks that enhance efficiency, reliability, and scalability. A key enabler of this vision is Open Radio Access Network (O-RAN), which integrates near-real-time (near-RT) and non-real-time (non-RT) Radio Intelligent Controllers (RICs) for deploying artificial intelligence (AI)-driven wireless applications \cite{AIRAN}. This architecture supports AI-RAN, where machine learning optimizes spectrum usage, interference management, and network orchestration. By embedding AI at its core, 6G will provide resilient, scalable, and intelligent wireless connectivity for future IoT ecosystems.
\color{black}
\color{black}Wireless \color{black} communication researchers have \color{black}therefore \color{black} made efforts to upscale the current infrastructure to manage this \color{black}additional \color{black} demand by increasing capacity, improving energy efficiency, and utilizing more of the electromagnetic spectrum \cite{gupta5G}. Ultimately, wireless communication providers must develop intelligent architectures and systems that can readily scale to meet increasing demand and be able to increase in effectiveness and efficiency as the amount of data \color{black}they process \color{black} increases \cite{13dongsurveyDL}.

\subsection{Physical Layer in Wireless Communication Systems}
\color{black}A \color{black} wireless communication system\color{black}'s physical layer \color{black} is responsible for converting data into radio or optical signals through modulation and encoding, and performing the reverse operation upon reception. The PHY consists of three distinct elements: the transmitter, the transmission medium, and the receiver. Traditionally, the PHY uses pre-defined protocols and techniques to transmit and receive data over a chosen channel \color{black}and relies \color{black} on well-defined mathematical models and algorithms to perform tasks such as modulation/demodulation, channel \color{black}encoding\color{black}/decoding, and channel \color{black}estimation. \color{black}These models are typically linear, stationary, and characterized by Gaussian statistics \cite{1hoydis}. However, real-world systems have many imperfections and nonlinearities, such as nonlinear power amplifiers and finite resolution quantization \color{black}that \color{black} can only be approximately captured by \color{black}these \color{black} models \cite{TschenkRF}.

\begin{figure}[t]
\centering{\includegraphics[width=\columnwidth]{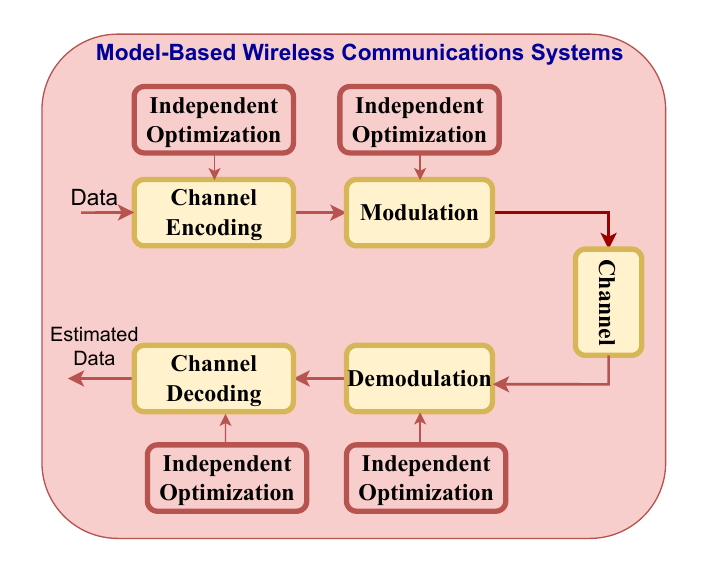}}
\caption{\color{black}Communication chain in a model-based system.\color{black}}
\label{fig:modelcomms}
\end{figure}

\color{black}
Furthermore, the transmitter and receiver that perform modulation, channel coding, channel estimation, and detection are often designed and implemented as modular functional blocks with well-defined interfaces, as shown in Fig.~\ref{fig:modelcomms}. While joint coding–modulation schemes such as coded modulation and space–time coding have been extensively studied, block-based modular design remains common in many practical and standardized systems\cite{Tarokh1998}.
\color{black}

The optimal performance of each PHY block is then typically simulated under specific theoretical conditions that do not always match real-world channel situations \cite{LuJiangHu2022}. As a result, traditional wireless communication systems \color{black}that make use of \color{black} the ``independent block" approach \color{black}can achieve only \color{black} ``local optimization" for each \color{black}block and are unable to guarantee \color{black} optimization of the entire communication system \cite{1hoydis}. The independent optimization of these wireless communication systems, which we \color{black}refer to as \color{black} ``model-based", and their discrepancies when compared to practical hardware and real-life channel conditions could potentially result in sub-optimal end-to-end system performance  when integrated and deployed~\cite{1hoydis,2channAEmag}. Furthermore, as data traffic and network usage continues to increase, the algorithms \color{black}that are \color{black} deployed in current systems must also increase in complexity to cope with such large and heterogeneous data traffic. While complex systems \color{black}are functional, they \color{black} are also more difficult to manage and predict, \color{black}and this problem is becoming more and more \color{black} prevalent as providers scale their systems to support the recent surge \color{black}in \color{black} low-cost wireless devices \cite{11santos}. 

Consequently, \color{black}research has shifted \color{black} towards \color{black}using \color{black} machine learning (ML) techniques in the PHY of wireless communication systems, specifically with end-to-end DL\color{black}-\color{black}based models, as they do not require a tractable mathematical model and \color{black}support the \color{black} joint optimization of all \color{black}PHY \color{black} components \cite{1hoydis}.

\subsection{Emergence of Deep Learning Methods}
\color{black}The emergence of DL has revolutionized numerous fields, offering powerful data-driven solutions that can extract complex patterns and optimize decision-making processes. Unlike traditional model-based approaches, which rely on predefined mathematical formulations and domain expertise, DL enables systems to learn from vast amounts of data, making them particularly effective for tasks involving high-dimensional, nonlinear relationships. However, despite its success, DL is not a one-size-fits-all solution. Field-proven model-based methods remain essential due to their interpretability, reliability, and efficiency in many scenarios. To leverage the strengths of both paradigms, a hybrid approach that combines expert knowledge with the power of DL is increasingly favored \cite{hybrid}. This fusion enhances performance, ensures robustness, and bridges the gap between theoretical understanding and data-driven adaptability, enabling more effective and trustworthy intelligent systems.

\color{black}

Several complementary learning paradigms such as federated learning, split learning, semantic communication, and reinforcement learning have also attracted significant attention in recent wireless research. Although these approaches enable distributed training, privacy preservation, task-oriented compression, or control-based adaptation, they generally operate at layers above the physical layer and preserve conventional waveform, coding, and detection structures. Surveys in these areas therefore concentrate on system-level intelligence or higher-layer tasks rather than on waveform-level co-optimization. In contrast, the focus of this survey is autoencoder-based physical-layer design, which integrates modulation, coding, and receiver processing into a single differentiable transceiver architecture. By unifying architectural principles, training methodologies, robustness considerations, and deployment pathways for AE-based systems, this work addresses a gap that is not covered by existing surveys on distributed or semantic communication frameworks and provides a dedicated, up-to-date tutorial on end-to-end learning for the physical layer.

\color{black}

A transformative approach in computational research, DL has gained significant attention as a specialized subset of machine learning. Its impact has been widely recognized, highlighted by the awarding of the 2024 Nobel Prize in Physics to John J. Hopfield and Geoffrey Hinton for their pioneering contributions to AI. \color{black}
The backbone of DL models is the deep neural network (DNN), \color{black}which is \color{black} a type of artificial neural network (ANN) that contains more than one hidden layer \cite{sze2017}.
DNNs allow for the automatic extraction of valuable information from data, known as feature extraction, which allows for the identification and classification of complex correlations that would typically be too intricate for humans to discern manually \cite{4mobileDLsurvey}. Therefore, DL has proven \color{black}itself to be very valuable \color{black} in fields such as natural language processing, computer vision, and speech recognition \color{black}because it is able \color{black} to solve complex problems that require discovering hidden patterns, or features, in data \cite{minaee2022}. Managing tasks that require advanced feature extraction using only rigid mathematical models or robust algorithms \color{black}would be very difficult, but DL is able to \color{black} accomplish these tasks at \color{black}accuracy levels far beyond what humans can achieve\color{black} \cite{deeprectifiers}. This feature extraction capability is what enables the performance of DL models to increase as the dataset they are trained on gets larger \cite{13dongsurveyDL}, meaning DL-based models, which we \color{black}refer to as \color{black} ``learning-based", \color{black}do \color{black} not suffer from the same complexity-related issues \color{black}that \color{black} traditional model-based systems do \color{black}when deployed in a PHY\color{black}.

Since DNNs are trained to minimize a loss function, implementing DL into the transmitter or receiver of a communication system \color{black}can serve to greatly improve its task-related \color{black} issues \cite{deepunfolding}\color{black}, \color{black} such as symbol detection in the presence of noise and interference in the case of the receiver \cite{sharmaUWB2021}, \color{black}and \color{black} modulation and coding in the case of the transmitter \cite{4mobileDLsurvey}. However, when the transmitter and receiver are trained together,  the entire communication system can be optimized for a common goal, such as minimizing the block error rate (BLER) or bit error rate (BER) \cite{1hoydis}. \color{black}Integrating \color{black} both the transmitter and \color{black}the receiver in \color{black} a ``black box" DL model that can train them simultaneously \color{black}makes it possible to optimize \color{black} the communication system in an end-to-end manner rather than in isolated blocks, which has shown performance improvements compared to conventional methods in the literature \cite{1hoydis, 4mobileDLsurvey, 68erpek}.

\begin{figure}[t]
\centering{\includegraphics[width=\columnwidth]{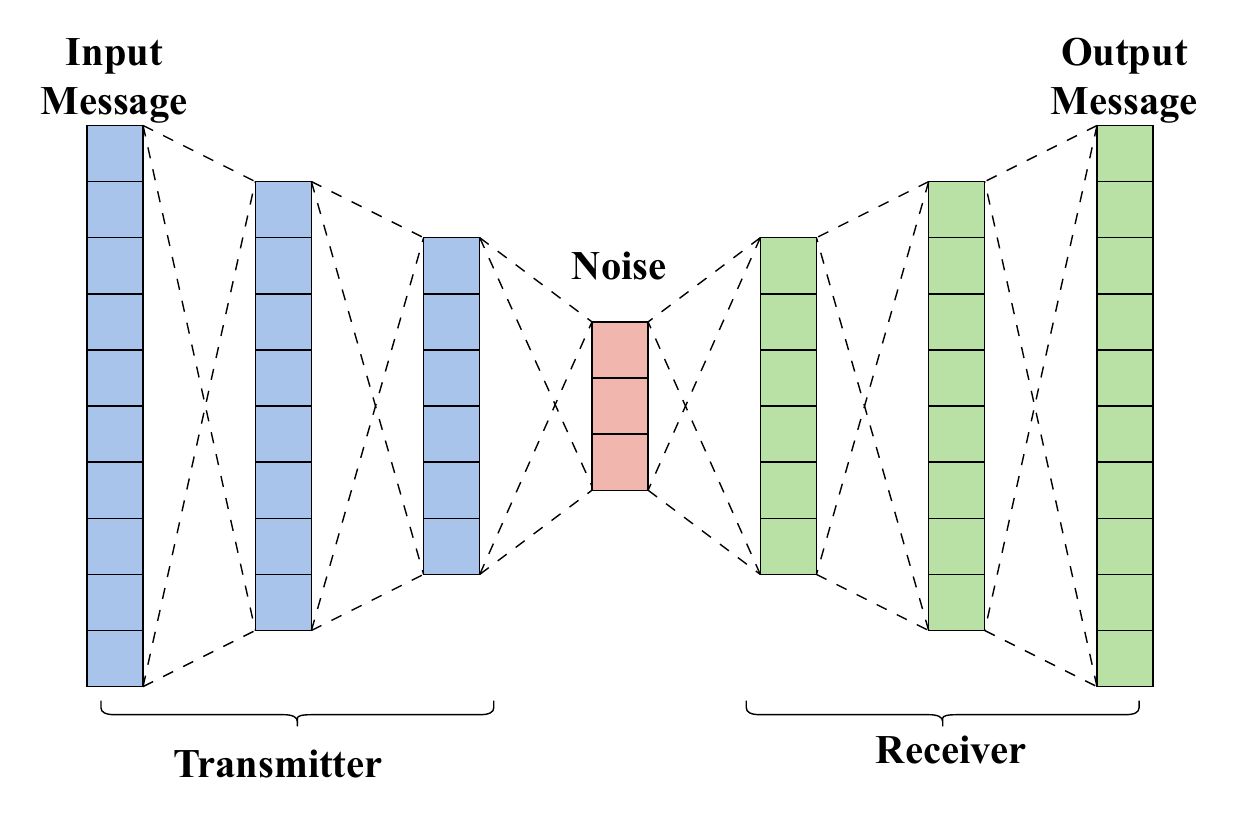}}
\caption{\color{black}Structure of an end-to-end DL based commmunications system modeled as an autoencoder.\color{black}}
\label{fig:basicAE}
\end{figure}

\color{black}An autoencoder (AE) model is typically used to apply \color{black} DL to the PHY \color{black}to optimize \color{black} end-to-end performance \color{black}because its structure \color{black} closely resembles the architecture of a simple communication system \cite{8goodfellow} \color{black}and \color{black} is shown in Fig. \ref{fig:basicAE}.
\FloatBarrier
\color{black}
\color{black}An AE's goal is typically \color{black} to compress an input vector into \color{black}a vector \color{black} with lower dimensionality, then decompress this lower dimension vector to reconstruct the input with a minimum amount of errors via minimizing its loss function. This structure \color{black}makes it possible for \color{black} the AE to learn how to prioritize which aspects (or features) of the input should remain in the compressed vector to maximize \color{black}the success of reconstruction\color{black} \cite{8goodfellow}. \color{black}Using\color{black} this approach \color{black}makes it possible to apply the model \color{black} to channels and situations where the optimal solution is unknown \cite{1hoydis}, \color{black}which is \color{black} especially useful in real-world scenarios where the channel’s transfer function can vary greatly.

While interest in applying DL for end-to-end PHY optimization is increasing, research contributions are still widely scattered across different areas and\color{black}, \color{black} to the best of the authors' knowledge, these research efforts have not been surveyed before. This paper closes this gap by offering an extensive and up-to-date review that connects learning-based AE models with their applications in the PHY to create a learned end-to-end communication system. Our ultimate goal is to offer \color{black}ML researchers \color{black}and wireless communication engineers who \color{black}wish \color{black} to use DL to address their \color{black}respective \color{black} problems of interest \color{black}a definitive guide to applying DL for end-to-end PHY optimization\color{black}.

\subsection{Survey Organization}
\begin{figure*}[ht!]
\begin{center}
\includegraphics[]{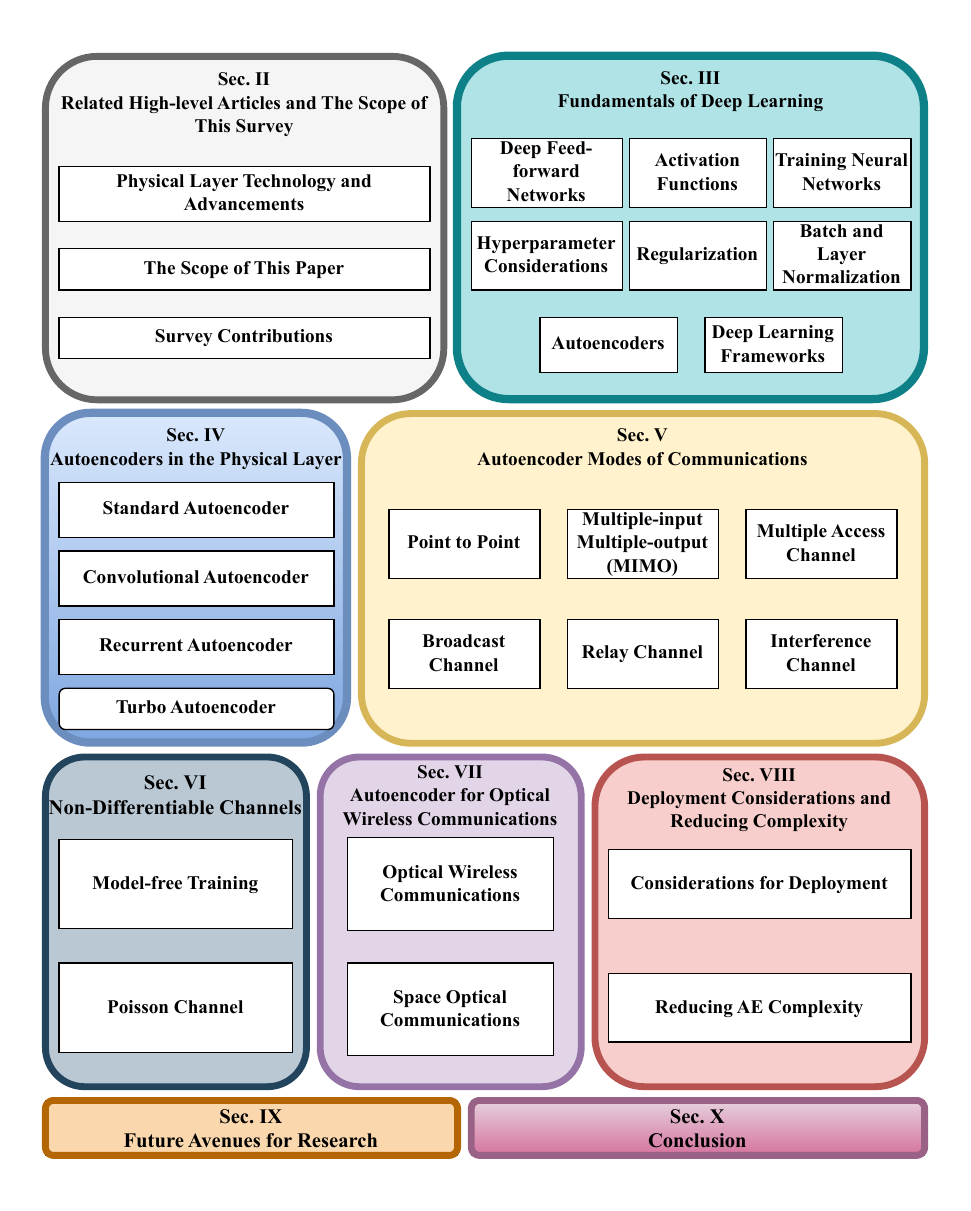}
\caption{\color{black}Diagramatic representation of section structure.}
\label{fig:outline}
\end{center}
\end{figure*}

The rest of this survey is structured in a top-down manner as shown in \color{black}Fig. \color{black} \ref{fig:outline}. 
In Section II\color{black}, \color{black} we discuss relevant surveys, books, magazines, and research papers\color{black}; determine \color{black} their relevance to the topics of DL, AEs, and end-to-end optimization\color{black}; and present \color{black} the scope and contributions of this paper. 
\color{black}The fundamentals of \color{black} DL and DNNs \color{black}that are \color{black} needed to better \color{black}understand \color{black} the topics discussed in the rest of the paper is provided in Section III\color{black}. \color{black} Section IV \color{black}focuses \color{black} on AE models and frameworks for the PHY.
In Section V, we go over state-of-the-art AE models used for several \color{black}different \color{black} modes of communication, along with comparing their relevant research papers and results.
In Section VI, we provide background on using AE models in the presence of non-differentiable channels, such as \color{black}a \color{black} Poisson channel, and state-of-the-art AE models \color{black}for these channels\color{black}. 
Section VII discusses using AEs in optical wireless communication (OWC) applications, focusing on applying such models for communication with spacecraft, known as space optical communication (SOC).
Section VIII provides a look into relevant research areas that still need to be expanded upon \color{black}in light of \color{black} real world deployment and complexity challenges.
Finally, we conclude the paper in Section IX with our closing remarks.

\section{Related High-Level Articles and \color{black}Survey Scope\color{black}}
We categorize relevant \color{black}works into three categories \color{black} \emph{(i)} tutorials and overviews focused solely on DL\color{black}, \color{black} \emph{(ii)} works \color{black}that review \color{black} recent advances in PHY technology \color{black}and relevant techniques, and \color{black} \emph{(iii)} works \color{black}that discuss \color{black} using DL for end-to-end \color{black}PHY optimization\color{black}. We provide a high-level summary of these \color{black}works \color{black} in Table~\ref{tab:surveys1} and use this section to discuss the most relevant publications from each category.


\begin{table*}[htb!]
	\centering
    \begin{threeparttable}
	\caption{Summary of existing surveys, magazine papers, and books related to deep learning and wireless communication. }
	\label{tab:surveys1}
	\renewcommand{\arraystretch}{1.3}
	\begin{tabular}{|l|Y|c|c|c|c|c|}
		\hline
		\multirow{3}{*}{\textbf{Publication}} & \multirow{3}{*}{\textbf{One-sentence summary \color{black}(Year)}} & \multicolumn{5}{c|}{\textbf{Scope}} \\
		\cline{3-7}
		& & \multicolumn{2}{c|}{\textbf{Deep Learning}} & \textbf{Layer} & \multicolumn{2}{c|}{\textbf{PHY Optimization}} \\
		\cline{3-7}
		& & \makecell{\textbf{Auto-} \\ \textbf{encoders}} & \makecell{\textbf{Other} \\ \textbf{Models}} &  \makecell{\textbf{PHY}} & \makecell{\textbf{Joint} \\ \textbf{or Global}} & \makecell{\textbf{Local} \\ \textbf{or Block}} \\
		\cline{3-7}
		\hline
		Goodfellow \textit{et al.} \cite{DLbookch14} & A foundational textbook on deep learning. \color{black}(2016)\color{black} & \checkmark & \checkmark &  &  &  \\
		\hline
		Dong \textit{et al.} \cite{13dongsurveyDL} & A comprehensive survey on deep learning applications. \color{black}(2021)\color{black} & \checkmark & \checkmark &  &  &  \\
		\hline
		Lecun \textit{et al.} \cite{51lecun} & A milestone overview of deep learning. \color{black}(2015)\color{black} & \xmark & \checkmark &  &  &  \\
		\hline
		Schmidhuber \textit{et al.} \cite{53schmidhuber} & A comprehensive timeline of deep learning research. \color{black}(2015)\color{black} & \checkmark & \checkmark &  &  &  \\
		\hline
		Yu \textit{et al.} \cite{54yu} & An essential deep learning textbook. \color{black}(2014)\color{black} & \checkmark & \checkmark & \xmark &  &  \\
		\hline
		Rumelhart \textit{et al.} \cite{70backprop} & Milestone overview on back-propagation. \color{black}(1986)\color{black} & \xmark & \xmark &  &  &  \\
		\hline
		Pouyanfar \textit{et al.} \cite{55pouyanfar} & A comprehensive tutorial on deep learning. \color{black}(2019)\color{black} & \xmark & \checkmark &  &  &  \\
		\hline
		Liu \textit{et al.} \cite{52liu} & A survey on deep learning architectures \& applications. \color{black}(2017)\color{black} & \checkmark & \checkmark &  &  &  \\
		\hline
		Ioffe \textit{et al.} \cite{3batchnorm} & Milestone introduction to batch normalization. \color{black}(2017)\color{black} & \xmark & \xmark &  &  &  \\
		\hline
		Berahmand \textit{et al.} \cite{66berahmand} & A survey on autoencoders and their applications. \color{black}(2024)\color{black} & \checkmark &  &  &  &  \\
		\hline
		Zimmermann \textit{et al.} \cite{67zimmermann} & An essential definition of the physical layer. \color{black}(1980)\color{black} &  &  & \checkmark &  &  \\
		\hline
		Jurdak \textit{et al.} \cite{64jurdak} & Foundational textbook on network layer technology. \color{black}(2007)\color{black} &  &  & \checkmark &  &  \\
		\hline
		Rgheff \textit{et al.} \cite{62rgheff} &  A book on the physical layer in 5G. \color{black}(2019)\color{black} &  &  & \checkmark &  &  \\
		\hline
		Jiang \textit{et al.} \cite{road6G} & Analysis on state of the art 6G advances. \color{black}(2021)\color{black} &  &  & \xmark & \xmark & \xmark \\
		\hline
		Liu \textit{et al.} \cite{reconfig_intelligent_surfaces} & Overview on RIS/IRS research for 6G. \color{black}(2021)\color{black} &  &  & \checkmark &  & \xmark \\
		\hline
		Wang \textit{et al.} \cite{G6_wireless_chann_measure} & A survey on wireless channel measurements for 6G. \color{black}(2020)\color{black} &  &  & \checkmark &  &  \\
		\hline
		Hemadeh \textit{et al.} \cite{mmWave_considerations} & Overview on mmWave PHY design and channel models. \color{black}(2018)\color{black} &  &  & \checkmark &  &  \\
		\hline
		Yang \textit{et al.} \cite{spatialdesign} & Considerations for spatial modulation in MIMO. \color{black}(2015)\color{black} &  &  & \checkmark &  &  \\
		\hline
		Junejo \textit{et al.} \cite{57junejo} & A survey on underwater physical layer techniques. \color{black}(2023)\color{black} &  &  & \checkmark &  &  \\
		\hline
		Makki \textit{et al.} \cite{NOMAstatus_makki} & Current status on NOMA for multiple access. \color{black}(2020)\color{black} &  & & \checkmark &  &  \\
		\hline
		O'Shea \textit{et al.} \cite{1hoydis} & Essential tutorial on learning-based physical layer optimization. \color{black}(2017)\color{black} & \checkmark &  & \checkmark & \checkmark &  \\
		\hline
		Zhang \textit{et al.} \cite{4mobileDLsurvey} & A survey on deep learning in mobile networks. \color{black}(2019)\color{black} & \xmark & \checkmark & \xmark & \xmark & \xmark \\
		\hline
		Erpek \textit{et al.} \cite{68erpek} & Extensive tutorial on deep learning for end-to-end optimization. \color{black}(2020)\color{black} & \checkmark & \checkmark & \checkmark & \checkmark & \xmark \\
		\hline
		Ozpoyraz \textit{et al.} \cite{28ozpoyraz} & A survey on learning-based physical layer architectures. \color{black}(2022)\color{black} & \xmark & \checkmark & \checkmark & \checkmark &  \\
		\hline
		Shi \textit{et al.} \cite{21shi} & Learning-based optimization techniques for 6G networks. \color{black}(2023)\color{black} &  & \checkmark & \checkmark & \checkmark & \xmark \\
		\hline
		Lv \textit{et al.} \cite{39lv} & Recent review on learning-based channel estimation. \color{black}(2023)\color{black} & \xmark & \checkmark & \checkmark & \xmark &  \\
		\hline
		Szott \textit{et al.} \cite{22szott} & A survey on machine learning applications for WiFi. \color{black}(2022)\color{black} &  & \xmark & \xmark & \xmark & \xmark \\
		\hline
		Ye \textit{et al.} \cite{42ye} & Recent survey on deep learning in the physical layer. \color{black}(2024)\color{black} & \checkmark & \checkmark & \xmark & \checkmark & \checkmark \\
		\hline
		Wang \textit{et al.} \cite{41wang} & A survey on deep learning techniques for the physical layer. \color{black}(2017)\color{black} & \checkmark & \checkmark & \checkmark & \checkmark & \xmark \\
		\hline
		Qin \textit{et al.} \cite{43qin} & Brief review on deep learning techniques for the physical layer. \color{black}(2019)\color{black} & \checkmark & \checkmark & \checkmark & \checkmark & \xmark \\
		\hline
		Huang \textit{et al.} \cite{44huang} & Review on deep learning for the physical layer in 5G networks. \color{black}(2020)\color{black} & \checkmark & \xmark & \checkmark & \checkmark &  \\
		\hline
		Akrout \textit{et al.} \cite{45akrout} & Review on deep learning research for the physical layer. \color{black}(2023)\color{black} & \xmark & \checkmark & \xmark & \xmark & \xmark \\
		\hline
		Tanveer \textit{et al.} \cite{47tanveer} & Survey on machine learning for the physical layer in 5G. \color{black}(2022)\color{black} & \xmark & \checkmark & \xmark & \xmark & \xmark \\
		\hline
		Huynh \textit{et al.} \cite{50huynh} & A survey on generative AI for the physical layer. \color{black}(2023)\color{black} & \checkmark & \checkmark & \checkmark &  & \xmark \\
		\hline
		Islam \textit{et al.} \cite{38islam} & Semantic applications from end-to-end PHY optimization. \color{black}(2024)\color{black} & \checkmark & \checkmark & \checkmark & \checkmark & \xmark \\
		\hline
		Kaur \textit{et al.} \cite{9kaur} & An overview on machine learning techniques in 6G networks. \color{black}(2021)\color{black} &  & \xmark & \xmark &  &  \\
		\hline
		Santos \textit{et al.} \cite{11santos} & A review of deep learning techniques in 5G networks. \color{black}(2020)\color{black} & \checkmark & \checkmark & \xmark &  & \\
		\hline
		Akyildiz \textit{et al.} \cite{12akyildiz} & An overview on 6G network advancements. \color{black}(2020)\color{black} & \xmark & \checkmark & \checkmark &  & \xmark \\
		\hline
		Mao \textit{et al.} \cite{16mao} & A survey on deep learning applications in network layers. \color{black}(2018)\color{black} & \xmark & \checkmark & \checkmark &  & \checkmark \\
		\hline
		Nguyen \textit{et al.} \cite{37nguyen} & Broad survey on AI techniques for next-gen wireless networks. \color{black}(2021)\color{black} & \xmark & \checkmark & \checkmark &  &  \\
		\hline
		\textbf{Elfikky \textit{et al.}} & \textbf{Our Work} & \checkmark & \checkmark & \checkmark & \checkmark & \checkmark\\
		\hline
	\end{tabular}
	\hspace*{\fill} 
\begin{tablenotes}
        \item \color{black} The symbol \checkmark\ indicates a publication is in the scope of a domain; \xmark\ marks papers that do not directly cover that area, but from which readers may retrieve some related insights. Publications related to both deep learning and mobile networks are shaded.

\end{tablenotes}
\end{threeparttable}
\end{table*}

\color{black}
\subsection{Tutorials and Overviews Focused Solely on Deep Learning}
The rapid advancement of deep learning over the past decade has been driven by both theoretical breakthroughs and the availability of large-scale data and computational resources. Early milestone works, such as the seminal overview by LeCun \textit{et al.} \cite{51lecun} and the historical perspective by Schmidhuber \textit{et al.} \cite{53schmidhuber}, established the foundations of neural network-based learning and identified key challenges that continue to influence modern architectures.

Comprehensive textbooks by Deng and Yu \cite{54yu} and Goodfellow \textit{et al.} \cite{DLbookch14} provide in-depth treatments of deep learning models, training algorithms, and optimization strategies, and remain standard references for both theoretical understanding and practical implementation. Building upon these foundational resources, a number of recent survey articles have captured the rapid evolution of deep learning paradigms and addressed emerging theoretical and algorithmic developments. For example, Wei \textit{et al.} \cite{Wei2024Mem} survey memorization and generalization behaviors in deep neural networks, while Gao \textit{et al.} \cite{Gao2024EDL} review evidential deep learning frameworks for uncertainty-aware modeling. In addition, surveys such as those by Gkarmpounis \textit{et al.} \cite{Gkarmpounis2024GNN} highlight the emergence of modern architectures, including graph neural networks, that extend beyond classical convolutional and recurrent models.

Despite their importance, these tutorials, textbooks, and surveys primarily study deep learning from a domain-agnostic perspective and do not explicitly account for the unique constraints and performance objectives of wireless communication systems, such as channel uncertainty, power and hardware limitations, or physical-layer metrics. Consequently, they serve as essential background for learning-based communication system design, rather than providing direct treatments of end-to-end physical-layer optimization.
\color{black}

\subsection{Physical Layer Technology and Advancements}
While the Open Systems Interconnection (OSI) framework for network protocols was introduced nearly 45 years ago \cite{67zimmermann}, work has continued over the decades to enhance the capacity, efficiency, and security of the PHY in wireless networks. 
Jurdak \textit{et al.} \cite{64jurdak} \color{black}present \color{black} a succinct introduction into the PHY and its structure, purpose, and prevalent technologies, with an emphasis on ad-hoc networks. \color{black}A \color{black} book by Abu-Rgheff \textit{et al.} \cite{62rgheff}\color{black}, on the other hand, \color{black} provides a more general overview \color{black}of \color{black} current PHY technologies \color{black}and focus \color{black} specifically on 5G networks. Wang \textit{et al.} \cite{G6_wireless_chann_measure} \color{black}summarize \color{black} channel measurement research trends for 6G networks and describe channel characteristics and models for different scenarios. 

Jiang \textit{et al.} \cite{road6G} discuss technological advancements in the PHY made towards 6G networks, covering many different communication modes such as millimeter wave (mmWave), terahertz (THz), visible light communication (VLC), and optical wireless communication (OWC).
Makki \textit{et al.} \cite{NOMAstatus_makki} \color{black}provide \color{black} a status report on research \color{black}into \color{black} non-orthogonal multiple access (NOMA) -- \color{black}which is a hot research topic \color{black} in academia for its potential in 6G systems -- \color{black}highlight \color{black} several research trends in NOMA\color{black}, \color{black} and \color{black}compare \color{black} their performance with \color{black}an \color{black} end goal of high power efficiency and low transmission delay.
Hemadeh \textit{et al.} \cite{mmWave_considerations} provide a detailed introduction \color{black}to \color{black} design guidelines for mmWave communication systems, \color{black}compare \color{black} different channel models and their accuracy in different scenarios\color{black}, \color{black} and \color{black}discuss \color{black} the recent technological advancements that \color{black}have enabled \color{black} mmWave communication, such as hybrid beamforming in \color{black}large-scale \color{black} multiple-input multiple-output (MIMO) systems. 
Yang \textit{et al.} \cite{spatialdesign} introduce spatial modulation for MIMO communication systems and summarize the existing literature to create design guidelines for the PHY. 
Junejo \textit{et al.} \cite{57junejo} dive into a comprehensive review of underwater acoustic (UWA) PHY technologies and issues arising with underwater channel communication. 
Liu \textit{et al.} \cite{reconfig_intelligent_surfaces} discuss re-configurable intelligent surfaces (RIS), \color{black}which are \color{black} also called intelligent reflective surfaces (IRS), and their applications for 6G, highlighting the necessity for the amalgamation of 6G technology and AI.

\color{black}
\subsection{Deep Learning for End-to-End Physical-Layer Optimization}
\label{subsec:e2e_phy_dl}

The application of deep learning to physical-layer optimization gained significant momentum following the introduction of the autoencoder abstraction for communication systems by O'Shea and Hoydis \cite{1hoydis}. This seminal work demonstrated that transmitter and receiver components can be jointly optimized in an end-to-end manner using neural networks, shifting the design paradigm from block-wise model-based processing toward data-driven transceiver architectures.

Subsequent surveys and tutorials have expanded this perspective. Erpek \textit{et al.} \cite{68erpek} provide a comprehensive overview of end-to-end neural transceivers, channel modeling, and training methodologies. Ozpoyraz \textit{et al.} \cite{28ozpoyraz} review learning-based PHY architectures with emphasis on emerging 6G scenarios, while Wang \textit{et al.} \cite{41wang} and Qin \textit{et al.} \cite{43qin} examine DL techniques for optimizing both individual and joint PHY components. Broader surveys addressing 5G and 6G systems include \cite{44huang,47tanveer,21shi,45akrout}, and emerging paradigms such as generative and semantic communication are discussed in \cite{50huynh,38islam}.

Despite their breadth, existing works typically focus either on specific PHY components (e.g., channel estimation or detection) or on general AI-enabled network optimization. A unified, architecture-driven tutorial centered specifically on autoencoder-based end-to-end PHY optimization across diverse communication modes remains limited. 

In contrast, this survey systematically organizes AE-based PHY optimization methods according to (i) architectural variants and inductive biases, (ii) training objectives and protocols, (iii) robustness under channel and hardware mismatch, and (iv) deployment-oriented considerations, thereby providing a consolidated reference for both ML researchers and communication engineers.

\color{black}

\begin{table}[htb!]
	\centering
	\caption{List of abbreviations in alphabetical order.}
	\label{tab:abbreviations}
	\begin{tabular}{|C{2.1cm}||C{5.7cm}|}
		\hline
		Acronym & Explanation \\\hline\hline
		5G / 6G & 5th/6th Generation \color{black}Mobile Networks\color{black} \\\hline
		AE & Autoencoder \\\hline
		AF & Amplify-and-Forward \\\hline
		AI & Artificial Intelligence \\\hline
		ANN & Artificial Neural Network \\\hline
		AWGN & Additive White Gaussian Noise \\\hline
		BC & Broadcast Channel \\\hline
		BER & Bit Error Rate \\\hline
		BLER & Block Error Rate \\\hline
		BN & Batch Normalization \\\hline
		BP & Back-Propagation \\\hline
		BPSK & Binary Phase Shift Keying \\\hline
		CCE & Categorical Cross-Entropy \\\hline
		CNN & Convolutional Neural Network \\\hline
		CSI & Channel State Information \\\hline
		DAE & Denoising \color{black}Autoencoder\color{black} \\\hline
		DF & Decode-and-Forward \\\hline
		DL & Deep Learning \\\hline
		DNN & Deep Neural Network \\\hline
		DBN & Deep Belief Network \\\hline
		DRL & Deep Reinforcement Learning \\\hline
		FD & Full Duplex \\\hline
		FNN & Feed-\color{black}Forward \color{black} Neural Network \\\hline
		GAN & Generative Adversarial Network \\\hline
		GAI & Generative AI \\\hline
		GD & Gradient Descent \\\hline
		GPU & Graphics Processing Unit \\\hline
		(Bi-)GRU & (Bidirectional) Gate Recurrent Unit \\\hline
		HD & Half Duplex \\\hline
		IC & Interference Channel \\\hline
		IM/DD & Intensity Modulation Direct Detection \\\hline
		IoT & Internet of Things \\\hline
		LDPC & Low Density Parity Check Codes \\\hline
		LN & Layer Normalization \\\hline
		LSTM & Long Short-Term Memory \\\hline
		MAC & Multiple Access Channel \\\hline
		ML & Machine Learning \\\hline
		MLP & Multilayer Perceptron \\\hline
		MLSE & Maximum Likelihood Sequence Estimation \\\hline
		MIMO & Multiple-Input Multiple-Output \\\hline
		MSE & Mean \color{black}Squared \color{black} Error \\\hline
		NLP & Natural Language Processing \\\hline
		NN & Neural Network \\\hline
		NOMA & Non-Orthogonal Multiple Access \\\hline
		NPU & Neural Processing Unit \\\hline
		OFDM &  Orthogonal Frequency-Division Multiplexing\\\hline
		OOK & On-Off Keying \\\hline
		OWC & Optical Wireless Communication \\\hline
		P2P & Point-to-Point Communication \\\hline
		PHY & Physical Layer \\\hline
		QAM & Quadrature Amplitude Modulation \\\hline
		RBM & Restricted Boltzmann Machine\\ \hline
		ReLU & Rectified Linear Unit\\ \hline
		RNN & Recurrent Neural Network\\ \hline
		SGD & Stochastic Gradient Descent \\ \hline
		SOC & Space Optical Communication \\ \hline
		SNR & Signal-to-Noise Ratio \\ \hline
		STK & System Tool Kit \\\hline
		TMAE & Transformer Masked Autoencoder \\\hline
		VAE & Variational Autoencoder \\ \hline
		VLC & Visible Light communication \\\hline
	\end{tabular}
\end{table}

\subsection{\color{black}Survey Scope\color{black}}

The objective of this survey is to provide a comprehensive \color{black}overview of \color{black} state-of-the-art DL models for end-to-end optimization of the PHY in wireless networks. To this end we aim to answer the following key questions:

\begin{enumerate}
	\item Why is DL promising for optimizing the PHY?
	\item \color{black}Which \color{black} cutting-edge DL models \color{black}are \color{black} relevant to end-to-end PHY optimization?
	\item What are the most recent successful DL applications in the PHY optimization domain?
	\item How can researchers tailor DL-based PHY optimization for specific modes of communication?
	\item Which are the most important and promising \color{black}future directions that are \color{black} worthy of \color{black}further \color{black} study?
\end{enumerate}

\color{black} The works reviewed in Section II-A and the related literature summarized in Table I only partially answer these questions. \color{black}This survey paper goes beyond these previous works and specifically focuses on the application of DL models for end-to-end optimization of the PHY.
\color{black}This \color{black} paper serves as a valuable resource for both ML researchers and communication engineers \color{black}because it provides a structured and in-depth survey\color{black}. It bridges the gap between theoretical advancements and practical implementation, \color{black}and offers \color{black} actionable insights to accelerate innovation in wireless communication systems.
\subsection{Survey Contributions}

This survey provides a comprehensive and systematic review of end-to-end DL framework \color{black}application \color{black} in the PHY of wireless communication systems, \color{black}and consolidates and categorizes \color{black} scattered research efforts. The key contributions of this work are as follows:
\begin{itemize}
	\item We categorize and analyze various AE architectures, such as standard, convolutional, recurrent, and TurboAEs, and discuss their specific use cases \color{black}for \color{black} PHY optimization. This includes detailed insights into their structure, performance metrics, and adaptability to \color{black}various \color{black} communication environments.
	\item We highlight \color{black}how \color{black} AE-based frameworks \color{black}contribute \color{black} addressing emerging challenges in communication modes such as point-to-point \color{black}communication \color{black} (P2P), multiple-input multiple-output (MIMO), multiple access channel (MAC), interference channel (IC), relay, and optical wireless communication (OWC). \color{black}We \color{black} provide a foundational understanding of how learning-based models can be deployed across different scenarios \color{black}by synthesizing state-of-the-art research\color{black}.
	\item We critically evaluate the limitations and challenges of deploying a DL-based PHY in real-world systems. Key areas \color{black}for \color{black} future research are identified, including computational complexity, scalability, hardware limitations, and the integration of advanced neural architectures like transformers.
\end{itemize}

This work is a definitive guide for leveraging end-to-end DL in the PHY of wireless communication systems, \color{black}and provides \color{black} a roadmap for future exploration \color{black}that enables \color{black} researchers and engineers to tackle the pressing demands of next-generation communication networks.

\section{Fundamentals of Deep Learning}
\color{black}DL is a subset of ML that focuses \color{black} on using \color{black}DNNs \color{black} with multiple layers to model complex patterns in data. \color{black}Networks, which are inspired by the \color{black}human brain's structure and function, are particularly effective at tasks such as image recognition, natural language processing (NLP), and speech recognition. This section provides an overview of the fundamental concepts and components that \color{black}underpin \color{black}DL.

\textit{Notations:} \color{black}We use bold lower-case letters to denote column vectors \( \mathbf{x} \), with $x_i$ representing the \textit{i}th element, \( \|\mathbf{x}\| \) being its Euclidian norm, \( \mathbf{x}^T \) being its transpose, and \( \mathbf{x} \odot \mathbf{y} \) being the element-wise product of vectors \( \mathbf{x} \) and \( \mathbf{y} \). We use bold upper-case letters to denote matrices $\textbf{X}$, with $X_{ij}$ representing the (\textit{i}, \textit{j}) element. The set of real numbers and the set of complex numbers \color{black}are represented by \( \mathbb{R} \) and \( \mathbb{C} \), respectively. The multivariate Gaussian and complex Gaussian distributions with mean vector \( \mathbf{m} \) and covariance matrix \( \mathbf{R} \) are denoted by \( \mathcal{N}(\mathbf{m}, \mathbf{R}) \) and \( \mathcal{CN}(\mathbf{m}, \mathbf{R}) \), respectively. The Bernoulli distribution with success probability \( \alpha \) is denoted by \( \text{Bern}(\alpha) \), and the gradient vector is denoted by \( \nabla \).

\subsection{Deep Feed-Forward Networks}
\label{sub:FNN}
NNs are the best known \color{black}type of DL model\color{black}, however, only NNs with a high enough number of ``hidden" layers can be regarded as ``deep" NNs (DNNs). DNNs \color{black}make \color{black} automatic feature extraction \color{black}possible and can represent functions of increasing complexity \color{black} by adding more layers and more units, or neurons, in a layer. \cite[Ch.~6]{8goodfellow}. 

Feed-forward NNs (FNNs) form the basis of DL. 
A FNN with $\mathit{L}$ layers describes the mapping of an input vector $\mathbf{x}_{0}$ $\in$ \( \mathbb{R}^{N_{0}} \) to an output vector $\mathbf{x}_{L}$ $\in$ \( \mathbb{R}^{N_{L}} \) given by \(f(\textbf{x}_{0}; \bm{\theta})\) : \( \mathbb{R}^{N_{0}} \) $\rightarrow$ \( \mathbb{R}^{N_{L}} \), where $\boldsymbol{\theta}$ is the set of all parameters in the network \(\boldsymbol{\theta}=\{\boldsymbol{\theta}_{1},\ldots,\boldsymbol{\theta}_{L}\}\). The FNN computes the mapping of each layer $\ell$ iteratively \color{black}as follows \color{black}:
\begin{equation}
	\mathbf{x}_{\ell} = f_{\ell}(\mathbf{x}_{\ell-1};\theta_{\ell}), \quad \ell = 1, \ldots, L
	\label{eq:hoydisLmap}
\end{equation}
where (\ref{eq:hoydisLmap}) is the mapping carried out at the $\ell$th layer.
 The $\ell$th layer's parameter set is \(\theta_{\ell}=\{\mathbf{W}_{\ell},\mathbf{b}_{\ell}\}\), where $\mathbf{W}_{\ell}$ is the matrix \color{black}that represents \color{black} the weights of the connections between \color{black}the neurons in \color{black} layers $\ell-1$ and $\ell$, and $\mathbf{b}_{\ell}$ is a vector \color{black}that represents \color{black} the bias of each neuron in the $\ell$th layer.

The \color{black}FNN's \color{black}goal is to select a mapping function \( \mathit{f} \) that approximates an ideal mapping \( \mathit{f^\star} \) \color{black}and learn 
 \color{black} the parameter \color{black}set \color{black} \( \bm{\theta} \) that outputs a predicted value $\mathbf{x}_{L}$ that is as close to the ideal output $\mathbf{x}_{L}^\star$ as possible, \color{black}which is  \color{black}known as minimizing the network's loss $L(\bm{\theta})$ \color{black}and \color{black} will be discussed shortly.

The \color{black}FNN's \color{black}hidden layers lie between the input layer (\(\ell=1\)) and output layer (\(\ell=L\)), and they are \color{black}what are responsible for applying  \color{black}the transformations and feature extraction from the input vector \( \mathbf{x}_{0} \). The number of hidden layers in the FNN is known as the ``depth" of the network, and generally speaking, having more hidden layers \color{black}enhances the FNN's ability to extract \color{black} higher-level information from the input data. However, having unnecessary hidden layers can lead to overfitting the model, \color{black}which means the model  \color{black}will learn the training data well, but not be able to generalize to new, unseen data. Overfitting is observed \color{black}when there is \color{black} a large difference between the training set error and the test set error \cite[Ch.~5]{8goodfellow}, and \color{black}is  \color{black} discussed later in this section.

\subsection{Activation Functions}
\label{sub:activations}
In order to apply DNNs to solve real-world problems, which often involve complex, non-linear relationships between inputs and outputs, we introduce non-linearity to the DNN by applying \textit{activation functions} \(\sigma(.)\) \color{black}to specific \color{black} hidden layers. 
\begin{equation}
	f_{\ell}(\mathbf{x}_{\ell-1};\theta_{\ell})=\sigma(\mathbf{W}_{\ell}\mathbf{x}_{\ell-1}+\mathbf{b}_{\ell})
	\label{eq:actweightbias}
\end{equation}
The general activation function (\ref{eq:actweightbias}) describes the mapping \color{black}of \color{black} the $\ell-1$ layer to the $\ell$th layer with an activation function. Table \ref{tab:actfunctions} provides a list of some commonly used activation functions.


\begin{table*}[t]
\centering
\caption{\color{black}Comparison of common activation functions in deep learning.\color{black}}
\label{tab:actfunctions}
\resizebox{\textwidth}{!}{%
\begin{tabular}{|l|c|c|l|l|l|}
\hline
\textbf{Activation Function} & \textbf{Formula}                     & \textbf{Range}          & \textbf{Pros}                                                    & \textbf{Cons}                                          & \textbf{Best Use Cases}                       \\ \hline
\textbf{Linear}              & \( u_i \)                            & \( (-\infty, \infty) \) & Retains raw values, useful for regression                        & No non-linearity; behaves like a linear model          & Output layers in regression tasks             \\ \hline
\textbf{ReLU}                & \( \max(0, u_i) \)                   & \( [0, \infty) \)       & Simple, computationally efficient, mitigates vanishing gradients & Can suffer from "dying ReLU" (inactive neurons)        & Hidden layers in CNNs and deep networks       \\ \hline
\textbf{Tanh}                & \( \tanh(u_i) \)                     & \( (-1, 1) \)           & Zero-centered output, better than sigmoid for deep networks      & Still susceptible to vanishing gradients               & RNNs, deep networks with balanced activations \\ \hline
\textbf{Sigmoid}             & \( \frac{1}{1 + e^{-u_i}} \)         & \( (0,1) \)             & Smooth gradient, useful for probability-based outputs            & Prone to vanishing gradient problem, not zero-centered & Binary classification, probability estimation \\ \hline
\textbf{Softmax}             & \( \frac{e^{u_i}}{\sum_j e^{u_j}} \) & \( (0,1) \)             & Outputs a probability distribution, ensuring sum equals 1        & Computationally expensive, not ideal for hidden layers & Multi-class classification in output layers   \\ \hline
\end{tabular}%
}
\end{table*}

The non-linearity \color{black}that is \color{black} introduced by activation functions is what \color{black}makes \color{black} stacking multiple hidden layers \color{black}advantageous \color{black}, as \color{black}it enables the \color{black} complex features \color{black}to be extracted \color{black} from the non-linear input data.

\subsection{Training Neural Networks}
\label{sub:training}
\color{black}In \color{black}the context of supervised learning, \color{black}NNs \color{black} are trained using training samples \color{black}that consist \color{black} of input-output pairs (\(\mathbf{x}_{0,i},\mathbf{x}_{L,i}^\star\)), where \(\mathbf{x}_{L,i}^\star\) is the desired output vector when \(\mathbf{x}_{0,i}\) is used as the input, and \(i=\{1,\ldots,S\}\) is the index of the training data sample. As 
\color{black}mentioned \color{black} earlier, the \color{black}NNs goal \color{black} is to minimize the loss $L(\bm{\theta})$, or reduce the \color{black}amount of \color{black} error between the desired output \(\mathbf{x}_{L,i}^\star\) and the observed output \(\mathbf{x}_{L,i}\) by adjusting the parameter \color{black}set \color{black} \(\bm{\theta}\):
\begin{equation}
	L(\bm{\theta})=\frac{1}{S}\sum_{i=1}^S l(\mathbf{x}_{L,i}^\star,\mathbf{x}_{L,i})
\end{equation}
 Common loss functions include \color{black}mean squared error \color{black} (MSE) for regression tasks and \color{black}categorical cross-entropy \color{black} (CCE) loss for classification tasks, as shown in Table \ref{tab:loss_functions}. 


\begin{table*}[t]
\centering
    \caption{\color{black}Comparison of common loss functions used in neural networks.\color{black}}
    \label{tab:loss_functions}
\resizebox{\textwidth}{!}{%
\begin{tabular}{|l|c|l|l|l|}
\hline
\textbf{Loss Function} &
  \textbf{Formula} &
  \textbf{Pros} &
  \textbf{Cons} &
  \textbf{Best Use Cases} \\ \hline
\textbf{Mean Squared Error (MSE)} &
  \( \| \mathbf{u} - \mathbf{v} \|^2 \) &
  Penalizes large errors, smooth gradient &
  Sensitive to outliers &
  Regression tasks \\ \hline
\textbf{Mean Absolute Error (MAE)} &
  \( \sum | u_i - v_i | \) &
  Robust to outliers &
  Non-smooth gradient, slow convergence &
  Regression with noise and outliers \\ \hline
\textbf{Huber Loss} &
\(\begin{cases} 
              \frac{1}{2} (u - v)^2, & |u - v| \leq \delta \\
              \delta (|u - v| - \frac{1}{2} \delta), & |u - v| > \delta
          \end{cases} \)&
  Balances MSE and MAE &
  Requires tuning \(\delta\) &
  Regression with moderate outliers \\ \hline
\textbf{Categorical Cross-Entropy} &
  \( -\sum_j u_j \log v_j \) &
  Works well for multi-class classification &
  Not ideal for regression tasks &
  Multi-class classification \\ \hline
\textbf{Binary Cross-Entropy} &
  \( - u \log v - (1 - u) \log (1 - v) \) &
  Effective for binary classification &
  Cannot handle multi-class problems &
  Binary classification \\ \hline
\textbf{Kullback-Leibler (KL) Divergence} &
  \( \sum_j u_j \log \frac{u_j}{v_j} \) &
  Measures information loss between distributions &
  Asymmetric, requires probability distributions &
  Variational autoencoders, probabilistic models \\ \hline
\end{tabular}%
}
\end{table*}

In training, the parameters are updated through a process called back-propagation, or the backward propagation of errors \cite{70backprop}, which consists of \color{black}a forward pass and a backward pass\color{black}. In the forward pass, the input data is propagated through the network layer by layer, \color{black}\color{black}with \color{black}\color{black} each layer \color{black}performing \color{black}\color{black} a linear transformation on the previous layer, then \color{black}going through \color{black}\color{black} a non-linear activation function. \color{black}The output of the final layer \(\mathbf{x}_{L,i}\) is then compared \color{black}with \color{black} the desired output \(\mathbf{x}_{L,i}^\star\) using the specified loss function which provides the loss value for that training sample.




In the backward pass, the partial derivatives of the gradient vector \(\nabla\) of the loss are computed with respect to each of the \color{black}network's \color{black} parameters, starting from the output layer and propagating back to the input layer. \color{black}The \color{black} partial derivatives are computed \color{black}using gradient descent \color{black} to determine how much each parameter contributes to the loss, and in what direction they should be adjusted \color{black}-- opposite the \color{black}direction of \( \nabla L(\bm{\theta}) \).

%
%
%

The \color{black}gradient descent variant that is most commonly used \color{black} in DL is called stochastic gradient descent (SGD). When \color{black}SGD is used\color{black}, the parameter \color{black}set \color{black}~\(\bm{\theta}=\{\mathbf{W},\mathbf{b}\}\) \color{black}is \color{black} initially chosen at random, with initial values \(\bm{\theta}=\bm{\theta}_{0}\), \color{black}and \color{black} \(\bm{\theta}\) is then updated iteratively.
\begin{equation}
	\bm{\theta}_{t+1}=\bm{\theta}_{t}-\alpha\nabla\tilde{\mathbf{L}}(\bm{\theta}_{t})
	\label{eq:SGDiteration}
\end{equation}
Equation (\ref{eq:SGDiteration}) represents an iteration of updating \(\bm{\theta}\), where \(\bm{\theta}_{t+1}\) is the next iteration of \(\bm{\theta}\), \(\alpha\) is the learning rate, and \(\tilde{\mathbf{L}}(\bm{\theta}_{t})\) is the approximated loss function. With SGD, the loss function is approximated every iteration over a mini-batch \(\mathcal{S}_{t}\subset\{1,2,\ldots,S\}\) of randomly chosen training samples, where the \color{black}size of the mini-batch \color{black}\(S_{t}\ll S\). The mini-batch loss \(\tilde{\mathbf{L}}(\bm{\theta}_{t})\) is calculated as the average loss over the samples in the mini-batch:
\begin{equation}
	\tilde{\mathbf{L}}(\bm{\theta}_{t})=\frac{1}{S_{t}}\sum_{i\in\mathcal{S}_{t}}^{S_{t}} l(\mathbf{x}_{L,i}^\star,\mathbf{x}_{L,i})
	\label{loss_minibatch}
\end{equation}

Using \color{black}SGD with \color{black} a small mini-batch is significantly less computationally complex than \color{black}using \color{black} full-batch gradient descent since the computations involve a much smaller number of training samples.

\subsection{Hyperparameter Considerations}




\label{sub:hype}
The hyperparameters of a DNN are the crucial settings chosen by the developer that dictate how the network will behave \color{black}and define \color{black} its architecture, training speed, and overall performance. \color{black}DNN \color{black} hyperparameters include the learning rate, \color{black}the \color{black} number of hidden layers, 
\color{black}the \color{black} number of hidden units in a layer, 
\color{black}the \color{black} activation functions, and \color{black}the \color{black} batch size.

The most important hyperparameter to tune is the learning rate. \color{black}As Goodfellow \textit{et al.} \cite{8goodfellow} \color{black}put \color{black} it, ``\textit{If you have time to tune only one hyperparameter, tune the learning rate}".\color{black} The learning rate is a scalar value that determines the \color{black} magnitude at which the parameter set \(\mathbf{\Theta}=\{\mathbf{W},\mathbf{b}\}\) \color{black}can change when they are updated. The time to convergence, \color{black}which is the amount of time it takes to reach the point at which \color{black} performance no longer improves significantly with further training, is heavily dependent on the learning rate.  A learning rate that is too small would gradually approach convergence, but the small \color{black}increments \color{black} in parameter change make it very easy to get trapped in \color{black}the \color{black} local minima of the loss gradient and would lead to a very \color{black}long \color{black} training time. On the other hand, a learning rate that is too \color{black}big \color{black} would cause the parameters to change in large \color{black}increments \color{black} and could lead to divergence or instability. 

Therefore, the best approach for optimal training is a balanced learning rate, as the training error with respect to the learning rate on a log scale. To alleviate this problem, popular ML frameworks such as PyTorch \cite{pytorch}, and \color{black}TensorFlow \color{black} \cite{tensorflow} provide built-in options for learning rate decay and 
\color{black}a \color{black} per-layer learning rate specification, which help to speed up training at the cost of having to determine an optimal decay schedule. 

Optimization techniques such as SGD are heavily dependent on the learning rate, as \color{black}is \color{black} shown in Equation (\ref{eq:SGDiteration}). Luckily, advanced SGD optimizers such as Adam \cite{69adam} are able to dynamically scale the base learning rate $\alpha$ for different layers, \color{black}which avoids \color{black} having to choose them manually. A comparison of optimization methods for backpropagation is shown in Table \ref{tab:optimization_methods}.

\begin{table*}[t]
\centering
    \caption{\color{black}Comparison of optimization methods for backpropagation.\color{black}}
    \label{tab:optimization_methods}
\resizebox{\textwidth}{!}{%
\begin{tabular}{|l|l|l|l|}
\hline
\textbf{Optimization Method} &
  \textbf{Key Features} &
  \textbf{Pros} &
  \textbf{Cons} \\ \hline
\textbf{Stochastic Gradient Descent (SGD)} &
  Updates weights using single samples &
  Computationally efficient, fast convergence &
  High variance, requires tuning learning rate \\ \hline
\textbf{Momentum SGD} &
  Adds a velocity term to update rule &
  Faster convergence, reduces oscillations &
  Still requires careful tuning of hyperparameters \\ \hline
\textbf{Adagrad} &
  Adaptive learning rate per parameter &
  No need for manual learning rate tuning &
  Accumulates past gradients, leads to vanishing updates \\ \hline
\textbf{RMSprop} &
  Uses moving average of squared gradients &
  Works well in non-stationary problems &
  Requires tuning decay parameter \\ \hline
\textbf{Adam (Adaptive Moment Estimation)} &
  Combines Momentum and RMSprop &
  Adaptive learning rates, widely used &
  Computationally expensive, can overfit \\ \hline
\textbf{Nadam (Nesterov-accelerated Adam)} &
  Adds Nesterov momentum to Adam &
  Faster convergence in some cases &
  More computational overhead \\ \hline
\end{tabular}%
}
\end{table*}

Another important hyperparameter to consider is the number of hidden layers used. As mentioned in Section \ref{sub:FNN}, having a greater number of hidden layers can increase the representational capacity of a DNN, at the cost of a greater chance of overfitting. Conversely, too few hidden layers could \color{black}result in insufficient \color{black}  representational capacity and lead to underfitting. This tradeoff between model capacity and under/overfitting is analogous to the relationship between training error and learning rate.


\color{black}There is clearly \color{black} a ``sweet spot" \color{black}when it comes to \color{black} choosing learning rate and model capacity, both of which must be optimized \color{black}to get \color{black} the best performance out of a DNN. Regardless of \color{black}the \color{black}techniques employed through optimizer or DL frameworks, care and effort must still be taken in \color{black}hyperparameter tuning \color{black} in order to achieve optimal performance \cite{empiricaloptimizers}.

\subsection{Regularization}
\label{subsection:regularization}

\color{black}Regularization techniques are used in conjunction with hyperparameter to prevent overfitting. \color{black} Generally, regularization can be defined as any modification made to the learning algorithm that reduces generalization error, but does not reduce training error \cite[Ch.~7]{8goodfellow}. This is usually done by adding a penalty term to the model's loss function.

This \color{black}increase in \color{black} generalization error \color{black}while \color{black} simultaneously decreasing training error \color{black}is \color{black} known as overfitting \color{black}and \color{black} results from a NN \color{black}having overly high \color{black} variance, \color{black}which means \color{black} that the network's outputs are very sensitive to the training samples that are \color{black}input. \color{black} DNNs in particular are very susceptible to overfitting due to the large number of parameters present in the hidden layers, and so it follows that as we increase the number of hidden layers and units, the model's variance and likelihood of overfitting \color{black}increase \color{black} as well.



Regularization techniques are used to decrease the model's variance by decreasing its representational \color{black}capacity\color{black}. \color{black}Commonly used techniques include \color{black}L1 regularization, L2 regularization, dropout regularization, and early stopping, which \color{black}is discussed later \color{black} in this section.

L1 regularization, \color{black}which is \color{black} also called norm or lasso regularization, adds a penalty term to the model's loss function \color{black}that corresponds \color{black} to the absolute value of the model's weights:
\begin{equation}
	\text{L1 loss} = l(\mathbf{u},\mathbf{v}) + \lambda \sum_{i,j=1} |\mathbf{W}_{i,j}|
\end{equation}
where \(\lambda\) is a hyperparameter \color{black}that defines \color{black} a regularization penalty ranging from \(0\) (no penalty) to \(1\) (full penalty). L1 regularization encourages sparsity in the model \color{black}when it comes to \color{black} small weights, \color{black}so they get \color{black} closer and closer to 0 in \color{black}each \color{black} training step given a high enough \color{black}value of \color{black} \(\lambda\). This helps prevent the model from learning irrelevant features that are only present in the training data. 

Conversely, L2 regularization\color{black}, which is known as ridge regression or weight decay, \color{black} reduces the prevalence of large weights, as it adds a penalty term corresponding to the squared magnitude of each weight:
\begin{equation}
	\text{L2 loss}=l(\mathbf{u},\mathbf{v})+\lambda \sum_{i,j=1}(\mathbf{W}_{i,j})^2
\end{equation}
In doing so, L2 regularization \color{black}helps \color{black} the model \color{black}to not assign \color{black} excessively large values to weights that represent the training data features well but \color{black} introduce generalization error\color{black}. L1 regularization is used when it is important to preserve sharp discontinuities in the training data, whereas L2 \color{black}regularization \color{black} introduces a smoothening effect as the magnitude of weights get spread out more evenly.

Another method for introducing sparsity in the model is called dropout, which is a form of regularization that randomly sets some units and weights to zero, or ``drops them out". This essentially forces the network to not rely solely on certain units, \color{black}which makes \color{black} it more robust to noise and \color{black}enables \color{black} it to better generalize to unseen data. When a \color{black}hidden or visible \color{black} unit is dropped out, \color{black}it and \color{black}all of its ingoing and outgoing connections are temporarily removed from the network. \color{black}Each \color{black} unit in the network has a probability \(p\) \color{black}of being \color{black} dropped out \color{black}at each training step\color{black}, where \(p\) \color{black}can either be \color{black} a chosen hyperparameter or 0.5, which is known to be optimal for a wide range of tasks \cite{JMLR:dropout}. In testing, the weights of each unit \color{black}were \color{black} scaled down by their ``keep probability", which is \(1-p\), to account for the reduced number of units, on average, used in training. \color{black}Despite being simple, \color{black} dropout regularization has become common practice \color{black} in DL research as it has proven \color{black}itself \color{black} to be a very effective way of preventing overfitting \color{black}either when \color{black} used alone or in combination with other regularization techniques. \color{black}  


\subsection{Batch and Layer Normalization}
\label{sub:normalization}

Training DNNs is challenging because as the model learns and updates its parameter \color{black}set \color{black} \(\theta_{\ell}=\{\mathbf{W}_{\ell},\mathbf{b}_{\ell}\}\), the data flowing through the network keeps changing its statistical properties, or distributions. Each layer's input data distribution changes because the previous layers' outputs are changing, which makes the training process more complex, \color{black}and requires smaller learning \color{black} rates, and consequently, \color{black}longer \color{black} training times. This problem is known as \textit{internal covariate shift} \cite{3batchnorm}\color{black}, and one way \color{black} to reduce this issue is \color{black}by normalizing \color{black} neuron activations.

The two main methods \color{black}used \color{black} for this normalization are batch normalization (BN) \cite{3batchnorm} and layer normalization (LN) \cite{LayerNormalization}.  
BN works by normalizing the neuron activations \(x\) in a layer over a mini-batch \(\mathcal{B}=\{x_{1},\ldots,x_{m}\}\) \color{black}that contains \color{black} \textit{m} samples to make each activation have a mean \(\mu_{\mathcal{B}}=0\) and variance \(\sigma_{\mathcal{B}}^2=1\) \color{black}as follows\color{black}:
\begin{equation}
	\begin{aligned}
		\mu_{\mathcal{B}} &\leftarrow \frac{1}{m} \sum_{i=1}^m x_{i}  \\
		\sigma_{\mathcal{B}}^2 &\leftarrow \frac{1}{m} \sum_{i=1}^m (x_{i}-\mu_{\mathcal{B}})^2 \\
		\hat{x_i} &\leftarrow \frac{x_{i}-\mu_{\mathcal{B}}}{\sqrt{\sigma_{\mathcal{B}}^2 +\epsilon}}
	\end{aligned}
\end{equation}
where \(\hat{x_i}\) is a batch-normalized activation and \(\epsilon\) is a constant \color{black}that is \color{black} added to the variance for numerical stability \cite{3batchnorm}. 
However, in order to maintain the original representation of the layers that \color{black}receive \color{black} the normalized inputs, two new  parameters, \(\gamma\) and \(\beta\), must be learned: 
\begin{equation}
	y_i \leftarrow \gamma \hat{x_i}+\beta
\end{equation}
where \(y_i\) is the scaled and shifted normalized value that \color{black}is \color{black} passed onto the next layer.
\color{black}It \color{black} is known as the BN transform \cite{3batchnorm} \color{black}since \color{black} \(y_i=BN_{\gamma,\beta}(x_i)\) is passed on to the next layer.

\color{black}BN's effectiveness at reducing \color{black}covariate shift is directly proportional to the batch size, \color{black}so \color{black} when small batch sizes are needed, or when dealing with sequential data, LN is used \color{black}instead\color{black}\cite{LayerNormalization}.
LN is essentially the transpose of BN, as \color{black}it \color{black} normalizes the activations of all the units in a layer for every sample \color{black} instead of normalizing each unit's activation independently over a range of samples. \color{black} This gives LN the advantage of being completely independent \color{black}of \color{black} batch size \color{black}and makes \color{black} it especially useful for recurrent neural networks (RNNs), where the summed inputs to the recurrent neurons often vary with the length of the sequence, and \color{black}for \color{black} online learning, where batch sizes are inherently required to be small. However, in networks with fixed models and offline learning, BN has been shown to be more effective \cite{3batchnorm},\cite{LayerNormalization}. A comparison of other common regularization techniques is given in Table \ref{tab:regularization_methods}.

\begin{table*}[t]
\centering
    \caption{\color{black}Comparison of common regularization techniques in deep learning.\color{black}}
    \label{tab:regularization_methods}
\resizebox{\textwidth}{!}{%
\begin{tabular}{|l|l|l|l|}
\hline
\textbf{Regularization Method}     & \textbf{Key Features}                        & \textbf{Pros}                                & \textbf{Cons}                     \\ \hline
\textbf{L1 Regularization (Lasso)} & Adds \( \lambda \sum |w| \) to the loss      & Induces sparsity, feature selection          & Can lead to weight instability    \\ \hline
\textbf{L2 Regularization (Ridge)} & Adds \( \lambda \sum w^2 \) to the loss      & Prevents large weights, stabilizes training  & Does not induce sparsity          \\ \hline
\textbf{Dropout}                   & Randomly deactivates neurons during training & Reduces overfitting, efficient               & Needs tuning dropout rate         \\ \hline
\textbf{Early Stopping} &
  Stops training when validation loss plateaus &
  Avoids overfitting without modifying model &
  Requires monitoring validation loss \\ \hline
\textbf{Batch Normalization} &
  Normalizes activations per mini-batch &
  Stabilizes training, speeds up convergence &
  Adds extra computation, may reduce diversity \\ \hline
\textbf{Data Augmentation}         & Artificially increases dataset size          & Reduces overfitting without extra parameters & Can be computationally expensive  \\ \hline
\textbf{Weight Pruning}            & Removes redundant weights post-training      & Reduces model size for deployment            & Requires careful threshold tuning \\ \hline
\textbf{Knowledge Distillation} &
  Trains a smaller model from a larger one &
  Reduces complexity while maintaining accuracy &
  Needs a high-performing teacher model \\ \hline
\end{tabular}%
}
\end{table*}

\subsection{Autoencoders}
\label{sub:autoencoders}
 \color{black} Autoencoders (AEs) are unsupervised learning models consisting of an encoder and a decoder. The encoder compresses data by reducing its dimensionality and capturing essential features, while the decoder performs the inverse operation to reconstruct the original data. The compression ratio directly impacts the model’s ability to retain essential features while discarding redundancy. A smaller bottleneck enforces a more compact representation, promoting generalization but potentially causing information loss and degraded reconstruction quality. On the other hand, a larger bottleneck preserves more details but may lead to overfitting and less effective feature extraction. Choosing the right bottleneck size is crucial for balancing compression efficiency and reconstruction fidelity, depending on the application requirements. \color{black}  The encoder can be represented as a mapping \textit{f} that maps the input $\mathbf{x}$ $\in$ \( \mathbb{R}^{N_{x}} \) to an internal representation $\mathbf{h}$~$\in$~\( \mathbb{R}^{N_{h}} \), where \({N_{x}}>{N_{h}}\):
\begin{equation}
	\mathbf{h}=f(\mathbf{x})
\end{equation}
The decoder is then described as a mapping \textit{g} that maps the internal representation \(\mathbf{h}\) to the reconstructed output \(\mathbf{\hat{x}}\in\mathbb{R}^{N_{\hat{x}}}\), where \({N_{\hat{x}}}={N_{x}}\):
\begin{equation}
	\mathbf{\hat{x}}=g(\mathbf{h})
\end{equation}

By making the internal representation \(\mathbf{h}\) have a smaller dimension than the input \(\mathbf{x}\), the AE is able to learn the most notable and important \color{black}features \color{black} of the input data that \color{black}enable it to accurately \color{black} reconstruct the input.
The formal \textit{autoencoder problem}, \color{black}which is \color{black} defined by the authors \color{black}of\color{black} \cite{baldiAE}, is to learn the mapping functions \(f: \mathbb{R}^{N_{x}} \rightarrow \mathbb{R}^{N_{h}}\) and \(g: \mathbb{R}^{N_{h}} \rightarrow \mathbb{R}^{N_{\hat{x}}}\) that satisfy:
\begin{equation}
	\arg\min_{f,g} E[\Delta(\mathbf{x}, g \circ f(\mathbf{x}))]
	\label{eq:AEproblem}
\end{equation}

\color{black}

\subsection{Design Guidelines for AE-Based PHY Systems}

Designing an autoencoder-based physical layer requires aligning the learning model with the operational constraints of the wireless interface. The process begins by fixing the communication requirements, including spectral efficiency, reliability, latency, and blocklength. For a message set of size \(M\), the number of information bits is
\begin{equation}
    k = \log_{2} M,
\end{equation}
and the corresponding code rate is
\begin{equation}
    R = \frac{k}{n},
\end{equation}
where \(n\) denotes the number of complex channel uses available to each transmitted codeword. These parameters are determined jointly by the intended channel family (AWGN, fading, frequency-selective, or Poisson optical channels), the mobility conditions, and hardware constraints such as peak power, quantization depth, and memory.

Once the system specification is established, the AE architecture and latent dimension are selected. The encoder typically maps the information-bearing vector \(\mathbf{m}\) into a complex-valued channel input,
\begin{equation}
    \mathbf{x} = f_{\theta_{\mathrm{enc}}}(\mathbf{m}) \in \mathbb{C}^{n},
\end{equation}
subject to energy constraints such as \(\|\mathbf{x}\|^{2} \leq n P\). Choosing the latent dimension, activation functions, and overall topology depends on the nature of the transmitted signals. Fully connected architectures are sufficient for small message sets, while convolutional or attention-based models are advantageous for OFDM-like grid structures or signals with spatial/temporal correlations. In many cases, inductive biases are embedded through fixed layers; for example, OFDM systems use
\begin{equation}
    \mathbf{x} = \mathbf{F}^{H} \mathbf{h},
\end{equation}
where \(\mathbf{F}^{H}\) represents a non-trainable IDFT. Optical channels may additionally enforce nonnegativity constraints implemented through activation functions such as ReLU. These architectural decisions constitute the primary degrees of freedom in AE parameter selection.

A suitable channel model is then placed in the AE bottleneck. Differentiable channels such as AWGN or standard fading are represented as
\begin{equation}
    \mathbf{y} = \mathcal{H}(\mathbf{x}) + \mathbf{w},
\end{equation}
with \(\mathbf{w} \sim \mathcal{CN}(0, N_{0} \mathbf{I})\). For non-differentiable or hardware-dependent channels, surrogate models, reinforcement learning estimators, or over-the-air (OTA) refinement strategies are required. The training objective must reflect the communication task. For message detection, categorical cross-entropy,
\begin{equation}
    \mathcal{L}_{\mathrm{CCE}} = - \sum_{i} m_{i} \log(\hat{m}_{i}),
\end{equation}
is commonly used, while joint source–channel coding employs distortion-oriented losses such as mean-squared error,
\begin{equation}
    \mathcal{L}_{\mathrm{MSE}} = \|\mathbf{x} - \hat{\mathbf{x}}\|^{2}.
\end{equation}

Training protocol choices determine the AE’s generalization performance. Practical configurations typically use mini-batch stochastic gradient descent with batch sizes \(S_{t} \in [32,256]\) and learning rates in the range \(\alpha \in [10^{-4},10^{-2}]\). To avoid overfitting the AE to an idealized channel, training SNRs are drawn from a range,
\begin{equation}
    \text{SNR}_{\mathrm{train}} \sim \text{Uniform}(\text{SNR}_{\min}, \text{SNR}_{\max}),
\end{equation}
and channel impairments such as carrier-frequency offset, sampling jitter, amplifier nonlinearities, and quantization effects are introduced through data augmentation. Regularization methods such as weight decay,
\begin{equation}
    \theta \leftarrow \theta - \alpha \nabla \mathcal{L}(\theta) + \lambda \theta,
\end{equation}
dropout, or early stopping improve robustness when the AE has high modeling capacity.

After training, the learned transceiver must be benchmarked against conventional PHY baselines—typically coded-modulation schemes with maximum-likelihood or near-optimal receivers. Evaluation should consider BER/BLER, computational complexity, decoding latency, and robustness to model mismatch. Performance degradation under mismatch may be quantified as
\begin{equation}
    \Delta_{\mathrm{mismatch}} = \mathrm{BER}_{\mathrm{real}} - \mathrm{BER}_{\mathrm{sim}},
\end{equation}
highlighting practical deployment considerations.  

These guidelines establish a clear set of steps for designing and training AE-based transceivers and directly support the application-specific architectures examined in Sections~IV–VIII.

\subsection{Robustness, Generalization, and Practical Validation of AE-Based PHY Systems}

A critical challenge in deploying AE-based transceivers is ensuring robustness when channel conditions deviate from the training distribution. Since the encoder and decoder are jointly optimized for the stochastic models used during training, mismatch between simulated and real channels can degrade performance. This motivates explicit strategies for improving generalization across SNRs, mobility regimes, hardware impairments, and propagation environments.

A common approach is to employ domain-randomized training, where the channel operator is sampled from a distribution of models rather than a fixed realization. For example, fading parameters such as Doppler spread, delay profiles, and Rician \(K\)-factors can be randomized across training batches; hardware effects such as nonlinear amplification, phase noise, quantization, and oscillator drift may be injected as stochastic augmentations. If the training set spans a sufficiently broad family of impairments, the resulting AE approximates a robust mapping
\begin{equation}
    f_{\theta}^{\star} = \arg \min_{\theta} \; \mathbb{E}_{\mathcal{H} \sim \mathcal{D}} 
    \big[ \mathcal{L}(g_{\theta}(\mathcal{H}(f_{\theta}(\mathbf{m}))), \mathbf{m}) \big],
\end{equation}
where \(\mathcal{D}\) denotes the distribution of channel variations expected in deployment.

Practical validation further requires evaluation under real measurements or over-the-air (OTA) testing. This can be accomplished via two complementary methods. The first uses recorded channel traces or hardware-in-the-loop testbeds to replace the differentiable channel model during the forward pass, enabling realistic but repeatable evaluation. The second employs online fine-tuning or model-free adaptation, where the AE parameters are updated using gradients estimated from empirical loss feedback, allowing the model to adjust to unmodeled distortions encountered in deployment.

Robustness is ultimately demonstrated by measuring generalization gaps across synthetic and real channels:
\begin{equation}
    \Delta_{\mathrm{gen}} = \mathrm{BER}_{\mathrm{OTA}} - \mathrm{BER}_{\mathrm{sim}},
\end{equation}
and by verifying stable performance under dynamic mobility, time-varying fading, and hardware mismatch. These considerations are essential for bridging the gap between simulation-trained AEs and commercially deployable 6G physical-layer systems, and will be revisited throughout Sections~IV–VIII when discussing AE architectures, training strategies, and evaluation methodologies.

\subsection{Relation to Adjacent Learning Paradigms}
AE-based end-to-end learning for the physical layer should be viewed in the broader context of recent learning paradigms for wireless networks. Federated and split learning provide communication-efficient and privacy-aware mechanisms for distributed training across many devices, and they have been widely studied in next-generation wireless and IoT settings~\cite{Guo2024FTLsurvey,Wazzeh2025DynamicSFL}. Semantic communication, on the other hand, seeks to transmit task-relevant meaning rather than raw bits, and recent surveys and vision articles have positioned it as a key enabler of AI-native 6G architectures~\cite{Xin2024SemanticSurvey,Chaccour2025NextGenSemCom,Ogenyi2025AINative6G}. These frameworks typically operate at layers above the physical layer and preserve conventional waveform, coding, and detection blocks. In contrast, the focus of this tutorial is on autoencoder-based physical-layer design, where modulation, coding, and receiver processing are jointly represented by a single differentiable transceiver. As a result, the design questions, architectural choices, and training methodologies considered in this survey are complementary to, rather than overlapping with, existing surveys on federated, split, and semantic communication, and they target waveform-level co-optimization that is not addressed in those works.

\color{black}



\subsection{Deep Learning Frameworks}
\label{sub:frameworks}

In recent years, much work has been done \color{black}to improve and optimize \color{black} ML libraries such as TensorFlow \cite{tensorflow}, PyTorch \cite{pytorch}, Caffe \cite{caffe}, and MXNet \cite{mxnet}, which \color{black}make it possible to use \color{black} high-level programming languages or configuration files to define the complex algorithms that enable computers to run DNNs. These libraries support the automated differentiation of loss functions required for training, as well as \color{black}the compilation of \color{black} the forward and backward passes into hardware-optimized dense matrix algebra kernels that can be computed concurrently. \color{black}Table \ref{tab:DLframes} presents various DL frameworks that researchers can utilize for end-to-end DL designs. \color{black}
\begin{table}[ht!]
	\centering
	\caption{\color{black}Specifications of different popular DL frameworks.\color{black}}
	\label{tab:DLframes}
	\begin{tabular}{|p{1.5cm}|c|c|}
		\hline
		\textbf{Framework} & \textbf{Core Language} & \textbf{Interface Support}  \\
		
		\hline
		
		PyTorch \cite{liuStandard} & C++ & Python, C++, \& Java   \\
		
		\hline
		
		Caffe \cite{harshithaysStandard} & C++ & Python \& MATLAB \\
		
		\hline
		
		MATLAB  \cite{baleviTurbo} & MATLAB & Python, ONNX, \& C/C++   \\
		
		\hline
		
		Keras  \cite{jiangTurbo} & Python & Python \\
		
		\hline
		
		Theano  \cite{wangTurbo} & Python & Python   \\
		
		\hline
		
		Scikit  \cite{scikitlearn} & Python, Cython, \& C/C++ & Python   \\
		
		\hline
		
		RStudio  \cite{rstudio_dl_github} & C++ \& Java & R   \\
		
		\hline
		
	\end{tabular}
\end{table}
On the other hand, the proliferation of powerful graphics processing units (GPUs) \color{black}has \color{black} been essential \color{black}for \color{black} DL research, as they are exceptionally well suited \color{black}to perform \color{black} multiple computations simultaneously. GPUs were originally designed to handle the parallel nature of graphics rendering, since tasks such as shading, texturing, and rendering multiple pixels simultaneously require handling thousands of operations in parallel. Similarly, running NNs \color{black}involves \color{black} many matrix multiplications, \color{black}which means \color{black} the activations for each neuron in a layer can be \color{black}computed \color{black} independently and concurrently, 
\color{black}and the process is \color{black} parallelizable. 
Furthermore, due to the recent surge \color{black}in \color{black} interest in DL research, computer hardware companies have developed \color{black}specific hardware components \color{black} that are better optimized for running NNs \cite{nvidia_tesla_t4}, \cite{googleTPU}.

\color{black} Although the end-to-end learning architectures reviewed in the subsequent sections are predominantly based on autoencoders, their performance and practical viability are fundamentally shaped by the underlying deep learning components discussed throughout this section. Choices of activation functions, loss functions, optimization algorithms, regularization techniques, and normalization strategies directly influence convergence behavior, robustness to channel variations, generalization under model mismatch, and implementation complexity of AE-based transceivers. As a result, the material presented in Sections III-A through III-F provides essential foundations for understanding the design tradeoffs, training procedures, and deployment challenges of autoencoder-based physical-layer systems examined in Sections IV–VIII. \color{black}

		
		
		
		
		
		
		
		
		
		
		
		
		
	
		
		
		
		

\section{Autoencoders in the Physical Layer}

As research progresses on learning-based end-to-end \color{black}PHY optimization\color{black}, work is being done \color{black}to develop \color{black} different AE architectures that are better suited for certain applications. 

\color{black}This \color{black} section describes different AE schemes 
\color{black}that are \color{black} being used for end-to-end PHY optimization, \color{black}their backgrounds, \color{black}and \color{black}the \color{black}motivation for each of their architectures. 

\subsection{Standard Autoencoders}
\begin{figure}[t!]
	\centering{\includegraphics[width=1\columnwidth]{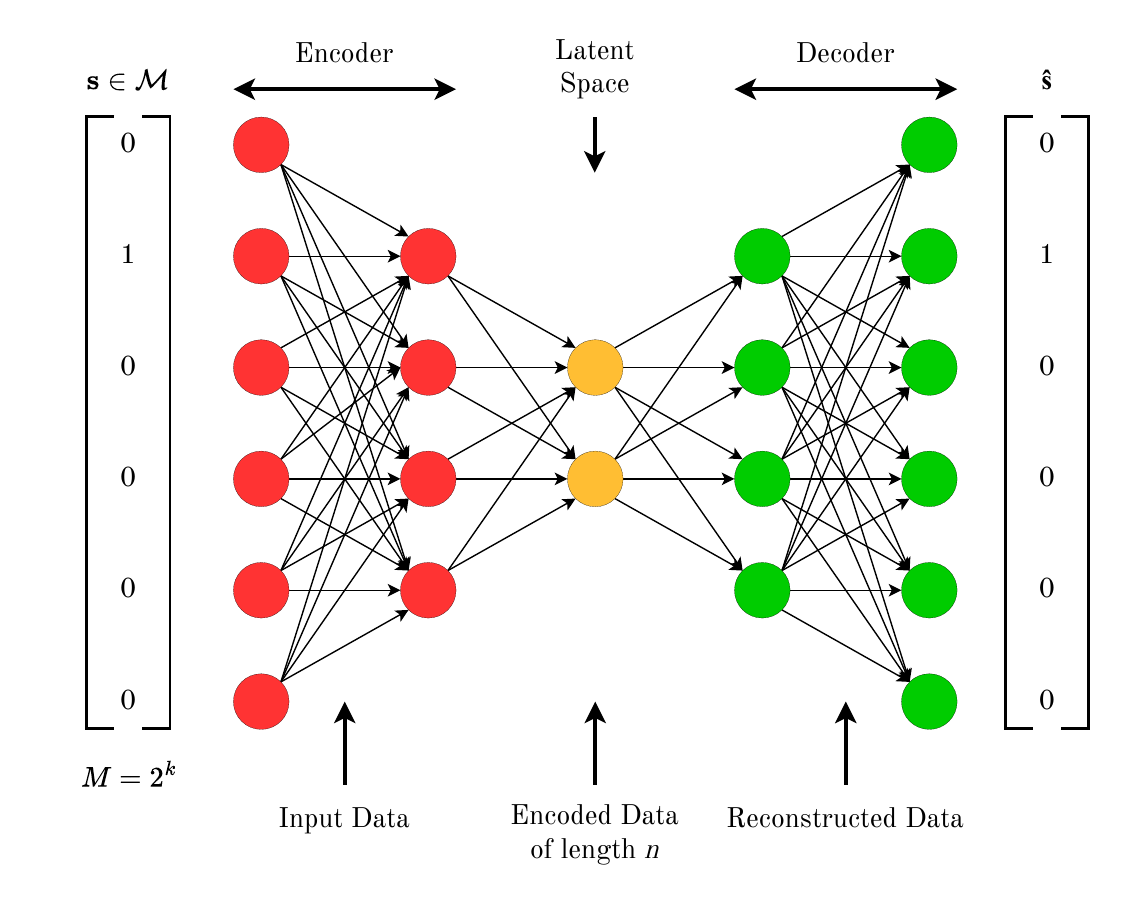}}
	\caption{\color{black}Diagram of basic autoencoder structure} 
	\label{fig:AEbasic}
\end{figure}

The standard \color{black}AE \color{black} architecture correlates well with \color{black}a \color{black} communication system, \color{black}since \color{black} the encoder, latent space, and decoder \color{black}correspond to \color{black} the transmitter, channel, and receiver, respectively, as shown in Fig. \ref{fig:AEbasic}. As of the time of writing, the standard AE model is the most \color{black}abundantly researched model \color{black} as it provides the basis for jointly learning the \color{black}entire \color{black} data link between \color{black}the \color{black}transmitter and receiver, including all \color{black}the signal processing required for tasks \color{black}such as detection, equalization and decoding \cite{mode16}, which \color{black}form the basis of \color{black}more advanced models.

%

\color{black}
The input \(\mathbf{s} \in \mathcal{M} = \{1,2,\dots, M\}\) is a \color{black}one-hot encoded \color{black} vector symbol \color{black}that corresponds \color{black} to one of \(M\) messages of length \(k\) bits, where \(k=log_{2}(M)\). \color{black}One-hot encoding enables multi-class classification, such that each input symbol can only correspond to one classification, making the network a lot more robust and simplified \cite{beyondonehot}\color{black}. One-hot encoding also \color{black}improves \color{black} feature learning performance, as the network \color{black}optimizes not only \color{black} for reconstruction but also \color{black}to destinguish \color{black} between classes.


\begin{figure*}[ht!]
\begin{center}
\includegraphics[width=1.6\columnwidth]{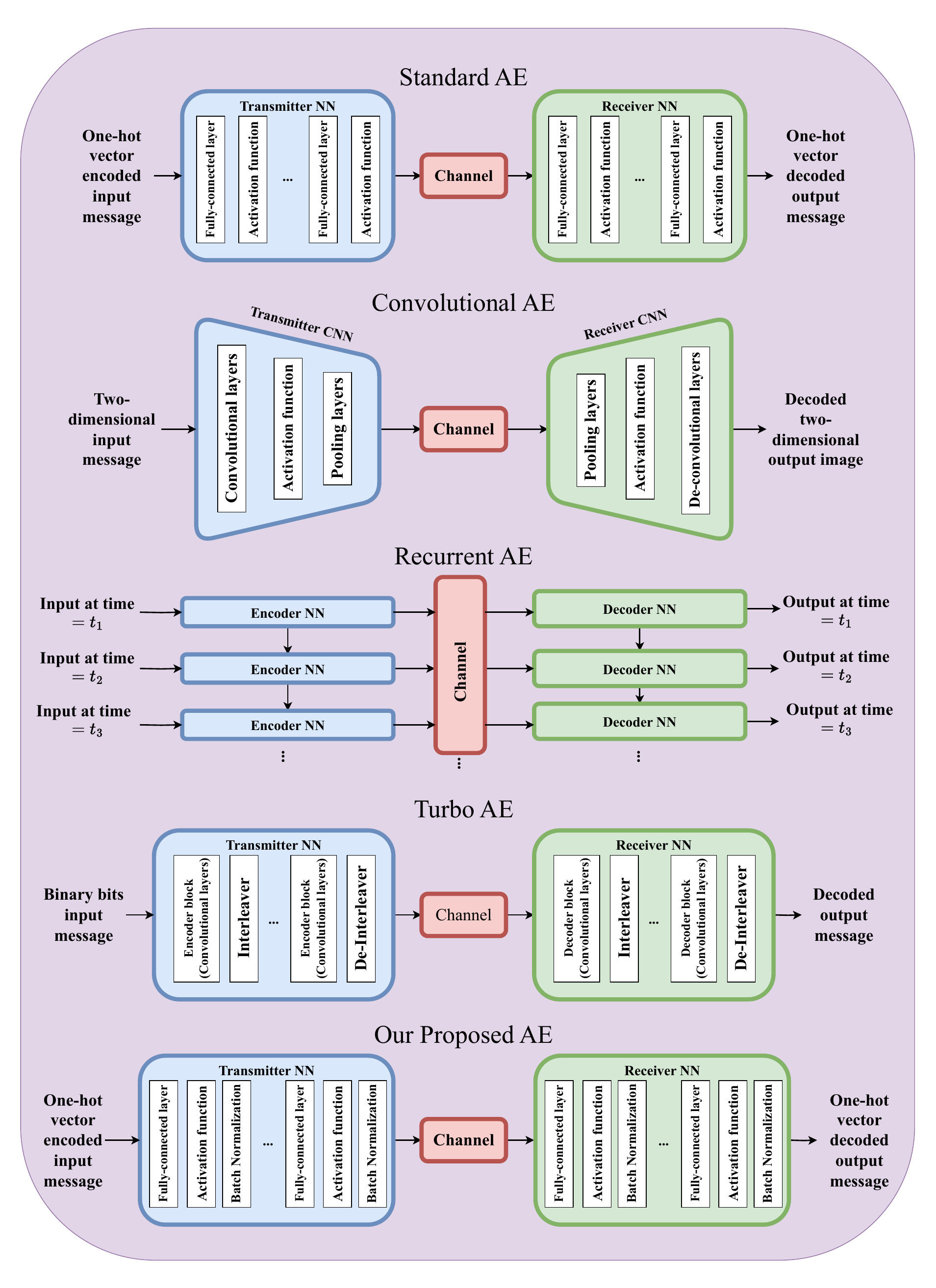}
\caption{Illustration of different AE types, including a Turbo Autoencoder, in which parallel concatenation and interleaving are implicitly realized through neural encoder blocks, \color{black} as well as the proposed Autoencoder introduced in our prior work~\cite{mode22}.\color{black}}
\label{fig:typesofAEs}
\end{center}
\end{figure*}

\color{black}
The latent space is a lower-dimension vector of length \(n\) \color{black}that \color{black} is then transmitted and received at the decoder. The communication rate, or code rate, of the AE communication system is \(R=k/n\) bits/channel use, where \(k\) bits are transmitted \color{black}during \color{black} \(n\) channel uses. The code rate for a given AE is often denoted as the pair \((n,k)\).
\color{black}Having \color{black} a latent space \color{black}whose \color{black} dimension \color{black}is lower than that of \color{black} the input \color{black}enables \color{black}the AE model \color{black}to learn \color{black} the optimal \color{black} input data representation for successful reconstruction and perform \color{black}both channel coding and modulation simultaneously. Reconstruction occurs at the decoder, where the AE learns the optimal way to map the \color{black}lower \color{black}dimensional latent representation back to the original input space \color{black}and \color{black} effectively try to ``undo" the compression performed by the encoder.

Due to the large amount of literature \color{black}that exists on \color{black} the standard AE model for PHY optimization, we find it most appropriate to discuss the works in this field with respect to their mode of communication \color{black}as specified \color{black}in Section \ref{sec:modes}.

\subsection{Convolutional Autoencoder}
Convolutional neural networks (CNNs), \color{black}which are \color{black}also known as~ConvNets, are a type of ANN that employs one or more convolutional layers \color{black}and is \color{black}specifically suited for processing data with grid-like topology, such as images. The convolutional layer applies a set of learned filters (\color{black}which are \color{black}also known as kernels) to the input data \color{black}by \color{black}performing element-wise multiplications and summing the results to produce a feature map.  Each filter in a convolutional layer can learn to detect different features, \color{black}which enables \color{black} CNNs to build a hierarchical representation of the input data. As the network goes deeper, it combines these features to form higher-level abstractions, which are essential for tasks \color{black}such as \color{black} classification, detection, \color{black}and \color{black} segmentation. 
The feature maps generated by the convolutional layers are often passed through a ReLU activation function \color{black}and \color{black} a pooling layer to reduce their spatial dimensions while retaining the most important features. In a pooling layer, a two-dimensional filter \color{black}is applied to \color{black} each channel of the feature map \color{black}to summarize \color{black} the information in the filter’s receptive field. One of the most common types of pooling is max pooling, \color{black}which selects \color{black} the maximum value in the region covered by the filter. This operation results in a downsampled feature map \color{black}and preserves \color{black} the most prominent features while reducing the computational complexity and \color{black}the \color{black}risk of overfitting.

\begin{figure}[t!]
	\centering{\includegraphics[width=1\columnwidth]{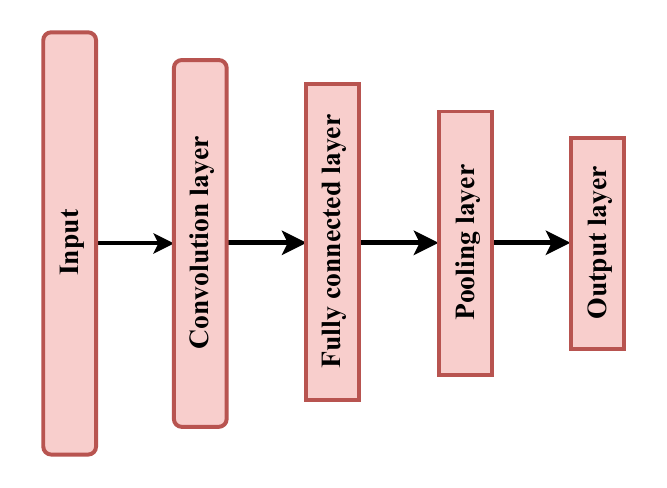}}
	\caption{\color{black}Example of CNN encoder.}
	\label{fig:cnn}
\end{figure}
\color{black}
In \color{black}CNN-based AEs \color{black} (CNN-AEs), the encoder \color{black}is typically comprised \color{black} of several convolutional layers that progressively reduce the spatial dimensions of the input while increasing the depth of the feature maps or the number of filters applied to the layer's input data. 
This process enables the network to capture complex hierarchical features in the latent representation. The decoder mirrors this structure with transposed convolutional layers (\color{black}which are \color{black}also known as de-convolutional layers) \color{black}that \color{black} up-sample the latent space back to the original input dimensions \color{black}and aim \color{black} to reconstruct the input data as accurately as possible.

\subsection{Recurrent Autoencoders}

\color{black} Recurrent neural networks (RNNs) are a class of neural models designed to process sequential data by maintaining an internal state that evolves over time. 
As illustrated in Fig. \ref{fig:typesofAEs}, this internal memory enables the network output at a given time instant to depend not only on the current input, but also on past observations. \color{black} RNNs are especially useful when dealing with time-dependent sequential data \color{black}whose \color{black} contextual information is important.

\color{black}A \color{black}key challenge \color{black}associated \color{black}with RNNs is \color{black}effective training. \color{black}Since RNNs rely on ``memory" of the previous weights, they can suffer from issues like vanishing or exploding gradients, which can hinder learning when dealing with long sequences of data. \color{black} To address this issue, specialized architectures such as \color{black}long short-term memory \color{black} (LSTM) networks and \color{black}gated recurrent units \color{black} (GRUs) \color{black}have been developed that \color{black}introduce mechanisms to better manage long-term dependencies and make RNNs more robust for complex sequence learning tasks.

The memory characteristics of RNNs are well suited for applications such as channel estimation \cite{RNN_channpred,RNN_channelest_vehicular}, sequence detection \cite{RNN_stella,LSTM_sigdetect}, turbo coding \cite{RNN_channcode} (see Section \ref{sub:turboAE}), and channel decoding \cite{RNN_lindec,RNN_commalgs}.
However, since end-to-end AE-based communications systems \color{black}generally use \color{black}one-hot vectors whose messages are not correlated \color{black}with \color{black} one another \color{black}as input data, \color{black} the memory characteristics of RNNs are not particularly useful for \color{black}these systems\color{black}. Nonetheless, there is a small body of work \color{black}that is \color{black} worthy of mention, e.g., \cite{LEARN_RNN, mode55, mode50}, \color{black}because it makes use of RNN-based AEs \color{black} (RNN-AEs) for end-to-end learned channel coding and modulation. 


\color{black}
\color{black}In light of the \color{black} low-latency constraints of 5G networks, the \color{black}low-latency efficient adaptive robust neural \color{black} (LEARN) codes \color{black}using an \color{black} RNN-AE \color{black}and \color{black} bi-directional GRUs (bi-GRUs) were used in \cite{LEARN_RNN}. 
In \cite{mode55}, an RNN-AE with bi-GRU layers was proposed for channels with intersymbol interference (ISI) \color{black}and proved to outperform \color{black} CNN-AEs as well as the maximum-likelihood sequence estimation (MLSE) ISI equalizer when perfect CSI is available. An RNN-AE with bi-GRU layers was also used in \cite{mode50} but \color{black}was \color{black}adapted to handle arbitrary code rates in ISI scenarios, \color{black}which is  \color{black}an advantageous trait for \color{black}the \color{black}5G standards \color{black}that are \color{black}used by quasi-cyclic low-density parity-check (LDPC) codes. \color{black}Their model uses dense layers to incorporate learned psuedo-puncture/depuncture modules, a code design method to achieve high-rate channel codes, to achieve these arbitrary code rates and outperformed short LDPC codes in the low SNR regime \cite{mode50}.\color{black}   

While the \color{black}works discussed above show \color{black} promising results, we currently believe that using RNN-based AEs for end-to-end PHY optimization is not a scalable solution due to RNNs' inherent high complexity, high memory constraints, and lack of parallelization capability. Recently, attention transformer models \cite{attentionisallyouneed} have \color{black}been \color{black}shown to outperform RNNs in \color{black}terms of being able \color{black} to process the entire input sequence simultaneously \color{black}and better remember long term dependencies\color{black}. AEs based on transformer architecture are a promising area for future research and \color{black}are further discussed \color{black} in Section \ref{section:future}.


\subsection{Turbo Autoencoders}
\label{sub:turboAE}

Turbo codes are a type of coding scheme that \color{black}pairs \color{black} parallel convolutional code \color{black}concatenation \color{black} with an interleaver that rearranges the bit order for each convolutional encoder~\cite{ogturbo}.

\begin{figure}[t!]
	\centering{\includegraphics[width=1\linewidth]{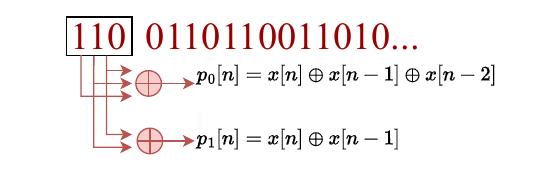}}
	\caption{\color{black}Example of a rate-1/3 convolutional encoder with constraint length 3, illustrating how encoder memory generates two parity bits in addition to the systematic bit for each input symbol}
	\label{fig:convcode}
\end{figure}

\color{black}Turbo codes are rooted in \color{black} the convolutional code, \color{black}which is \color{black}a type of error-correcting code \color{black}that is \color{black}used in digital communication. Unlike block codes, which encode fixed-size blocks of data, convolutional codes process input bits in a \color{black}continuous stream and generate \color{black} redundant output bits to help detect and correct errors. Figure \ref{fig:convcode} shows the encoding logic \color{black}of \color{black} a convolutional code \color{black}that has \color{black} each input bit \color{black}assigned to \color{black} two parity bits, for a constraint length of \(3\).


\color{black}
Turbo codes are used because they can achieve performance close to the Shannon limit, \color{black}which means \color{black} they provide excellent error correction efficiency with minimal redundancy \cite{ogturbo}. The encoding process begins with an input data stream that is passed through two or more convolutional encoders in parallel \color{black}that are \color{black}separated by an interleaver—a device that scrambles the input bits in a pseudo-random \color{black}way \color{black} before feeding them to the second encoder. This interleaving \color{black}step \color{black}introduces diversity into the coded output \color{black}to ensure \color{black}that the encoded data \color{black}produced by the encoders \color{black}is less correlated, which is crucial for effective burst error correction. \color{black}During \color{black} the decoding process, \color{black}the \color{black}turbo codes employ an iterative decoding algorithm, typically using soft-decision decoders like the Bahl-Cocke-Jelinek-Raviv (BCJR) algorithm \cite{Jiang2019TurboAE}. Each decoder operates on its \color{black}own \color{black} portion of the codeword and exchanges probabilistic information (known as "extrinsic information") with the other decoder \cite{ogturbo2}. \color{black}The \color{black} decoders refine their estimates of the original data \color{black}over multiple iterations, which \color{black}significantly \color{black}improves \color{black} the accuracy of error correction. 




\color{black}Recent advances in AE technology have built \color{black} on the ``turbo concept" \cite{turboprinciple} \color{black}with \color{black} interleaved encoding and iterative decoding, \color{black}and \color{black} introduced the turbo autoencoder (TurboAE), \color{black}which was first proposed \color{black} by Jiang \textit{et al.} \cite{Jiang2019TurboAE}. 
\color{black}The main motivation \color{black}behind TurboAE \color{black}\color{black} was to deal with the \color{black}issue of channel coding being optimal only \color{black}\color{black} for block lengths \color{black}that approach \color{black}\color{black} infinity, whereas in practice, finite block lengths are used and \color{black}channel coding is therefore \color{black}\color{black}not guaranteed to be optimal. 
Unlike standard AEs or CNN-AEs, \color{black}TurboAE \color{black}\color{black} uses three learnable encoder blocks made of \color{black}1D \color{black}\color{black} convolutional layers \color{black}whose encoders are \color{black}\color{black} denoted by \(f_{i,\theta}(.)\), where \(i\in\{1,2,3\}\) and \(h(.)\) \color{black}denotes \color{black}\color{black} a power constraint block. For \(i=1,2\), the \color{black}encoders' inputs \color{black}\color{black} are copies of the input message sequence \(u\), \color{black}whereas \color{black}\color{black} for \(i=3\), the \color{black}encoder's \color{black}\color{black} input is an interleaved permutation of \(u\) \color{black}that is \color{black}\color{black} denoted by \(u^{\pi}\), where \(\pi\) denotes the interleaving operation and \(\pi^{-1}\) denotes the de-interleaving operation. \color{black}
\color{black}The TurboAE encoder is illustrated in Fig. \ref{fig:typesofAEs}\color{black}. 

Since the encoders process the input sequence in parallel, this gives \color{black}TurboAE \color{black} a code rate of \(R=1/3\), with three sequences (\(\mathbf{x}_{1},\mathbf{x}_{2},\mathbf{x}_{3}\)) being transmitted over the channel. To decode the message, TurboAE uses two sequential decoders, \color{black}which are denoted by \color{black} \(g_{\phi_{i},1}\) and \(g_{\phi_{i},2}\), \color{black}where \(i=\{1,\dots,6\}\) for interleaved and de-interleaved order for the \(i\)-th iteration\color{black}. The first decoder, \(g_{\phi_{i},1}\), processes the received signals \(y_{1},y_{2}\) \color{black}and \color{black}the de-interleaved prior \(p\) \color{black}whose shape is \color{black} \((K,F)\), where K is the message length and F represents the size of the feature information for each code bit. This produces a posterior \(q\) \color{black}that retains \color{black} the shape \((K,F)\). The second decoder, \(g_{\phi_{i},2}\), takes the interleaved sequence \(\pi(y_{1}),y_{3},\) and the interleaved prior \(p\) \color{black}and generates \color{black} another posterior \(q\). The \color{black}second \color{black}posterior \(q\) then acts as the prior \(p\) for the next stage. The first iteration starts with a prior \(p=0\), and by the final iteration, the posterior is reshaped to \((K,1)\). \color{black}It is then \color{black}decoded using the sigmoid function to output \(\hat{u}=sigmoid(q)\). A high-level illustration of the TurboAE and its respective layers is shown in Fig. \ref{fig:typesofAEs}. \color{black}TurboAE exhibited \color{black} an increase in reliability as block length \(K\) increased, \color{black}which is \color{black}known as blocklength gain \cite{ogturbo}, and outperformed the CNN-AE from \cite{jointCNN} in the AWGN channel. Furthermore, when compared \color{black}with \color{black} conventional Turbo codes and a dedicated Turbo-decoder NN known as DeepTurbo \cite{jiang2019deepturbodeepturbodecoder} in non-AWGN channels, \color{black}TurboAE outperformed them both \color{black}in the low SNR regime.

\color{black}

TurboAE is improved upon in \cite{jiangTurbo}, where transfer learning is applied for learned modulation and demodulation. The new model is being called TurboAE-MOD. TurboAE-MOD has been proven to perform comparably to conventional modulation techniques, showing that TurboAE can effectively learn joint modulation and channel coding. 

In \cite{turbo7Clausius}, the TurboAE architecture is studied in a serial concatenation configuration with the two encoders arranged in series instead of in parallel to study the scalability of TurboAEs.
Papers \cite{baleviTurbo}, \cite{jiangTurbo}, \cite{wangTurbo}, and \cite{xuTurbo} also propose AE models that integrate turbo codes. 
These TurboAEs achieve better performance than conventional turbo codes with one-bit CSI quantization \cite{baleviTurbo}. 
In \cite{jiangTurbo}, the TurboAE model proposed outperformed traditional modulation and code schemes in non-AWGN channels. 
Additionally, the rate-matched TurboAE model proposed in \cite{wangTurbo} proved itself to be effective across multiple code rates and outperform mismatched TurboAEs at higher code rates \cite{wangTurbo}. 
In \cite{xuTurbo}, a TurboAE was used for denoising rather than transmitting data, and it performed better than traditional OFDM schemes.

From a PHY perspective, different AE architectures carry inherent inductive biases that make them better suited to particular end-to-end communication tasks. Standard fully connected autoencoders are well suited to short-block point-to-point transmissions and proof-of-concept studies, but their flat structure scales poorly with increasing message size. CNN-based autoencoders are preferable whenever the data exhibits spatial or grid-like structure, or when correlations across subcarriers can be exploited, as in joint source–channel coding for images or OFDM-based systems, since they provide parameter sharing and a degree of translation invariance~\cite{Ferdous2024CNNAE}. Recurrent and sequence-based autoencoders are more appropriate for streaming or time-correlated scenarios, such as channels with intersymbol interference or time-varying fading, where sequence detection and equalization benefit from temporal memory. Attention- and transformer-based autoencoders are promising when long-range dependencies in time or frequency must be modeled, or when large context windows and semantic tasks are considered; these models are generally more computationally demanding but are particularly compelling for large-block and task-oriented communication settings~\cite{Sindal2024SparseAE,Zayat2024TMAE,Owfi2025OMLCAE}. The subsequent sections on autoencoder modes of communication build on this mapping and highlight how these architectural choices translate into performance and implementation trade-offs for each scenario.

\color{black}

\begin{table*}[]
\centering
    \caption{\color{black}Comparison of Autoencoder Architectures in End-to-End Transceiver Design.\color{black}}
    \label{tab:autoencoder_comparison}
\resizebox{\textwidth}{!}{%
\begin{tabular}{|l|l|l|l|}
\hline
\textbf{Autoencoder Type} &
  \textbf{Key Features} &
  \textbf{Pros} &
  \textbf{Cons} \\ \hline
\textbf{Standard Autoencoder (AE)} &
  Fully connected layers &
  Simple, effective for general data &
  Not optimized for spatial or sequential data \\ \hline
\textbf{Convolutional Autoencoder (CNN-AE)} &
  Uses CNN layers for encoding/decoding &
  Excels in image and spatial feature extraction &
  High computational cost \\ \hline
\textbf{Recurrent Autoencoder (RNN-AE)} &
  Uses RNNs or LSTMs for sequence encoding &
  Ideal for time-series and sequential signals &
  Prone to vanishing gradients, complex training \\ \hline
\textbf{Turbo Autoencoder (TurboAE)} &
  Inspired by turbo codes, optimized for communication channels &
  Improves reliability and error correction &
  Limited to wireless applications, requires iterative decoding \\ \hline
\textbf{Variational Autoencoder (VAE)} &
  Probabilistic encoding with latent space constraints &
  Robustness to noisy channels, regularized latent space &
  Requires complex training and KL divergence computation \\ \hline
\textbf{Denoising Autoencoder (DAE)} &
  Learns to reconstruct signals from noisy input &
  Effective in channel estimation, noise suppression &
  Can distort signal if noise assumptions are incorrect \\ \hline
\textbf{Sparse Autoencoder (SAE)} &
  Applies sparsity constraints on latent representation &
  Enhances feature extraction and reduces overfitting &
  May discard useful information during sparsification \\ \hline
\textbf{Attention-Based Autoencoder (AAE)} &
  Integrates attention mechanisms for adaptive feature extraction &
  Enhances long-range dependencies in sequences &
  Computationally expensive, requires fine-tuning \\ \hline
\end{tabular}%
}
\end{table*}

\begin{table*}[ht!]
	\centering
	\caption{Autoencoder Modes of Communication}
	\label{tab:AEmodes}
	\begin{tabular}{|>{\centering\arraybackslash}m{4cm}|>{\centering\arraybackslash}m{6cm}|}
		\hline
		\textbf{References} & \textbf{Mode of Communication}  \\ \hline
		\cite{1hoydis,mode2,mode3,mode16,mode43_JSCCImage,mode18,mode19,mode21,mode22} & Point to Point (P2P) \\ \hline
		\cite{mode39,mode40,mode54,mode24,mode30,mode1,mode23,mode29,mode37,mode33} & Multiple Input Multiple Output (MIMO) \\ \hline
		\cite{mode16,OFDM_AE_OG,OFDM_AE_ensemble,OFDM_AEtrimmingfat,PDNOMA_powerallocation,DLconstellation_downlinkNOMA,PDNOMA_weightedAE,CDAEbasedNOMA,ogSCMA,CDNOMA_DLaided_SCMA,CDNOMA_ElsevierAE_SCMA,CDNOMA_sparseAE_SCMA,CDNOMA_sergienko_SCMA,E2E_SCMA,CDNOMA_residualCNN_SCMA,mac_Minsig_CDNOMA_2020,mac_Minsig_CDNOMA_2022,CDNOMA_6GWGAN_2023_SCMA,CDNOMA_MCMA,MAC_SWIPT_RSMA,mode18,mode20} & Multiple Access Channel (MAC) \\ \hline
		\cite{mode25,mode28,mode29,mode30} & Broadcast Channel (BC) \\ \hline
		\cite{mode4,mode8,mode9,mode10,mode11,mode12,mode13,mode14,mode15,mode7,mode45_semanticrelay} & Relay Channel (FD and HD, AF and DF) \\ \hline
		\cite{1hoydis,mode53,mode31,mode6.5,mode6,mode57,mode56,mode47,mode48.5,mode48,mode20,mode51,mode49,mode52} & Interference Channel (IC) \\ \hline
	\end{tabular}
\end{table*}

\section{Autoencoder Modes of Communication}
\label{sec:modes}

Autoencoders (AEs) have emerged as a powerful tool for end-to-end physical-layer (PHY) optimization in wireless communication systems. By learning optimal transmission and reception strategies directly from data, AE-based transceivers are able to jointly optimize modulation, coding, detection, and estimation without relying on explicit analytical channel models. This data-driven capability enables AEs to address complex and highly nonlinear communication scenarios where traditional model-based designs become suboptimal or analytically intractable.

This section surveys the application of AEs across a range of communication modes, including point-to-point (P2P), multiple-input multiple-output (MIMO), multiple access channel (MAC), broadcast channel (BC), relay channel, and interference channel (IC) networks. 

\textcolor{black}{The organization adopted in this section is based on network-level channel topologies rather than physical-layer transmission techniques. Specifically, P2P, MAC, BC, relay, and IC models are treated as distinct information-theoretic communication settings that define the number of transmitters, receivers, and their interaction patterns. Physical-layer techniques such as MIMO signaling, beamforming, OFDM, or spatial modulation are regarded as enabling mechanisms that may be employed within these topologies. As a result, a P2P system may utilize MIMO transmission, and a relay channel may operate under either single-antenna or multi-antenna configurations. Each work is therefore categorized according to its underlying channel topology, while architectural extensions are discussed within the corresponding subsection to avoid redundancy.}

Table~\ref{tab:AEmodes} summarizes the representative works discussed in this section according to their modes of communication.

\subsection{Point-to-Point Communication}

\begin{figure}[t!]
	\centering{\includegraphics[width=1\linewidth]{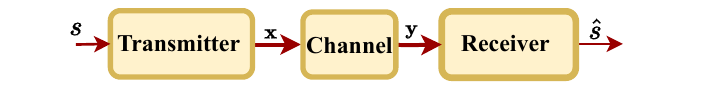}}
	\caption{Block diagram of a point-to-point communication system.}
	\label{fig:P2P}
\end{figure}

Point-to-point (P2P) communication represents the most fundamental communication topology, where a single transmitter sends information to a single receiver over a wireless channel, as illustrated in Fig.~\ref{fig:P2P}. Despite its conceptual simplicity, the P2P setting serves as a critical testbed for evaluating the feasibility and performance limits of AE-based end-to-end learning frameworks.

The landmark work in \cite{1hoydis} was the first to demonstrate that an AE can be used to jointly learn modulation and channel coding in a P2P communication system. In particular, the authors showed that a $(7,4)$ AE model achieved the same block error rate (BLER) as a traditional system employing binary phase-shift keying (BPSK) modulation and a $(7,4)$ Hamming code, while outperforming an uncoded $(8,8)$ BPSK system over the entire $E_b/N_0$ range. Moreover, the AE achieved comparable performance to an uncoded $(2,2)$ BPSK system, indicating that the AE implicitly learned a joint coding and modulation strategy that provides coding gain without explicitly enforcing conventional channel coding structures.

Subsequent work in \cite{mode2} extended the P2P AE framework by introducing convolutional neural network (CNN)-based architectures. By exploiting local correlations and reducing the number of trainable parameters, the proposed CNN-AE outperformed uncoded BPSK at code rate $R=1$ and quadrature phase-shift keying (QPSK) at $R=2$. These results highlighted the advantages of convolutional structures for improving robustness and generalization in AE-based transceivers.

In \cite{mode3}, the authors modified the standard AE architecture to operate directly on binary data rather than one-hot encoded symbols. Pre-processing and post-processing blocks were introduced to map binary input data to one-hot vectors and convert output probability vectors into soft binary words. By employing binary cross-entropy (BCE) loss instead of the commonly used categorical cross-entropy (CCE), the AE was explicitly optimized for bit error rate (BER) minimization rather than symbol error rate (SER). The resulting bit-wise AE exhibited superior BER performance compared to both standard AEs and conventional M-QAM systems. To ensure Gray coding behavior, the transmitter and receiver were first trained separately before performing joint end-to-end training.

While many early studies focused on simulation-based performance evaluation, \cite{mode16} investigated the practical deployment of an AE-based P2P system using software-defined radio (SDR). Since the real wireless channel is unknown and non-differentiable, the authors constructed a differentiable channel model incorporating pulse shaping, constant sample-time offset, constant phase offset, carrier frequency offset, and additive white Gaussian noise. The deployed AE achieved performance within approximately 1~dB of a GNU Radio differential QPSK baseline at a BLER of $10^{-4}$. This work provided valuable insights into addressing practical impairments such as inter-symbol interference (ISI), synchronization errors, and hardware non-idealities in real-world AE-based communication systems.

AE-based P2P communication has also been extended to joint source–channel coding (JSCC) applications. In \cite{mode43_JSCCImage}, a CNN-AE was proposed for wireless image transmission, where source compression and channel coding were jointly optimized in an end-to-end manner. The proposed JSCC scheme outperformed JPEG and JPEG2000 compression algorithms in the low-SNR regime under both AWGN and Rayleigh slow-fading channels, demonstrating the effectiveness of AEs for transmitting structured data.

Beyond radio-frequency (RF) systems, AEs have been applied to optical wireless communication (OWC) and space optical communication (SOC) P2P links. The authors of \cite{mode18} compared the BLER performance of their OWC AE against state-of-the-art model-based OWC systems and showed that both $(4,4)$ and $(2,2)$ AE models outperform uncoded on–off keying (OOK). In \cite{mode19}, the AE framework was evaluated under AWGN and log-normal fading channels to model turbulence effects, assuming both perfect and noisy channel state information (CSI). The results showed that a $(7,4)$ AE outperformed Hamming-coded $(7,4)$ OOK with hard-decision decoding and performed slightly worse than soft-decision decoding.

\textcolor{black}{Collectively, these works demonstrate that P2P communication serves as a foundational setting for AE-based transceiver design, enabling systematic investigation of architectural choices, loss-function design, and practical deployment considerations before extending AE frameworks to more complex multi-user and multi-antenna scenarios.}

\subsection{Multiple-Input Multiple-Output (MIMO)}

\begin{figure}[t!]
	\centering{\includegraphics[width=1\columnwidth]{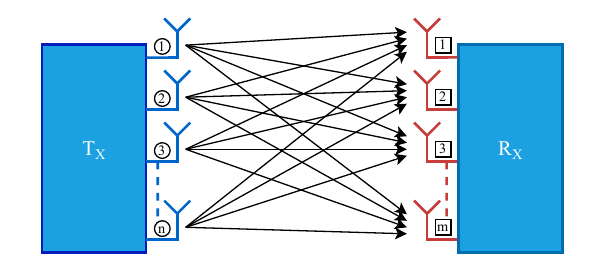}}
	\caption{Block diagram of a simple MIMO communication system.}
	\label{fig:exmimo}
\end{figure}

Multiple-input multiple-output (MIMO) systems are widely employed in modern wireless communications to improve spectral efficiency, reliability, and coverage by exploiting multiple transmit and receive antennas. A representative MIMO communication system is illustrated in Fig.~\ref{fig:exmimo}.

Similar to the point-to-point case, a MIMO channel with $N_t$ transmit antennas and $N_r$ receive antennas can be modeled as
\begin{equation}
	\mathbf{y} = \mathbf{H}\mathbf{x} + \mathbf{n},
\end{equation}
where $\mathbf{y}$ denotes the received signal vector, $\mathbf{x}$ is the transmitted signal vector, $\mathbf{H}$ is the channel matrix, and $\mathbf{n}$ is the additive noise vector. The channel matrix $\mathbf{H}$ is given by
\begin{equation}
	\mathbf{H} =
	\begin{bmatrix}
		h_{11} & h_{12} & \cdots & h_{1N_t} \\
		h_{21} & h_{22} & \cdots & h_{2N_t} \\
		\vdots & \vdots & \ddots & \vdots \\
		h_{N_r1} & h_{N_r2} & \cdots & h_{N_rN_t}
	\end{bmatrix},
\end{equation}
where each element $h_{ij}$ represents the complex channel gain between the $j$-th transmit antenna and the $i$-th receive antenna. MIMO systems exploit the spatial dimensions of the channel through precoding and spatial multiplexing to improve throughput and link robustness.

Effective precoding in MIMO systems relies heavily on accurate channel state information (CSI), making CSI acquisition, feedback, and utilization central challenges. A significant body of work has therefore focused on using AEs to jointly optimize precoding and decoding under various CSI assumptions \cite{mode39,mode40,mode54}.

One of the earliest AE-based MIMO designs was proposed in \cite{mode24}, where a specialized AE architecture was introduced for single-user closed-loop MIMO communication with limited feedback. To address CSI feedback constraints, the channel was quantized into $2^v$ discrete states, enabling the transmitter to learn a finite set of precoding strategies. The proposed AE was shown to match and, in some cases, outperform conventional space–time block coding (STBC) and singular value decomposition (SVD)-based precoding schemes, particularly at moderate-to-high SNRs.

In \cite{mode1}, the authors proposed a two-stage AE training framework aimed at bridging the gap between simulation and deployment. The AE was first trained on a Rayleigh fading channel and subsequently fine-tuned using real in-phase and quadrature (IQ) samples collected after deployment. CSI was estimated at the receiver, quantized, and fed back to the transmitter, where it was concatenated with the input message. This approach outperformed conventional zero-forcing and MMSE precoding schemes. Furthermore, the proposed $(7,4)$ AE outperformed Hamming-coded $(7,4)$ systems with both hard-decision and maximum-likelihood decoding, with performance improving as the code rate decreased.

In massive MIMO systems, hardware complexity becomes a critical bottleneck, particularly due to the high-resolution analog-to-digital converters (ADCs) required at the receiver. To address this issue, \cite{mode23} proposed an AE-based MIMO framework that jointly optimizes transceiver parameters and ADC quantization. The ADCs were implemented as trainable quantization layers within the AE, enabling the use of low-resolution (3-bit) ADCs while maintaining acceptable performance. This approach significantly reduces hardware cost and power consumption in massive MIMO deployments.

A unified AE-based MIMO framework was presented in \cite{mode29}, where the same model was evaluated across multiple scenarios, including open-loop MIMO, closed-loop MIMO, multi-user broadcast MIMO, and interference channels. The results demonstrated that AE-based designs are capable of learning optimal mappings from messages to transmit vectors and achieving robust decoding across diverse MIMO configurations.

Millimeter-wave (mmWave) massive MIMO systems introduce additional challenges due to hardware constraints and sparse propagation environments. In \cite{mode37}, an AE-based hybrid beamforming architecture was proposed, where analog beamformers were modeled as constrained layers enforcing phase-only adjustments, while digital beamforming weights were learned jointly. Trained using the Saleh--Valenzuela channel model, the proposed AE achieved superior bit error rate (BER) performance compared to conventional linear hybrid beamforming methods.

To address interference and detection challenges in spatial modulation (SM) systems, \cite{mode33} proposed several AE-based SM frameworks. These included architectures trained to jointly decode transmitted symbols and antenna indices, as well as models that embed phase-shift keying (PSK)-based antenna signatures. The proposed AE designs achieved notable gains in block error rate and power efficiency, particularly in scenarios with high antenna correlation.

\textcolor{black}{Overall, AE-based MIMO systems primarily focus on learning precoding strategies, CSI representation and feedback mechanisms, and hardware-aware transceiver designs. The incorporation of explicit physical constraints, such as quantization, phase-only beamforming, and feedback limitations, is essential for ensuring that learned MIMO solutions remain practically deployable.}

\subsection{Multiple Access Channels}
\label{subsec:mac}

\begin{figure}[t!]
	\centering{\includegraphics[width=1\columnwidth]{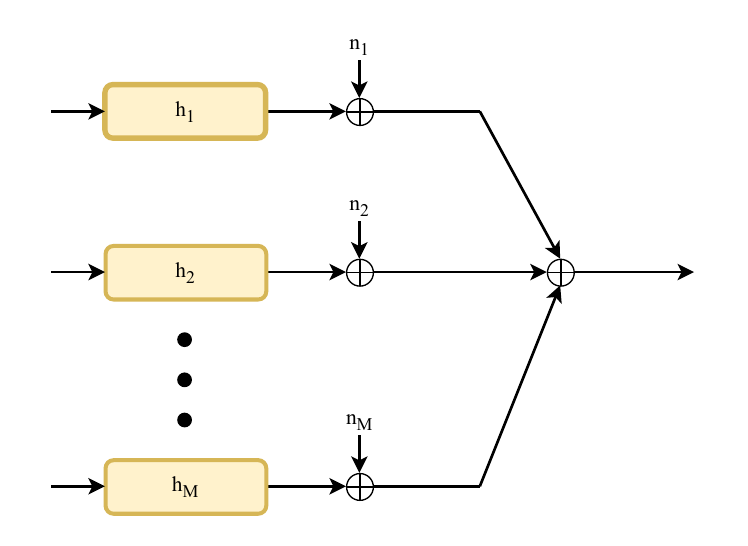}}
	\caption{Block diagram of a multiple access channel.}
	\label{fig:exmac}
\end{figure}

In many wireless communication systems, multiple transmitters simultaneously send information to a single receiver, such as in the uplink of a cellular network. This communication setting is referred to as a multiple access channel (MAC) and is illustrated in Fig.~\ref{fig:exmac}.

For an $M$-user MAC, the received signal at the receiver can be expressed as
\begin{equation}
	y = h_{1}x_{1} + n_{1} + h_{2}x_{2} + n_{2} + \dots + h_{M}x_{M} + n_{M},
\end{equation}
where $x_m$, $h_m$, and $n_m$ denote the transmitted signal, channel coefficient, and additive noise associated with the $m$-th user, respectively. In order to reliably recover each user's information at the receiver, appropriate multiple access schemes must be employed.

Traditional MAC designs rely heavily on orthogonal multiple access (OMA) schemes, such as frequency-division multiple access (FDMA), time-division multiple access (TDMA), code-division multiple access (CDMA), orthogonal frequency-division multiple access (OFDMA), and single-carrier FDMA (SC-FDMA). The primary advantage of these orthogonal schemes is the elimination of inter-user interference, which simplifies receiver design and decoding. However, the rigid resource partitioning inherent in OMA limits spectral efficiency, particularly in dense and heterogeneous networks.

\begin{figure*}[t!]
	\centering
	\includegraphics[width=2\columnwidth]{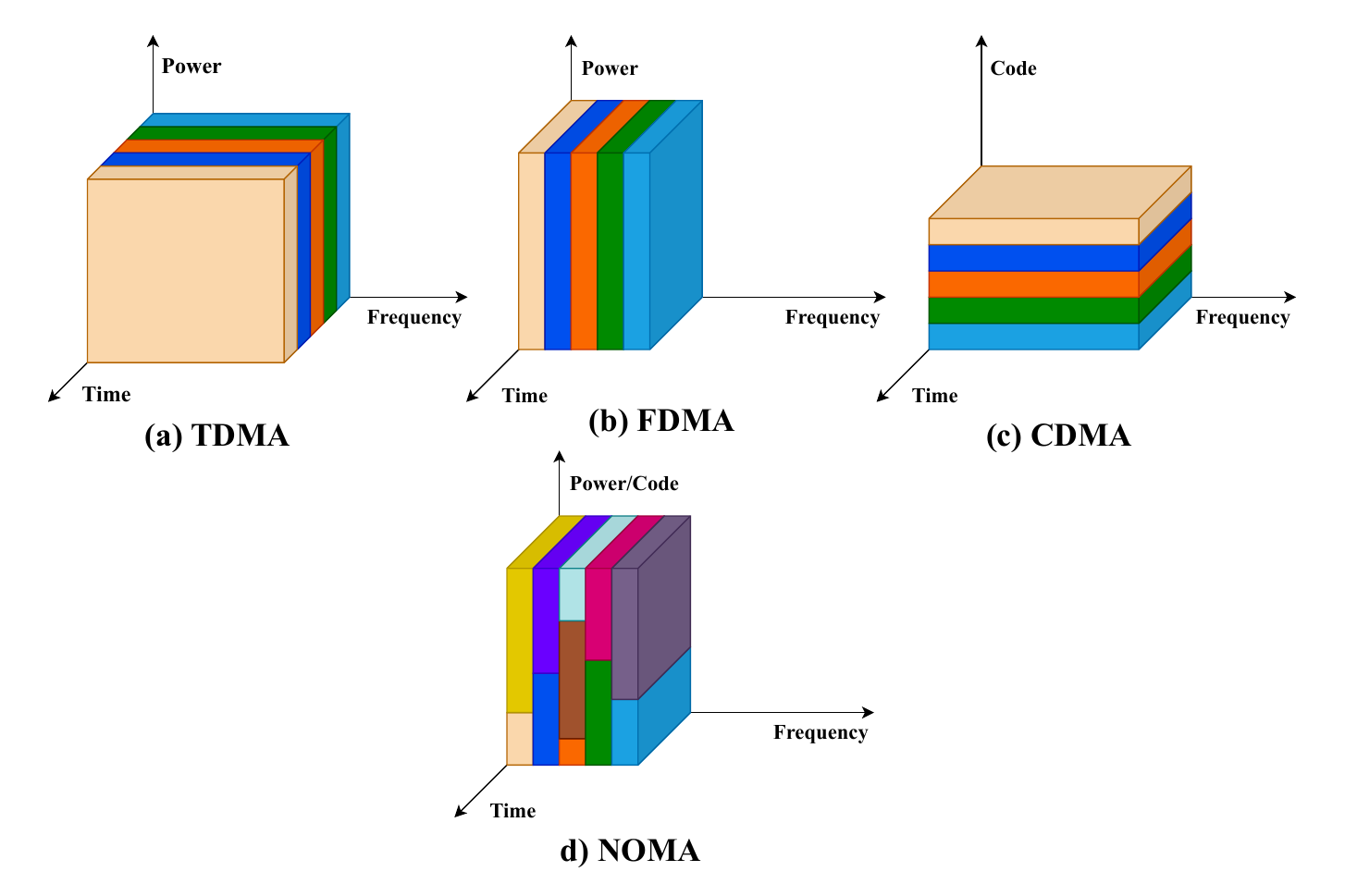}
	\caption{Comparison of different multiple access schemes: (a) TDMA, (b) FDMA, (c) CDMA, and (d) NOMA.}
	\label{fig:multipleaccess}
\end{figure*}

Recent research has therefore explored AE-based approaches to enhance both orthogonal and non-orthogonal multiple access schemes. Early work extended AE-based transceivers to OFDM and OFDMA systems. In \cite{OFDM_AE_OG}, the AE framework originally proposed in \cite{mode16} was adapted to an OFDM setting using a cyclic prefix to mitigate synchronization errors and multipath effects. The resulting AE-based OFDM system achieved performance comparable to QPSK with MMSE equalization in AWGN channels.

An alternative approach was presented in \cite{OFDM_AE_ensemble}, where a community-based AE architecture was proposed. Inspired by \cite{communityAE}, this framework employed multiple CNN-based encoders and decoders trained using the simultaneous perturbation stochastic approximation (SPSA) algorithm instead of back-propagation, enabling training in scenarios where exact channel models are unavailable. By randomly selecting encoder–decoder pairs during training and sharing weights across the community, the proposed system outperformed conventional 64-QAM in AWGN channels.

In \cite{OFDM_AEtrimmingfat}, the authors proposed an OFDM AE design that eliminates both pilot symbols and the cyclic prefix. The transmitter employed a centering and normalization module with trainable parameters, while the receiver used a residual CNN. Evaluated over time- and frequency-selective fading channels, the proposed approach demonstrated that removing pilots and the cyclic prefix resulted in negligible BER degradation while achieving significant throughput gains.

Beyond orthogonal schemes, considerable attention has been devoted to non-orthogonal multiple access (NOMA) techniques, which allow multiple users to share the same time–frequency resources. NOMA schemes can be broadly categorized into power-domain NOMA (PD-NOMA) and code-domain NOMA (CD-NOMA). In PD-NOMA, users are multiplexed using different power levels, enabling successive interference cancellation (SIC) at the receiver.

In \cite{PDNOMA_powerallocation}, an AE-based PD-NOMA framework was proposed for service-based transmission, where multiple services for a single user are transmitted over the same resources. The work in \cite{PDNOMA_weightedAE} further extended this idea by introducing a weighted loss function that allows balancing error probabilities among users or sub-blocks with different importance levels. By incorporating the SICNet receiver \cite{SICnet}, the proposed AE improved the performance of conventional PD-NOMA constellation designs.

Code-domain NOMA techniques distinguish users by assigning non-orthogonal codes or multidimensional constellations. One such technique is low-density spreading CDMA (LDS-CDMA), for which an AE-based design was proposed in \cite{CDAEbasedNOMA}. In this approach, the AE jointly optimizes spreading sequences and detection, improving performance over conventional LDS-CDMA systems.

Sparse code multiple access (SCMA) has emerged as a prominent CD-NOMA scheme due to its favorable trade-offs between performance and complexity. SCMA employs sparse multidimensional codebooks, enabling low-complexity detection using message passing algorithms (MPA). However, the iterative nature of MPA and the difficulty of designing optimal codebooks motivate learning-based approaches. In \cite{CDNOMA_DLaided_SCMA}, an AE was proposed to jointly optimize SCMA codebooks and decoding, achieving lower BER than conventional list sphere decoding methods. Subsequent works proposed AE-SCMA \cite{CDNOMA_ElsevierAE_SCMA}, sparsity-regularized AEs \cite{CDNOMA_sparseAE_SCMA}, and residual CNN-based decoders \cite{CDNOMA_residualCNN_SCMA} to further improve performance and reduce complexity.

Additional extensions explored multi-task learning \cite{E2E_SCMA}, generalized AE frameworks for both SCMA and DCMA \cite{mac_Minsig_CDNOMA_2020,mac_Minsig_CDNOMA_2022}, and generative adversarial network (GAN)-assisted training for SCMA under challenging channel conditions \cite{CDNOMA_6GWGAN_2023_SCMA}. Moderate-density code multiple access (MCMA), an extension of SCMA designed to improve spectral efficiency and reduce latency in IoT applications, was introduced in \cite{CDNOMA_MCMA}, where an AE learns factor graph edge weights to enable non-iterative multi-user detection.

AE-based MAC optimization has also been applied to emerging paradigms such as simultaneous wireless information and power transfer (SWIPT). In \cite{MAC_SWIPT_RSMA}, an AE integrated with rate-splitting multiple access (RSMA) was proposed to jointly optimize power allocation and decoding under interference constraints, achieving near-optimal power efficiency with reduced computational complexity.

\textcolor{black}{Overall, AE-based MAC designs focus on learning interference-aware representations, flexible resource-sharing strategies, and low-complexity receivers. The ability of AEs to jointly optimize constellation design, decoding, and power allocation makes them particularly attractive for dense multi-user and IoT-oriented uplink scenarios.}

\subsection{Broadcast Channels}
\label{sub:broadcast}

\begin{figure}[t!]
	\centering{\includegraphics[width=1\columnwidth]{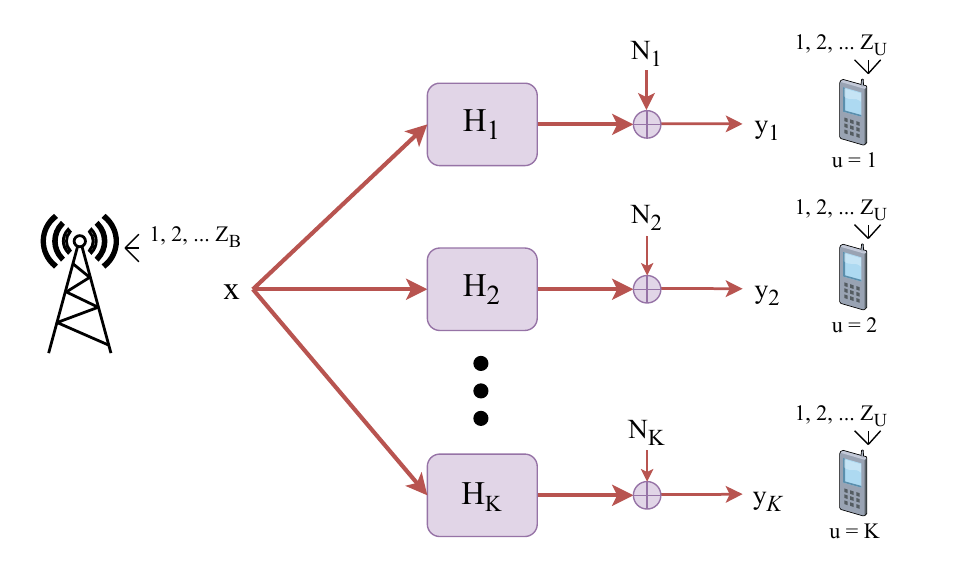}}
	\caption{Block diagram of a broadcast channel.}
	\label{fig:exbroad}
\end{figure}

A broadcast channel (BC) models downlink communication scenarios in which a single transmitter sends information to multiple receivers, as illustrated in Fig.~\ref{fig:exbroad}. This setting is representative of many practical systems, such as cellular downlink transmission from a base station to multiple user equipments (UEs).

By comparing the structures of the BC and MAC shown in Fig.~\ref{fig:exbroad} and Fig.~\ref{fig:exmac}, respectively, it can be observed that these two channel models exhibit a form of duality, as also supported by the theoretical results reported in \cite{mode27}. While the MAC focuses on multiple transmitters communicating with a single receiver, the BC addresses the inverse scenario, which introduces unique challenges related to fairness, power allocation, and heterogeneous channel conditions among receivers.

In practical broadcast scenarios, receivers often experience different levels of channel degradation due to factors such as distance from the transmitter, shadowing, or hardware limitations. Such scenarios are commonly referred to as degraded broadcast channels. A substantial body of classical information-theoretic work has focused on developing optimal transmission strategies for degraded BCs; however, these approaches typically rely on accurate channel models and precise CSI.

The standard AE-based communication framework can be extended to degraded BCs by employing a shared encoder at the transmitter and separate decoder networks at each receiver. In \cite{mode25}, the encoder and all receiver decoders were jointly trained in an end-to-end manner, enabling the AE to implicitly learn power and rate allocation strategies that accommodate heterogeneous channel conditions. The proposed AE-based BC outperformed conventional schemes in terms of fairness across receivers.

Channel output feedback (COF) was incorporated into AE-based BC designs in \cite{mode28}, where a recurrent neural network (RNN)-based AE was proposed. In this framework, feedback information regarding channel dynamics and decoding performance from previous transmissions is fed back to the transmitter. A two-phase training scheme was employed, in which the transmitter encodes symbols based on both the information bits and the COF, while the receiver decodes the information bits after collecting all received symbols. This approach demonstrated improved robustness to channel variations and receiver heterogeneity.

AE-based designs have also been applied to multiple-input multiple-output (MIMO) broadcast channels. In \cite{mode30}, a precoding layer was integrated into the AE encoder to explicitly handle multi-antenna transmission. Under the assumption of perfect CSI, the proposed AE-based precoding scheme outperformed classical linear precoding techniques, including zero-forcing precoding (ZFP), minimum mean square error (MMSE) precoding, and Tomlinson--Harashima precoding (THP).

\textcolor{black}{Overall, AE-based broadcast channel designs leverage joint encoder--decoder training to implicitly manage user heterogeneity, power allocation, and precoding. These properties make AEs particularly well suited for downlink scenarios with diverse receiver conditions and dynamic channel environments.}

\subsection{Relay Channel}

\begin{figure}[t!]
	\centering{\includegraphics[width=1\columnwidth]{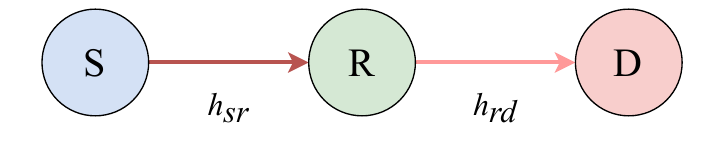}}
	\caption{Block diagram of a one-way relay channel.}
	\label{fig:exonerelay}
\end{figure}

Relay channels enhance communication reliability and coverage by introducing intermediate nodes that assist information transmission between a source and a destination. In a simple one-way relay channel, shown in Fig.~\ref{fig:exonerelay}, the relay node forwards the source signal using either amplify-and-forward (AF) or decode-and-forward (DF) protocols.

In AF relaying, the relay amplifies the received signal and forwards it to the destination without explicit decoding. While this approach does not require accurate CSI at the relay, it also amplifies noise and interference. In contrast, DF relaying avoids noise amplification by decoding and re-encoding the source message at the relay, but requires reliable CSI and introduces additional processing complexity.

Relay nodes can operate in half-duplex (HD) or full-duplex (FD) modes. In HD relaying, the relay alternates between reception and transmission, whereas FD relays transmit and receive simultaneously, potentially doubling spectral efficiency. However, FD relaying introduces severe self-interference, which must be mitigated through advanced cancellation techniques.

\begin{figure}[t!]
	\centering{\includegraphics[scale=0.5]{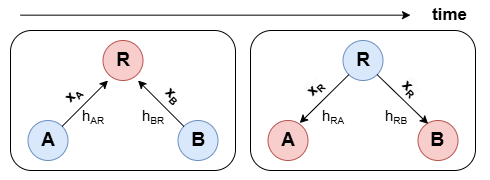}}
	\caption{Block diagram of a two-way relay channel.}
	\label{fig:extwaf}
\end{figure}

Two-way relay channels, illustrated in Fig.~\ref{fig:extwaf}, enable bidirectional communication between two terminals via a relay node and are widely used in modern wireless networks.

A substantial body of work has investigated AE-based optimization of relay channels, with primary emphasis on one-way half-duplex (OWHD) and two-way half-duplex (TWHD) configurations \cite{mode4,mode8,mode9,mode10,mode11,mode12,mode13,mode14,mode15}. These studies typically employ joint training of source, relay, and destination neural networks to learn end-to-end transmission strategies that outperform conventional relaying schemes.

Full-duplex AE-based relaying has received comparatively less attention. The work in \cite{mode7} remains one of the few studies addressing AE-based one-way full-duplex (OWFD) relay networks. In this work, AE-based block coded modulation (BCM) and differential BCM were employed at the source and destination to mitigate the effects of residual self-interference following imperfect cancellation. The results demonstrated improved robustness compared to conventional AF relaying under self-interference conditions.

In \cite{mode8}, a deep learning architecture was proposed for a two-way AF relay network that employs a modulator neural network at the terminals, a physical network coding (PNC) modulator at the relay, and a demodulator neural network at the receivers. These networks were jointly trained to minimize cross-entropy loss between transmitted and decoded binary data, achieving significantly higher achievable sum rates than conventional TWAF protocols.

Building on this framework, \cite{mode9} proposed AE-based designs for both one-way AF and two-way AF relay networks operating over block-fading Rayleigh channels. This work was extended in \cite{mode10}, where the neural network at the relay was replaced with a conventional AF relay to reduce implementation complexity, reflecting practical constraints where relay nodes may lack CSI. The authors also introduced a bit-wise AE employing multi-label classification, demonstrating superior BER performance compared to conventional AF relaying with Hamming coding.

Hardware impairments were explicitly considered in \cite{mode11}, where AE-based two-way AF relaying was studied under in-phase and quadrature imbalance (IQI). Separate encoder and decoder networks were used at each terminal to mitigate IQI effects, resulting in performance gains of approximately 3~dB compared to maximum-likelihood detection without IQI compensation. This line of work was further extended in \cite{mode12} and \cite{mode13} to account for additional impairments, including multi-branch and multi-hop relaying, using ensemble decoding strategies.

Stacked autoencoder (SAE) architectures were proposed in \cite{mode14} and \cite{mode15} for one-way and two-way DF relay networks under Rayleigh fading. These designs employed layered encoder–decoder stacks to support both coded and differential coded modulation, depending on CSI availability. The proposed SAE-based relays outperformed conventional DF schemes and demonstrated improved robustness to channel uncertainty.

A CNN-based AE was introduced in \cite{mode4} to facilitate physical network coding in a two-way HD DF relay system. By jointly learning modulation, coding, and decoding, the proposed approach outperformed both conventional PNC schemes and standard AE-based PNC across multiple modulation orders.

Recently, AE-based relay architectures have been extended to semantic communication. In \cite{mode45_semanticrelay}, a semantic communication relay network was proposed in which sentence-level semantics are encoded and decoded using transformer-based architectures integrated with AEs. This design enables the relay to operate in the semantic domain, improving robustness and efficiency for language-based communication tasks.

\textcolor{black}{Relay channels have also been linked to intelligent reflective surfaces (IRSs), which can be viewed as a special case of passive relaying. Recent AE-based designs jointly optimize transmitter, IRS configuration, and receiver parameters to adaptively manipulate the propagation environment, offering promising gains in spectrum and energy efficiency.}

\subsection{Interference Channel}
\label{sub:interferencechannel}

Interference channels arise when multiple transmitter--receiver pairs communicate over shared spectral resources, resulting in mutual interference at the receivers. A representative two-user interference channel is illustrated in Fig.~\ref{fig:exinter}.

\begin{figure}[t!]
	\centering{\includegraphics[width=1\columnwidth]{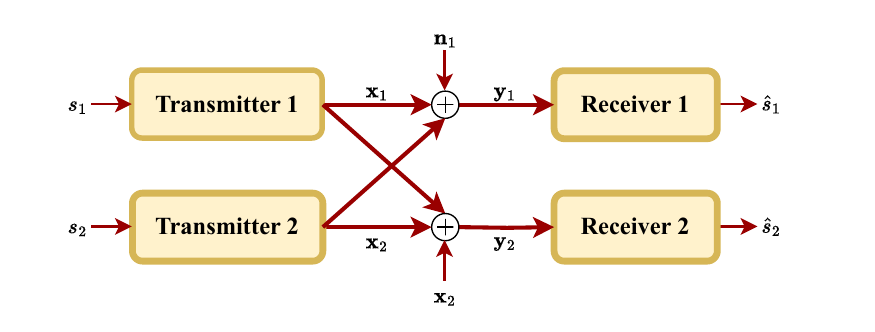}}
	\caption{Block diagram of a two-user interference channel.}
	\label{fig:exinter}
\end{figure}

For a two-user interference channel, the received signals at the two receivers can be expressed as
\begin{equation}
	\begin{aligned}
		y_{1} &= h_{11} x_{1} + h_{12} x_{2} + n_{1}, \\
		y_{2} &= h_{22} x_{2} + h_{21} x_{1} + n_{2},
	\end{aligned}
\end{equation}
where $y_{1}$ and $y_{2}$ denote the received signals at Receiver~1 and Receiver~2, respectively; $x_{1}$ and $x_{2}$ are the transmitted signals; $h_{11}$ and $h_{22}$ represent the direct-link channel coefficients; $h_{12}$ and $h_{21}$ correspond to the cross-link interference coefficients; and $n_{1}$ and $n_{2}$ denote additive noise terms. Interference channels are prevalent in dense wireless networks, particularly in cellular, device-to-device, and heterogeneous network deployments.

The foundational work in \cite{1hoydis} extended the standard AE framework to a single-input single-output (SISO) two-user interference channel by modeling the two transmitter--receiver pairs as coupled autoencoders with conflicting objectives. To balance the performance of both links, the authors proposed minimizing a weighted sum of the individual losses,
\begin{equation}
	\tilde{L} = \alpha \tilde{L}_{1} + (1-\alpha)\tilde{L}_{2},
\end{equation}
where $\alpha \in [0,1]$ is a weighting parameter. By dynamically updating $\alpha$ at each training iteration according to
\begin{equation}
	\alpha_{t+1} = \frac{\tilde{L}_{1}(\boldsymbol{\theta}_{t})}{\tilde{L}_{1}(\boldsymbol{\theta}_{t}) + \tilde{L}_{2}(\boldsymbol{\theta}_{t})},
\end{equation}
with $\alpha_{0}=0.5$, the proposed AE achieved identical block error rate (BLER) performance for both transmitter--receiver pairs. This work demonstrated that AE-based designs can implicitly learn interference-aware transmission strategies without explicit coordination.

The AE-based interference channel framework was extended to multiple-input multiple-output (MIMO) settings in \cite{mode53}, where the authors investigated $2\times2$ and $4\times4$ MIMO interference channels under Rayleigh fading. By jointly optimizing the encoders and decoders, the proposed AE-based MIMO IC designs outperformed conventional baselines, highlighting the scalability of AE-based interference management to multi-antenna systems.

Dynamic and time-varying interference scenarios were studied in \cite{mode31} and \cite{mode6.5}. In these works, standard AE architectures were augmented with adaptive deep learning mechanisms at the receiver, enabling online learning in response to changing interference conditions. Pilot symbols were employed to estimate an interference coupling parameter, which was then used by the receiver neural network to update decoding parameters in real time. These adaptive AEs demonstrated improved performance in channels with strong and very strong interference.

A practical multi-cell interference scenario was considered in \cite{mode6}, where multiple 5G base stations interfere with one another. The encoders at each base station were trained concurrently, with each encoder update modifying the interference landscape for the other users. This iterative training process enabled the AEs to converge to interference-aware constellations without explicit coordination among base stations.

Special cases of interference channels have also been investigated using AE-based designs. The two-user single-sided interference channel, commonly referred to as the Z-interference channel (ZIC), was studied in \cite{mode57,mode56,mode47}. In a ZIC, only one transmitter interferes with the unintended receiver. The proposed ZIC-DAE architectures employ bit-wise inputs, power normalization subnetworks, and binary cross-entropy loss functions to explicitly optimize BER. These designs achieved improved spectral and power efficiency compared to conventional orthogonal access schemes.

\textcolor{black}{Despite substantial progress, most AE-based interference channel studies focus on two-user scenarios. Extending these frameworks to dense multi-user interference networks with scalable training complexity and stability remains an open research challenge.}

\subsection{Interference Channel}
\label{sub:interferencechannel}

Interference channels arise when multiple transmitter--receiver pairs communicate over shared spectral resources, resulting in mutual interference at the receivers. A representative two-user interference channel is illustrated in Fig.~\ref{fig:exinter}.

\begin{figure}[t!]
	\centering{\includegraphics[width=1\columnwidth]{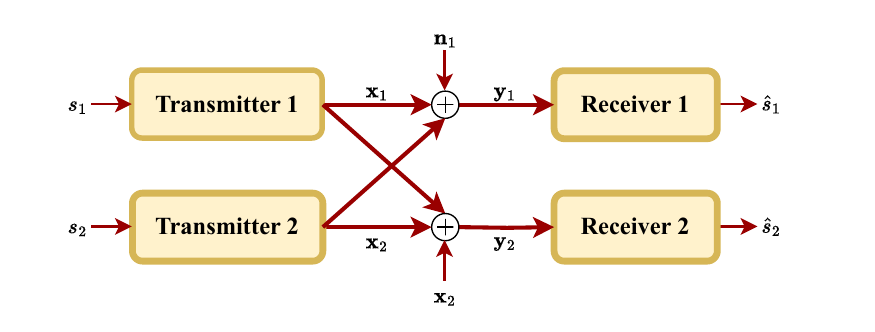}}
	\caption{Block diagram of a two-user interference channel.}
	\label{fig:exinter}
\end{figure}

For a two-user interference channel, the received signals at the two receivers can be expressed as
\begin{equation}
	\begin{aligned}
		y_{1} &= h_{11} x_{1} + h_{12} x_{2} + n_{1}, \\
		y_{2} &= h_{22} x_{2} + h_{21} x_{1} + n_{2},
	\end{aligned}
\end{equation}
where $y_{1}$ and $y_{2}$ denote the received signals at Receiver~1 and Receiver~2, respectively; $x_{1}$ and $x_{2}$ are the transmitted signals; $h_{11}$ and $h_{22}$ represent the direct-link channel coefficients; $h_{12}$ and $h_{21}$ correspond to the cross-link interference coefficients; and $n_{1}$ and $n_{2}$ denote additive noise terms. Interference channels are prevalent in dense wireless networks, particularly in cellular, device-to-device, and heterogeneous network deployments.

The foundational work in \cite{1hoydis} extended the standard AE framework to a single-input single-output (SISO) two-user interference channel by modeling the two transmitter--receiver pairs as coupled autoencoders with conflicting objectives. To balance the performance of both links, the authors proposed minimizing a weighted sum of the individual losses,
\begin{equation}
	\tilde{L} = \alpha \tilde{L}_{1} + (1-\alpha)\tilde{L}_{2},
\end{equation}
where $\alpha \in [0,1]$ is a weighting parameter. By dynamically updating $\alpha$ at each training iteration according to
\begin{equation}
	\alpha_{t+1} = \frac{\tilde{L}_{1}(\boldsymbol{\theta}_{t})}{\tilde{L}_{1}(\boldsymbol{\theta}_{t}) + \tilde{L}_{2}(\boldsymbol{\theta}_{t})},
\end{equation}
with $\alpha_{0}=0.5$, the proposed AE achieved identical block error rate (BLER) performance for both transmitter--receiver pairs. This work demonstrated that AE-based designs can implicitly learn interference-aware transmission strategies without explicit coordination.

The AE-based interference channel framework was extended to multiple-input multiple-output (MIMO) settings in \cite{mode53}, where the authors investigated $2\times2$ and $4\times4$ MIMO interference channels under Rayleigh fading. By jointly optimizing the encoders and decoders, the proposed AE-based MIMO IC designs outperformed conventional baselines, highlighting the scalability of AE-based interference management to multi-antenna systems.

Dynamic and time-varying interference scenarios were studied in \cite{mode31} and \cite{mode6.5}. In these works, standard AE architectures were augmented with adaptive deep learning mechanisms at the receiver, enabling online learning in response to changing interference conditions. Pilot symbols were employed to estimate an interference coupling parameter, which was then used by the receiver neural network to update decoding parameters in real time. These adaptive AEs demonstrated improved performance in channels with strong and very strong interference.

A practical multi-cell interference scenario was considered in \cite{mode6}, where multiple 5G base stations interfere with one another. The encoders at each base station were trained concurrently, with each encoder update modifying the interference landscape for the other users. This iterative training process enabled the AEs to converge to interference-aware constellations without explicit coordination among base stations.

Special cases of interference channels have also been investigated using AE-based designs. The two-user single-sided interference channel, commonly referred to as the Z-interference channel (ZIC), was studied in \cite{mode57,mode56,mode47}. In a ZIC, only one transmitter interferes with the unintended receiver. The proposed ZIC-DAE architectures employ bit-wise inputs, power normalization subnetworks, and binary cross-entropy loss functions to explicitly optimize BER. These designs achieved improved spectral and power efficiency compared to conventional orthogonal access schemes.

Following the success of TurboAE for point-to-point communication, DeepIC \cite{mode48.5} and its successor DeepIC+ \cite{mode48} were proposed for learning channel codes in two-user interference channels. DeepIC+ incorporates systematic interleaving and a two-phase training strategy in which encoder--decoder pairs are first pretrained on point-to-point channels and subsequently fine-tuned jointly on symmetric interference channels. This approach enabled DeepIC+ to outperform conventional interference management schemes.

AE-based interference handling has also been explored in optical and space communication contexts. In \cite{mode20}, the MAC AE model developed for space optical communication was adapted to a two-user interference channel by allowing encoded vectors to interfere at the receiver. The resulting AE with time sharing outperformed LDPC coding with time sharing across the entire SNR range.

For interference channels involving more than two users, interference alignment (IA) has been widely adopted to manage interference by confining it to lower-dimensional subspaces. In \cite{mode51}, an AE-based IA framework was proposed for a five-user interference channel, achieving better performance than classical IA algorithms such as Max-SINR and TIM-TIN. An alternative approach was presented in \cite{mode49}, where separate AEs were trained using a shared loss function for a three-user interference channel, achieving lower symbol error rates than FDMA and TDMA under Rayleigh fading.

Finally, interference channels with unknown or non-differentiable channel models were addressed in \cite{mode52}, where a generative adversarial network (GAN) was employed to emulate the channel during training. By decoupling AE training from explicit channel models, this approach enabled robust interference-aware communication in scenarios where accurate channel modeling is infeasible.

\textcolor{black}{Despite substantial progress, most AE-based interference channel studies focus on two-user scenarios. Extending these frameworks to dense multi-user interference networks with scalable training complexity and stability remains an open research challenge.}

\section{Non-Differentiable Channels}
\label{sec:nondiff}

\begin{table*}[]
\centering
\caption{\color{black}Comparison of the state-of-the-art techniques used for non-differentiable channel conditions.}
\label{tab:nondifferentiable}
\begin{tabular}{|l|l|l|l|l|l|l|}
\hline
{\color{black} \textbf{Technique Type}}                                                                                                                                     & {\color{black} \textbf{Approach}}                                                                                                                                                                    & {\color{black} \textbf{\begin{tabular}[c]{@{}l@{}}Channel\\ Type\end{tabular}}}                      & {\color{black} \textbf{\begin{tabular}[c]{@{}l@{}}Optimization\\ Method\end{tabular}}}            & {\color{black} \textbf{Strengths}}                                                                                                                                                                                                                      & {\color{black} \textbf{Limitations}}                                                                                                                                                                               & {\color{black} \textbf{Best Use Cases}}                                                                                                                                                              \\ \hline
{\color{black} \begin{tabular}[c]{@{}l@{}}Alternating\\ Training (AT)\\ \cite{aoudiaWithoutChannelModel},\\ \cite{deep_deterministic_zhang_modelfree}\end{tabular}}         & {\color{black} \begin{tabular}[c]{@{}l@{}}Iteratively\\ alternates\\ between training\\ the transmitter\\ and receiver NN\end{tabular}}                                                              & {\color{black} \begin{tabular}[c]{@{}l@{}}AWGN,\\ Rayleigh\\ fading,\\ Rician\\ fading\end{tabular}} & {\color{black} \begin{tabular}[c]{@{}l@{}}SGD using\\ cross-entropy\\ loss function\end{tabular}} & {\color{black} \begin{tabular}[c]{@{}l@{}}Outperforms fully\\ supervised training\end{tabular}}                                                                                                                                                         & {\color{black} \begin{tabular}[c]{@{}l@{}}Inefficient since it\\ cannot jointly\\ optimize both NNs;\\ requires dedicated\\ feedback channel\\ between receiver\\ and transmitter\\ for training\end{tabular}}     & {\color{black} \begin{tabular}[c]{@{}l@{}}Messaging;\\ classification and\\ reconstruction of\\ raw data\end{tabular}}                                                                               \\ \hline
{\color{black} \begin{tabular}[c]{@{}l@{}}Cubature Kalman\\ Filter (CKF)\\ \cite{jovanovicModelfree}\end{tabular}}                                                          & {\color{black} \begin{tabular}[c]{@{}l@{}}Use CKF for\\ optimization\\ during training,\\ as opposed\\ to gradient-based\\ techniques\end{tabular}}                                                  & {\color{black} AWGN}                                                                                 & {\color{black} CKF}                                                                               & {\color{black} \begin{tabular}[c]{@{}l@{}}Suitable for channels\\ with finite memory\end{tabular}}                                                                                                                                                      & {\color{black} (not discussed)}                                                                                                                                                                                    & {\color{black} \begin{tabular}[c]{@{}l@{}}Training AEs,\\ e.g. for\\ Geometric\\ Constellation\\ Shaping (GCS)\\ or with recurrent\\ NNs; suitable for\\ channels with\\ finite memory\end{tabular}} \\ \hline
{\color{black} \begin{tabular}[c]{@{}l@{}}Reinforcement\\ Learning (RL)\\ \cite{deep_deterministic_zhang_modelfree},\\ \cite{modelfree_REINFORCEMENTLEARNING}\end{tabular}} & {\color{black} \begin{tabular}[c]{@{}l@{}}Use actor-critic\\ approach, where\\ the transmitter is\\ the actor; or use\\ Deep Q-Learning\\ approach\end{tabular}}                                     & {\color{black} --}                                                                                   & {\color{black} \begin{tabular}[c]{@{}l@{}}Stochastic\\ Policy\\ Gradient\end{tabular}}            & {\color{black} \begin{tabular}[c]{@{}l@{}}Outperforms other\\ DL-based training\\ schemes\end{tabular}}                                                                                                                                                 & {\color{black} \begin{tabular}[c]{@{}l@{}}Slower\\ convergence\\ than non-RL\\ schemes,\\ convergence is\\ not guaranteed\\ at all in some\\ cases\end{tabular}}                                                   & {\color{black} \begin{tabular}[c]{@{}l@{}}Semantic\\ communications\end{tabular}}                                                                                                                    \\ \hline
{\color{black} \begin{tabular}[c]{@{}l@{}}GANs\\ \cite{oshea2018_overtheair_GAN},\\ \cite{approximatingthevoid}\end{tabular}}                                               & {\color{black} \begin{tabular}[c]{@{}l@{}}Represent channel\\ model as a NN\\ that is trained\\ jointly with the\\ encoder and\\ decoder NN\\ during training\end{tabular}}                          & {\color{black} \begin{tabular}[c]{@{}l@{}}black-box\\ channel\end{tabular}}                          & {\color{black} Adam}                                                                              & {\color{black} \begin{tabular}[c]{@{}l@{}}Does not require\\ pre-training to\\ match channel\\ conditions; can\\ adapt to stochastic\\ impairments\end{tabular}}                                                                                        & {\color{black} \begin{tabular}[c]{@{}l@{}}Future work includes\\ transferring a stream\\ of symbols, adapting\\ to changing channel\\ conditions, and\\ scalability for larger\\ block sizes\end{tabular}}         & {\color{black} Stochastic channels}                                                                                                                                                                  \\ \hline
{\color{black} \begin{tabular}[c]{@{}l@{}}Online Meta-\\ Learning\\ \cite{metalearning_modelfree}\end{tabular}}                                                             & {\color{black} \begin{tabular}[c]{@{}l@{}}During training,\\ transmitter sends\\ pilot messages to\\ receiver, and a\\ feedback link\\ enables\\ communication\\ between the two\\ NNs\end{tabular}} & {\color{black} \begin{tabular}[c]{@{}l@{}}Rayleigh\\ fading\end{tabular}}                            & {\color{black} \begin{tabular}[c]{@{}l@{}}SGD using\\ cross-entropy\\ loss function\end{tabular}} & {\color{black} \begin{tabular}[c]{@{}l@{}}Reduces the amount\\ of pilot signals\\ required; can\\ modify encoder\\ and decoder to adapt\\ to new channel\\ conditions; does not\\ require feedback link\\ during testing and\\ deployment\end{tabular}} & {\color{black} \begin{tabular}[c]{@{}l@{}}Large channel\\ correlation leads\\ to meta-overfitting,\\ while small\\ correlation reduces\\ the number of\\ samples available\\ during meta-\\ training\end{tabular}} & {\color{black} \begin{tabular}[c]{@{}l@{}}Channel\\ conditions that\\ require large\\ amounts of pilot\\ blocks; situations\\ where no feedback\\ link is available in\\ deployment\end{tabular}}    \\ \hline
\end{tabular}
\end{table*}

The standard AE architecture for end-to-end learning of the PHY and its variants interpret the transmitter, channel, and receiver as a single DNN \color{black}and jointly trains them using back-propagation\color{black}. 
However, practical implementation is hindered by the fact that the gradient of the instantaneous channel transfer function must be known in order to perform \color{black}back-propagation\color{black}, \color{black}and the fact that \color{black} as the channel is treated as a black box in practice \color{black}means this is often impractical \color{black} \cite{aoudiaModelfree}. 
Furthermore, a practical channel generally encapsulates hardware constraints such as quantization, which is non-differentiable by definition.

To this end, methods to train the AE communication systems without the need for a differentiable channel model, hence called \textit{model-free training} methods, are an important area of research.
\color{black}Current \color{black} state-of-the-art models can address specific channel models, such as the Poisson channel\color{black}, \color{black} and several model-free training methods are effective across several channel types.

Non-differentiable channels represent a critical yet \color{black}under-explored \color{black} area \color{black}of \color{black} communication systems, as their inherent complexity poses significant challenges for researchers.  
Despite \color{black}how difficult gradient estimation is \color{black} in non-differentiable settings, advancing research in this area is crucial for developing realistic channel models that better capture the behavior of practical communication systems. 
We encourage researchers to tackle these challenges and develop innovative gradient approximation techniques, like the approach we proposed in \cite{poissongradient}, to unlock new opportunities \color{black}to enhance \color{black} system performance and robustness.


\subsection{Model-free Training}

\color{black}The authors of \cite{aoudiaWithoutChannelModel} \color{black} introduced an AE training method for unknown or non-differentiable channel models \color{black}and \color{black} proposed an alternating algorithm that switches between training the receiver NN to minimize the loss gradient \color{black}with respect to \color{black} the receiver parameters and training the transmitter NN to minimize an approximated loss gradient obtained by sampling a distribution fitted to the channel input. 
\color{black}The authors \color{black} continued this work in \cite{aoudiaModelfree}, where they \color{black}performed \color{black} both simulation-based experiments \color{black}and \color{black} hardware \color{black}implementations \color{black} using a \color{black}software-defined \color{black} radio (SDR). Comparing their training scheme \color{black}with a \color{black} standard AE \cite{1hoydis} in both AWGN and Rayleigh block-fading, \color{black}and both schemes achieved \color{black} identical performance. In \color{black}the simulations\color{black}, the proposed scheme also outperformed QPSK \color{black}in \color{black} both AWGN and Rayleigh block-fading channels \color{black}and came close \color{black} to the near-optimal sphere packing algorithm developed by Erik Agrell \cite{agrell_database}. In the hardware deployment tests, the model-free training scheme outperformed both QPSK and Agrell\color{black}'s near-optimal sphere packing algorithm \color{black} \cite{agrell_database} \color{black}in \color{black} wireless \color{black}channels and \color{black} coaxial cable \color{black}channels\color{black}. 

\color{black}In \cite{jovanovicModelfree}, the authors \color{black} propose an alternative method for non-differentiable channels \color{black}that makes use of a \color{black} gradient-free \color{black}cubature Kalman filter (CKF)-based \color{black} training method. 
The CKF is a non-linear filter that can approximate probability distributions without performing any gradient calculations, \color{black}which makes \color{black} it suitable for non-differentiable channel models. 
The authors proved CKF-based communication \color{black}is viable \color{black} by achieving the same results on differentiable channels as the standard \color{black}back-propagation \color{black} algorithm. 
\color{black}Their training algorithm showed that learned AE constellations can outperform conventional constellation shapes in a non-differentiable phase noise channel by \color{black} expressing the AE weights and system as state-space models and then applying CKF for encoder and decoder weight estimation.

A deep deterministic policy gradient (DDPG) method is proposed in \cite{deep_deterministic_zhang_modelfree} \color{black}that \color{black} uses reinforcement learning to jointly train \color{black}a \color{black} CNN-AE. DDPG is an off-policy actor-critic algorithm that leverages deep Q-learning and \color{black}a \color{black} policy gradient to learn policies in high-dimensional, continuous action spaces \cite{continuouscontroldeepreinforcement}. \color{black}The authors compare the \color{black} DDPG scheme \color{black}with \color{black} the alternating training (AT) \color{black}schemes \color{black} proposed in \cite{aoudiaModelfree} and \cite{aoudiaWithoutChannelModel}, \color{black}and \color{black} show that their scheme outperforms the AT scheme in both Rayleigh and Rician fading channels \cite{deep_deterministic_zhang_modelfree}. 
RL is also used in \cite{modelfree_REINFORCEMENTLEARNING}, \color{black}but with \color{black} a stochastic policy gradient (SPG) rather than a deterministic one. In \cite{modelfree_REINFORCEMENTLEARNING}, SPG is used to design a semantic communication system that optimizes the transmitter and receiver NN \color{black}separately \color{black} to forego \color{black}requiring \color{black} a known channel model. Their model \color{black}performed comparably \color{black} to 
the model-aware semantic communication model proposed in \cite{modelaware_semantic}.

To overcome the shortcomings of SPG \color{black}and \color{black} DDPG schemes for non-differentiable channels not scaling well with the number of channel uses, long coherence time needed of the channel compared to the rate at which examples can be processed during training, poor sample efficiency during training, slow convergence for larger constellations, and poor generalization, the authors \color{black}of \color{black} \cite{causal_policy_gradient_modelfree} propose leveraging the causal relationships \color{black}that exist \color{black} between the state (transmitted symbols), the action (constellation shaping), and the reward (loss between received and transmitted symbols). 
\color{black}Using causal \color{black} reinforcement learning (CRL) \cite{towardCRL} \color{black}made it possible \color{black} to achieve \color{black}nearly \color{black} a 90\% decrease in convergence time\color{black}, \color{black} and, given full CSI, their model trained with CRL \color{black}gave \color{black} a BER close to the DDPG \color{black}scheme from \cite{deep_deterministic_zhang_modelfree} and the \color{black} and SPG scheme from \cite{aoudiaModelfree}.

An online learning method \color{black}that leverages \color{black} model-agnostic meta-learning (MAML) \cite{MAML} \color{black}is \color{black} proposed in \cite{metalearning_modelfree}\color{black}. It leverages\color{black} meta-training \color{black}to train \color{black} the decoder and joint training \color{black}for \color{black} the encoder\color{black}, with \color{black} a feedback link in the (meta-)training phase.   
The receiver meta-trains the decoder to quickly adapt to new channel conditions based on pilot bits \color{black}that precede \color{black} the payload in each frame\color{black}, \color{black} while the transmitter simultaneously trains the encoder using the feedback link.
In testing, the feedback link is removed and the pilots are used to dynamically update the decoder \color{black}in accordance with \color{black} channel conditions.
This \color{black}serves to \color{black} overcome the limitation in joint training \cite{aoudiaModelfree} of not being able to adapt to new channel conditions.
In contrast, online meta-learning is able to guarantee an \color{black}optimized average \color{black} performance for a variety of channel conditions \cite{metalearning_modelfree}.


The authors \color{black}of \color{black} \cite{residual_aids_jiang_modelfree} use a residual network-aided GAN (RA-GAN)\color{black}-based \color{black} training scheme to mitigate the gradient vanishing and overfitting \color{black}problems \color{black} that arise when using a traditional GAN. Their model outperformed the standard AE with GAN training scheme in the AWGN, Rayleigh fading, and DeepMIMO \cite{DeepMIMO_dataset} dataset channels.

Using GANs for end-to-end optimization over unknown channels \color{black}is \color{black} introduced in \cite{oshea2018_overtheair_GAN} and expanded on in \cite{approximatingthevoid}, where they implemented a variational GAN generator for channel estimation in \color{black}a \color{black} standard AE model. The variational GAN \color{black}approximates \color{black} an accurate conditional channel probability distribution to be used \color{black}when \color{black} training the standard AE\color{black}. The authors \color{black} quantified their results by comparing the predicted channel distribution PDF \color{black}with \color{black} ground truth PDFs for the \color{black}different types of \color{black} channels. 
A conditional GAN \color{black}is \color{black} also used in \cite{Hao_channelagnostic}\color{black}, which expands upon \color{black} the system \color{black}from \color{black} \cite{approximatingthevoid} \color{black}to make it \color{black} more applicable to time-varying channels. The authors \color{black}of \color{black} \cite{Hao_channelagnostic} also employ the use of pilot bits as part of the conditioning information for real-time channel information.
In \cite{Hao_conditionalGANS}, the authors improve \color{black}upon \color{black} this work by adding convolutional layers to increase the number of input bits \color{black}and \color{black} training their network to function better under practical conditions such as \color{black}in \color{black} channels with \color{black}ISI\color{black}.
A pilot-free end-to-end communication system \color{black}is developed \color{black} over unknown channels in \cite{Hao_bilinear}\color{black}. This work makes use of \color{black} a CNN-AE \color{black}and represents \color{black} the channel \color{black}with \color{black} an adversarial convolutional layer. The receiver also consists of a channel feature extractor module along with the data recovery decoder, showing significant improvements over ZF and MMSE over the frequency-selective and MIMO fading channels.
\color{black}The work in \cite{GAIforPHY_Survey2024} expands upon GAI and \color{black} GAN for PHY optimization\color{black}, and covers \color{black} applications \color{black}of signal classification/modulation to \color{black} channel modeling, channel estimation, channel equalization\color{black}, \color{black} IRS\color{black}, \color{black} and beamforming.




\subsection{Poisson Channel}

\begin{figure}[pt!]
	\centering{\includegraphics[scale=0.53]{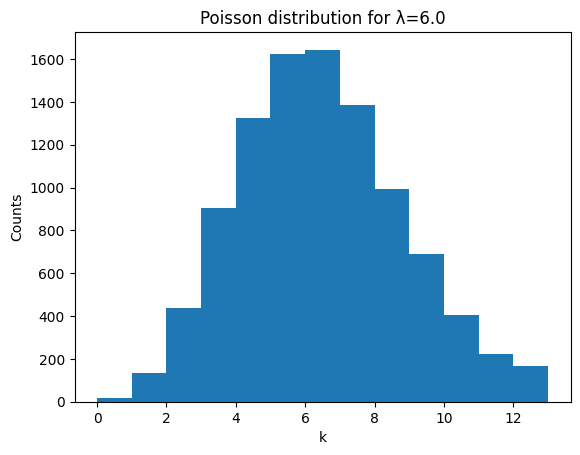}}
	\caption{Poisson distribution for photon count with $\lambda=6.0$ where $\lambda$ is the dark current rate of the photodetector and \(k\) is the number of events.}
	\label{fig:poisson}
\end{figure}

The discrete-time Poisson channel model is used in optical wireless communication (OWC) and space optical communication (SOC) to model the behavior of photons. \color{black}A \color{black} Poisson distribution is a probability distribution \color{black}that describes \color{black} the expected number of photons in a time frame and is shown in (\ref{eq:poisson})\color{black}:\color{black}
\begin{equation}
	f(k;\lambda)=\frac{\lambda^{k}e^{-\lambda}}{k!}
	\label{eq:poisson}
\end{equation}
where \(k\) is the number of photons received in a given time frame (non-negative integer), \(\lambda\) is the average or expected number of photons received per unit \color{black}of \color{black} time (the mean of the Poisson distribution), and \(e\) is Euler's number.
Although this channel model is highly accurate for SOC, \color{black}a \color{black} Poisson distribution is discrete and therefore non-differentiable\color{black}, \color{black} as shown in Fig. \ref{fig:poisson}.
 
\color{black}The authors of \color{black} \cite{poissongradient} \color{black}propose \color{black} a gradient approximation method for training AEs \color{black}in a \color{black} Poisson channel \color{black}that involves using \color{black} a non-gradient-based optimization scheme \color{black}that makes use of \color{black} a covariance matrix adaptation evolution strategy (CMA-ES) to address the non-differentiable channel \color{black}and adds \color{black} batch and layer normalization layers to the AE to increase the speed of convergence during training. The AE with CMA-ES outperforms a standard AE that uses particle swarm optimization (PSO) in place of gradient descent, \color{black}which shows that \color{black} that CMA-ES is a viable strategy for dealing with the gradient problem in non-differentiable channels.  

\color{black}In \color{black} \cite{poissondnn} \color{black}the authors propose an AE that is \color{black} dubbed SR-AE \color{black}that has been \color{black} adapted for \color{black}a \color{black} Poisson channel \color{black}and \color{black} uses the square root transform to approximate the gradient of the Poisson distribution. Additionally, the authors propose a double NN model \color{black}that \color{black} consists of two separate NNs in the transmitter and receiver for joint optimization \color{black}in \color{black} any channel whose conditional PDF or PMF is differentiable with respect to the channel input. This double NN model \color{black}performs \color{black} slightly better than the SR-AE \color{black}model \color{black} but has more network parameters since it has to compute two loss functions instead of one.
\color{black}A further \color{black} discussion on AEs for OWC and SOC \color{black}is provided \color{black} in the following section.

\section{Autoencoder for Optical Wireless communication}
\label{sec:owc}

\color{black}OWC \color{black} employs lasers to transmit data using carrier frequencies ranging from \(120-385\) THz \color{black}and provides \color{black} an alternative to RF communication technology. OWC \color{black}has \color{black} many advantages over RF, including a faster data rate \color{black}and \color{black} an unlicensed spectrum. The simplest and most commonly used modulation technique \color{black}for \color{black} OWC is intensity modulation and direct detection (IM/DD), in which the transmitted signal is modulated by the intensity of light and demodulated at the receiver \color{black}in accordance with the intensity detected\color{black}. However, unlike RF systems, OWC also has non-negativity, average power, and peak power constraints, so traditional RF techniques for the PHY cannot be applied directly to OWC. DL is able to work with these constraints \color{black}and \color{black} learn unknown channel models, \color{black}which makes it suitable for \color{black} designing end-to-end OWC systems.



\color{black}Optical fiber communication systems laid the \color{black} groundwork for applying AEs \color{black}in \color{black} optical communication and was introduced in \cite{owc5fiber}, where \color{black}the authors proposed \color{black} an AE for IM/DD and validated \color{black}it using \color{black} an experimental hardware testbed. The AE \color{black}performed as well as \color{black} a 2-level pulse amplitude modulation (PAM-2) system operating at 42 Gbaud, which paved the way for future research into \color{black}using AEs in \color{black} optical fiber systems. This \color{black}opened the door for \color{black} many subsequent works that explored diverse aspects including decoding optimization, OFDM-based modulation, and advanced Wasserstein-based AEs for optical fiber communication \cite{owc6fiber, owc7fiber, owc9fiber, owc15fiber, owc18fiber, owc3, owc8decoding, owc10OFDM, owc16wasserstein}.

As for the wireless domain, AEs have been proposed for various OWC and visible light communication (VLC) systems. For instance, in \cite{colorVLC}, AEs \color{black}are \color{black} introduced for multi-colored VLC systems, and in \cite{soltaniOWC}, AE-based models \color{black}are \color{black} applied to both P2P and MAC scenarios in OWC. These models demonstrated improvements over traditional on-off keying (OOK) and time-sharing schemes, respectively. An AE adapted to log-normal fading channels with noisy CSI was \color{black}also \color{black} proposed in \cite{zhuOWC}, achieving robust performance under these conditions.


\color{black}In \color{black} the realm of free space optical communication (FSO), \color{black}which is \color{black} a subset of OWC, significant progress has been made in leveraging AEs to address system challenges. In \cite{owc2}, an AE was developed to optimize unmanned aerial vehicle (UAV)-to-ground communication \color{black}using \color{black} infrared laser-based FSO with atmospheric turbulence modeled using the Gamma-Gamma  distribution. This model achieved a notable BER improvement over the hard decision forward error correction (HD-FEC) baseline at approximately 22 dB. Extending this work, \cite{owc4} introduced a CNN-AE designed for FSO communication under unknown CSI and Gamma-Gamma fading conditions. This model, \color{black}which is \color{black} termed FSO-E2E, demonstrated robust performance even under strong turbulence scenarios.

\color{black}The \color{black} application of AEs in OFDM-based OWC systems has also been explored. For example, \color{black}in \color{black} \cite{owc12} an OFDM AE \color{black}that was \color{black} initially developed for RF communication \color{black}was adapted \color{black} to OWC systems specifically for asymmetrically-clipped optical OFDM (ACO-OFDM). \color{black}When trained \color{black} on AWGN and multi-path fading channels, \color{black}the model \color{black} achieved superior BER performance \color{black}than \color{black} traditional QAM-based ACO-OFDM systems.

Another promising domain for AE application is underwater wireless optical communication (UWOC), which offers advantages such as high data rates and low latency. However, UWOC systems face unique challenges, including absorption and scattering in water, multi-path channel fading, and ISI. \color{black}The work in \cite{owc13underwater} is inspired by the \color{black} work in \cite{uwocchannelGAN}, \color{black}and proposes \color{black} a hybrid AE and GAN (AE-GAN) framework to mitigate these challenges. The AE-GAN framework \color{black}employs \color{black} adaptive one-bit quantization thresholds to reduce receiver complexity and power consumption, achieving superior BER performance compared to traditional UWOC systems using convolutional codes and AEs with known CSI.

Recent advancements have also incorporated learning-based channel modeling in AE frameworks for OWC. In particular, \cite{owc14biLSTM} \color{black}introduces \color{black} a bidirectional LSTM (BiLSTM)-based channel model to approximate realistic FSO channels. This model \color{black}incorporates \color{black} geometric loss, atmospheric attenuation, and turbulence effects during training \color{black}and achieves \color{black} lower BER and extended transmission distances \color{black}when integrated in an AE framework than \color{black} conventional schemes like pulse position modulation (PPM) and OOK.

In the domain of visible light communication (VLC), AEs have been utilized to address dimming constraints and multi-path noise components. For instance, \color{black}in \color{black} \cite{binaryVLC} an AE framework \color{black}is proposed that uses \color{black} a novel stochastic binarization method to generate binary codewords with universal dimming support. \color{black}The system transmits \color{black} messages using OOK while ensuring compliance with dimming requirements and \color{black}accounts \color{black} for both line-of-sight and multi-path noise. The proposed model \color{black}outperforms \color{black} traditional VLC frameworks \cite{binaryVLCbaseline} in terms of SER performance under both known and unknown CSI scenarios.

Soltani \textit{et al.} compare the BLER of their OWC AE to state of the art model-based OWC systems \cite{mode18}. The authors in \cite{mode18} find that both their (4,4) and (2,2) AE models outperform uncoded OOK. \color{black}In \color{black} \cite{mode19}\color{black}, the authors \color{black} implement their model \color{black}in both \color{black} an AWGN channel \color{black}and \color{black} a log-normal fading channel to account for degradation caused by turbulence, \color{black}which is \color{black} a common problem in OWC. \color{black}They \color{black} assume both perfect and noisy CSI \color{black}and find \color{black} that their (7,4) model outperforms Hamming-coded (7,4) OOK with hard-decision decoding, while performing slightly worse for Hamming-coded (7,4) OOK with soft-decision decoding \color{black}in all cases \color{black} \cite{mode19}.

Given the unique challenges OWC systems face, including non-negativity constraints, peak power limitations, and shot-noise effects, \color{black}the authors of \color{black} \cite{owc1} propose \color{black}a \color{black} differential signaling\color{black}-based AE\color{black}. This approach allows the encoder to produce vectors \color{black}that contain \color{black} both positive and negative elements using two optical sources operating at different wavelengths. The model \color{black}proves to outperform \color{black} traditional PAM-4 and PAM-8 schemes as well as AE models developed for RF and SOC systems, such as those in \cite{1hoydis} and \cite{soc2}.

Researchers should expand their focus beyond OOK modulation in optical wireless communication AEs \color{black}because the \color{black} low spectral efficiency of OOK limits its potential for high-capacity communication systems. 
Exploring advanced techniques, such as amplitude\color{black}- \color{black} or frequency-based modulation, can improve data transmission rates, adaptability, and performance in varying environments. 
AEs can optimize these techniques by learning the characteristics of the communication channel \color{black}to enable \color{black} more efficient and reliable systems for future communication technologies.

\subsection{Space Optical communication}

\begin{figure}[t!]
	\centering{\includegraphics[width=1\columnwidth]{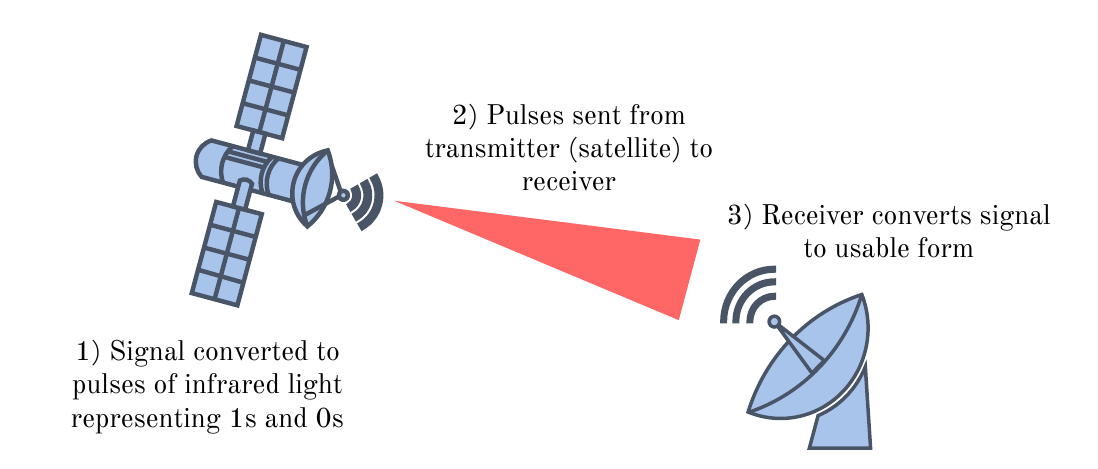}}
	\caption{Block diagram of simple downlink SOC system.}
	\label{fig:SOCdownlink}
\end{figure}

\color{black}SOC \color{black} is similar to OWC in that both use lasers to transmit signals. A key difference between the two is that SOC is used \color{black}for \color{black} very long-distance communication, such as \color{black}to send \color{black} data between a satellite and \color{black}a \color{black} ground station \cite{soc3}. State-of-the-art AE models for SOC are able to outperform traditional systems. The AE BER is further \color{black}enhanced \color{black} with the introduction of models \color{black}containing \color{black} multiple decoders \cite{soc1}. Adding batch and layer normalization to the model can also improve the BER. Additionally, SOC AEs can be used for multiple-access scenarios \cite{soc2}.

\color{black}The authors of \cite{mode21} expand \color{black} upon the work done in OWC by applying an AE model for use in space optical communication (SOC), considering both the AWGN and log-normal fading channel. The authors implement a novel multi-decoder architecture in their AE and find that their model outperforms both model-based convolutional codes \color{black}and \color{black} state-of-the-art AEs \color{black}for code rates \(R=1/2\) and \(R=1/3\)\color{black} \cite{mode21}. In \cite{mode22}, \color{black}the authors \color{black} design a novel AE with a channel estimator NN for use in a downlink SOC channel under log-normal or G-G fading. The authors find that their channel estimator NN outperforms \color{black}state-of-the-art \color{black} learning-based models and achieves the same performance as the MMSE estimator \color{black}\cite{mode22}\color{black}. Furthermore, the \color{black}AE proposed \color{black} in \cite{mode22} \color{black}outperforms \color{black} both \color{black}state-of-the-art \color{black} AEs and model-based convolutional codes used in SOC.

\section{Deployment Considerations and Reducing Complexity}
\label{section:future}

\subsection{Considerations for Deployment}

Deploying an AE-based communication systems on real hardware is a key \color{black}research area \color{black} that needs to be discussed. 

The first proof-of-concept AE-based communication system implemented on real hardware \color{black}is \color{black} discussed in \cite{mode16}, \color{black}and \color{black} the authors deployed their P2P AE model on a \color{black}software-defined \color{black} radio (SDR). 
As \color{black}is \color{black} stated in \cite{mode16}, the main challenges related to hardware implementation were 1) having an unknown channel transfer function, 2) accounting for hardware imperfections in AE design, and 3) sending continuous transmissions of small messages to account for the exponential growth in complexity as message size increases.

To account for Challenge 1, the authors proposed a two-phase training strategy \color{black}in which \color{black} the AE \color{black}is pre-\color{black}trained end-to-end on an AWGN channel model \color{black}prior to deployment\color{black}, and then the receiver \color{black}is \color{black} fine-tuned \color{black}in accordance with \color{black} IQ samples received \color{black}over \color{black} the channel using a form of transfer learning \color{black}after deployment\color{black} \cite{transferlearning}. 
To account for Challenge 2, the AWGN channel model that the AE was first trained \color{black}on \color{black} was modified to include upsampling \color{black}and \color{black} pulse shaping, constant sample time offset, constant phase offset, and carrier frequency offset (CFO). 
The sending of continuous transmissions in Challenge 3 requires timing synchronization to determine which \color{black}samples received \color{black} from the sample stream should be passed through the receiver NN to be decoded, which makes the system vulnerable to sampling frequency offset (SFO) \color{black}and \color{black} ISI from pulse shaping at the transmitter. To account for this, a sequence decoder is implemented in the receiver part of the AE as a combination of multiple NNs. The sequence decoder includes a phase estimator NN that \color{black}specifically \color{black} estimates phase offset over multiple subsequent messages, as well as a feature extractor NN to facilitate transfer learning and a frame synchronization NN to tackle the SFO issue. 
The authors \color{black}of \color{black} \cite{mode16} were able to successfully implement an AE-based P2P communication system with the PHY being entirely made up of NNs, with their fine-tuned model performing only \(1 dB\) worse than the GNU \color{black}radio \color{black} DQPSK (GR-DQPSK) baseline. 

Through \color{black}their \color{black} experiments in \cite{mode16}, the authors learned that the main difficulty when it \color{black}comes to the \color{black} hardware deployment of AE models \color{black}is that \color{black} the actual channel \color{black}differs from \color{black} the stochastic channel model used for training, which led to significant performance loss compared \color{black}with \color{black} their simulated results. As discussed in Section \ref{sec:nondiff}, training AEs to function well \color{black}with \color{black} unknown or non-differentiable channel models is an important topic to be researched further\color{black}. It \color{black} is what motivated the authors \color{black}of \color{black} \cite{aoudiaModelfree} to determine the deployment viability of their model-free training algorithm on real hardware.

\color{black}

A number of practical challenges arise when deploying AE-based transceivers in operational wireless systems. Beyond computational load, a key constraint is real-time processing: PHY-layer inference must satisfy stringent per-symbol or per-slot latency budgets, requiring carefully optimized parallel implementations and hardware-aware model compression. Hardware mismatch represents another bottleneck, as distortions introduced by RF chains, quantized phase shifters, nonlinear power amplifiers, and synchronization imperfections can significantly degrade performance if not explicitly modelled during training or corrected through online calibration. Robustness evaluation also becomes essential; AE-based receivers must be validated over large channel ensembles, mobility profiles, and interference conditions to avoid brittleness outside their training distribution. Furthermore, integrating learned PHY blocks into 5G/6G protocol stacks requires compatibility with standardized interfaces and control loops, including potential deployment within near-RT or non-RT RIC environments for ML-driven RAN optimization. Finally, model life-cycle management—covering monitoring, versioning, online updates, and fallback mechanisms—must be incorporated to ensure reliability over hardware aging, software revisions, and evolving spectrum environments. Emerging studies and prototyping efforts underscore the importance of these considerations for sustainable deployment of learning-based PHY components~\cite{Cammerer2023NeuralRx,Owfi2025OMLCAE,Bonati2024ORANML}.

\color{black}
E2E learning for the physical layer and AI-native radio access networks (RANs) is also attracting increasing attention from standardization bodies and industry. On the standardization side, 3GPP has introduced dedicated study items on AI/ML for the NR air interface (TR~38.843) and on AI/ML management in the 5G system (TR~28.908), which formalize terminology, operational workflows, and requirements for model training, deployment, inference, and life-cycle management across RAN and core network functions~\cite{3gpp_tr_38843,3gpp_tr_28908,lin_ai_5g_adv}. In parallel, the O-RAN Alliance specifies a disaggregated and programmable RAN architecture with non-RT and near-RT RAN Intelligent Controllers (RICs) and open interfaces that are explicitly designed to host AI/ML-based xApps and rApps for data-driven RAN optimization~\cite{bonati_intelligence_oran,yungaicela2024misconfig}. These activities are complemented by ITU-T and ETSI work on ML sandboxes and management frameworks, which address aspects such as model governance, monitoring, and security~\cite{itu_y3181}. On the industrial side, equipment vendors and test-and-measurement suppliers have demonstrated neural receivers and learned constellations in 5G-compliant hardware-in-the-loop testbeds, providing early evidence that AE-based PHY blocks can be integrated into realistic RF chains~\cite{cammerer2023neural,rohde2023neural}. Collectively, these efforts indicate a rapidly maturing ecosystem for trialing AE-based transceivers, managing their life cycle at scale, and standardizing interfaces for exchanging ML artifacts in future mobile networks.

\color{black}






\subsection{Reducing AE Complexity}

Researchers are encouraged to focus on reducing computational complexity to ensure the practical deployment of models addressing non-differentiable channels. Simplifying algorithms without compromising accuracy is essential for enabling their integration into real-world systems with limited computational resources, such as IoT devices, space communication platforms, and energy-constrained networks. Developing lightweight, efficient solutions will significantly enhance the scalability and accessibility of these advanced techniques.
One of the main problems when it comes to practically implementing AE-based PHY communication systems is \color{black}that complexity grows exponentially \color{black} as message length \(M\) increases when using a one-hot vector input \(b\), as the one-hot vector is of length \(M=2^k\). Currently, the maximum feasible message length is \(k<100\) bits, i.e., \(M=2^{100}\) messages \cite{1hoydis}.

One way to \color{black}overcome this is to omit \color{black} the use of a one-hot vector\color{black}, pass \color{black} binary data \color{black}to be transmitted \color{black} through the encoder, \color{black}and then convert \color{black} the transmitted bits into one-hot vector form at the decoder (receiver). This would allow the decoder model to remain a multi-classification problem rather than a multi-label problem \color{black}and reduce \color{black} the complexity exponentially. However, since transmitting binary data would cause significant degradation \color{black}of \color{black} the BER, we need to mitigate this degradation by a sufficient modification of the encoder structure, perhaps by including more fully-connected or convolutional layers.

\begin{table}[t] 
	\caption{Comparison of encoding schemes and their respective encoder parameters.} 
	\label{tab:onehottwohot} 
	\centering
	\begin{tabular}{|c|c|} 
		\hline
		Encoding Scheme & Number of Learnable Parameters \\
		\hline
		One-hot Encoding & 15,505 \\
		\hline
		Two-hot Encoding & 4,163 \\
		\hline
	\end{tabular}
	
\end{table}

\begin{figure}[t]
	\includegraphics[width=\columnwidth]{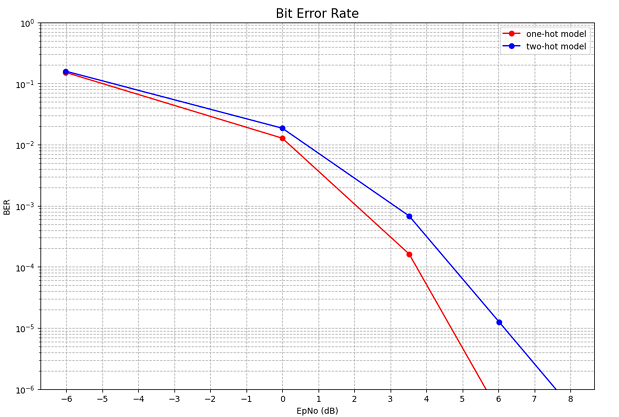}
	\caption{BER performance of one-hot encoding versus two-hot encoding in AWGN channel using \((7,21)\) AE.}
	\label{fig:onehottwohot}
\end{figure}

We also experimented with the AE using a \textit{two-hot vector} input, where the message \(b\) is encoded into a vector of length \(\alpha\) that consists of exactly two values that are equal to \(1\), and the rest \color{black}that are \color{black} zeroes.
To ensure that each message \(b\) can be mapped to a unique vector, the length of the vector is selected such that \(^{\alpha} C_{2} \geq M\).
For \(M=128\), the smallest value for $\alpha$ that satisfies this condition is \(17\), whereas if one-hot encoding is used, the input length must be \(128\).

In Fig. \ref{fig:onehottwohot}, both encoding schemes are compared using a \((7,21)\) AE \color{black}in \color{black} an AWGN channel and the same peak intensity \(A=3\).
It can be \color{black}observed \color{black} that one-hot encoding outperforms two-hot encoding by \(1.8\) \(dB\) at BER \(10^{-6}\). 
Table \ref{tab:onehottwohot} shows the number of learnable parameters in the encoder \color{black}when using each \color{black} encoding scheme.
The encoder \color{black}that uses \color{black} two-hot encoding for the input has \(73\%\) fewer parameters than the encoder that uses one-hot encoding, which makes it more suitable for mobile, low-power devices.
 


\section{Future Avenues for Research}


\color{black}

Looking forward, several research directions remain critical for advancing AE-based physical-layer systems toward commercial viability. A central challenge is achieving strong generalization across channels, mobility regimes, and hardware platforms, potentially requiring physics-informed inductive biases, domain-randomized training pipelines, and meta-learning strategies capable of rapid post-deployment adaptation. Another promising direction is the development of AI-native air interfaces in which modulation, coding, and waveform design are co-optimized with the neural transceiver architecture, supported by scalable differentiable channel models and system-level differentiability. Reducing computational cost while maintaining reliability calls for joint algorithm–hardware co-design, including quantized and low-rank AEs tailored to emerging accelerators. The integration of AEs with RAN intelligence frameworks opens opportunities for real-time OTA refinement, closed-loop coordination with xApps/rApps, and task-oriented communication services. Finally, rigorous robustness certification, explainability tools, and standardized procedures for ML model exchange and auditing will be essential to ensure safety, interoperability, and regulatory acceptance of learned PHY layers as 6G systems mature.

Another important avenue for future research concerns the interaction between AE-based PHY learning and emerging distributed-learning paradigms. Federated and split learning offer structured mechanisms for coordinating large-scale training across multiple devices under privacy, bandwidth, and compute constraints, and have been extensively studied in wireless and IoT settings~\cite{Chen2022FLIoT,Wu2023SplitSL}. However, their convergence behavior when optimizing PHY-level objectives such as BLER, EVM, and waveform fidelity is still insufficiently understood. In a similar spirit, semantic communication and task-oriented sensing operate at a higher abstraction layer and aim to transmit meaning rather than raw symbols, which raises new questions on how semantic encoders and decoders can be jointly aligned with AE-based PHY representations~\cite{Zhang2024SemCom6G,Ma2023TOESemCom}. These intersections suggest that AE-based transceivers should not be viewed as independent of broader learning frameworks, but as complementary components in a unified cross-layer design. Exploring how distributed learning, semantic compression, and self-supervised representation learning can be integrated with AE-based waveform generation and detection remains an open and promising direction.

\color{black}

                                                                  
Zayat \textit{et al.} \cite{transformermaskedAE} discuss transformer-masked autoencoders (TMAE\color{black}s\color{black}) and their applications in wireless communication for 5G and beyond. Transformers offer several advantages over classic DNNs, such as \color{black}being able to use \color{black} the attention mechanism to extract time dependencies from data; \color{black}being able \color{black} to use a pre-trained model for various specific tasks, \color{black}which is \color{black} known as transfer learning; and \color{black}being able to \color{black} use parallel processing\color{black}, \color{black} which increases efficiency. Due to these features, transformers have shown strong performance in the fields of computer vision and natural language processing.
The main principle \color{black}behind \color{black} the transformer is the self-attention mechanism, in which segments of input data are encoded to correspond to different positions in a sequence. This can also be extended to multi-head attention by applying the self-attention process in parallel multiple times, which allows the model to extract more complex relationships from the data. A TMAE, based on transformers, uses this principle to reconstruct the transmitted original data from incomplete observed segments. This reduces the amount of data a transmitter must send to the receiver, as the receiver \color{black}is \color{black} able to predict any missing data from the transmitted segments. Presently, the main challenges \color{black}that are \color{black} faced by the TMAE are costly hardware and energy requirements and a lack of pre-trained transformer models suited for wireless communication applications.
The authors \color{black}of \color{black} \cite{transformermaskedAE} present the numerical results of training a TMAE for JPEG image compression, which helps to improve the rate of data transfer in wireless communication. The JPEG-TMAE \color{black}outperforms \color{black} the standard JPEG compression scheme and can even outperform other state-of-the-art models. In addition, TMAEs have many other applications in 5G and beyond, including channel coding, channel estimation, and encryption.

\color{black}
The authors of \cite{frombitstosemantics} \color{black} provide an overview of semantic communication and how it can be integrated with AI. In contrast to traditional wireless communication, which focuses on the correctness of  transmitted bits, semantic communication places emphasis on maintaining the accuracy of the signal’s semantic meaning. The encoder and decoder each have a knowledge base that stores semantic information, \color{black}and is used to extract only the data features relevant to the current task. \color{black}This method improves the efficiency of transmission by removing redundant data. Additionally, for changing information, semantics are defined as the timeliness of information over time, which can be measured using the age of information (AoI) metric.

In traditional wireless communication, sensing and sampling means taking samples of a signal at a fixed rate, and creating a large amount of data that is then sent to the receiver. Compressed sampling combines this process with data compression to reduce the amount of samples required. \color{black}Semantic-aware sampling outperforms traditional techniques in terms of numerical results, and AoI-based semantic sampling is suitable for the real-time transmission of correlated data sequences. \color{black}

In a similar fashion, semantic communication can improve  source and channel coding, which consists of compressing source data into bit sequences and then adding correction codes that help with error detection and correction. \color{black} Semantic-aware joint source and channel coding can be integrated with AEs by connecting a semantic encoder to the channel encoder, \color{black}and a semantic decoder for the channel decoder. \color{black}DL-based joint semantic channel coding (JSCC) \color{black}outperforms  traditional source \color{black}and channel coding in fields such as speech transmission. \color{black}However, \color{black}the DNN model has to be retrained for every new task, which requires a large amount of data. With the rise of DL, semantic-based modulation methods have also become more efficient and hold potential for future research. The strongest semantic modulation scheme is semantic-aware joint coding-modulation, which combines JSCC with digital modulation and can outperform other semantic-aware digital modulation techniques.

Semantic communication has numerous applications in end-to-end communication, including data reconstruction, image classification, and IOT. AI plays a major role in improving semantic communication, especially with the rise of large language models, which excel in preserving natural language. However, DL has high computational complexity, so it would require a large amount of power and computational resources to implement. There is interest in using computer networks to share data more efficiently between devices, which can help mitigate this issue. \color{black}One possible area for future research of semantic communication includes the Shannon information theory, which determines the theoretical capacity of a physical information channel, but this and other similar performance metrics are focused on traditional communication techniques and may not accurately model the characteristics of semantic communication and the virtual semantic channel. It is also important to investigate the tradeoffs between increasing security \color{black}with \color{black} extra bits and maintaining an efficient transmission speed. Lastly, the power of semantic communication also makes it suitable for applications in 6G networks and beyond.

\section{Conclusion}

In this survey, we explored the transformative potential of end-to-end DL \color{black}frameworks for optimizing the PHY of wireless communication systems. \color{black} By delving into various AE architectures and their applications across different communication scenarios, we highlighted how DL transcends traditional model-based approaches, \color{black}and enables seamless \color{black} integration of transmitter and receiver components for joint optimization. The \color{black} works surveyed demonstrate that end-to-end frameworks \color{black}significantly improve system performance metrics such as BER and BLER, even in challenging environments like fading and interference-laden channels.

Moreover, this paper systematically reviewed advancements in AE-based designs, from standard architectures to specialized convolutional and recurrent models, and innovative approaches like TurboAE. These models not only enhance spectral and energy efficiency but also pave the way for robust communication in emerging paradigms like OWC and 6G systems.

Looking ahead, there are several promising avenues for future research. The integration of advanced neural architectures, such as transformer-based AEs, holds significant promise for addressing challenges like scalability and real-time processing. Furthermore, the deployment of \color{black} end-to-end DL frameworks in real-world systems demands attention to issues such as computational efficiency, \color{black} hardware constraints, and adaptability to dynamic channel conditions. As the demand for low-latency, high-reliability communication systems continues to rise, the role of end-to-end learning in PHY optimization is set to become even more critical.

Lastly, we emphasize the importance of interdisciplinary collaboration between ML researchers and communication engineers to realize the full potential of learning-based approaches. By bridging these domains, the research community can drive innovation, \color{black} and transform the next generation \color{black}of wireless communication systems into more intelligent, efficient, and resilient networks.

\bibliographystyle{IEEEtran}
\bibliography{refv2}

\begin{thebibliography}{100}
\providecommand{\url}[1]{#1}
\csname url@samestyle\endcsname
\providecommand{\newblock}{\relax}
\providecommand{\bibinfo}[2]{#2}
\providecommand{\BIBentrySTDinterwordspacing}{\spaceskip=0pt\relax}
\providecommand{\BIBentryALTinterwordstretchfactor}{4}
\providecommand{\BIBentryALTinterwordspacing}{\spaceskip=\fontdimen2\font plus
\BIBentryALTinterwordstretchfactor\fontdimen3\font minus
  \fontdimen4\font\relax}
\providecommand{\BIBforeignlanguage}[2]{{%
\expandafter\ifx\csname l@#1\endcsname\relax
\typeout{** WARNING: IEEEtran.bst: No hyphenation pattern has been}%
\typeout{** loaded for the language `#1'. Using the pattern for}%
\typeout{** the default language instead.}%
\else
\language=\csname l@#1\endcsname
\fi
#2}}
\providecommand{\BIBdecl}{\relax}
\BIBdecl

\bibitem{IMT2030REF1}
A.~Kaushik \emph{et~al.}, ``Toward integrated sensing and communications for
  {6G}: Key enabling technologies, standardization, and challenges,''
  \emph{IEEE Communications Standards Magazine}, vol.~8, no.~2, pp. 52--59,
  2024.

\bibitem{AIRAN}
Y.~S. Junejo, F.~K. Shaikh, B.~S. Chowdhry, and W.~Ejaz, ``Role of ai and open
  ran in {6G} networks: Performance impact and key technologies,'' in
  \emph{2024 IEEE International Conference on Advanced Telecommunication and
  Networking Technologies (ATNT)}, vol.~1, 2024, pp. 1--4.

\bibitem{gupta5G}
A.~Gupta and R.~K. Jha, ``A survey of 5g network: Architecture and emerging
  technologies,'' \emph{IEEE Access}, vol.~3, pp. 1206--1232, 2015.

\bibitem{13dongsurveyDL}
\BIBentryALTinterwordspacing
S.~Dong, P.~Wang, and K.~Abbas, ``A survey on deep learning and its
  applications,'' \emph{Computer Science Review}, vol.~40, p. 100379, 2021.
  [Online]. Available:
  \url{https://www.sciencedirect.com/science/article/pii/S1574013721000198}
\BIBentrySTDinterwordspacing

\bibitem{1hoydis}
T.~O’Shea and J.~Hoydis, ``An introduction to deep learning for the physical
  layer,'' \emph{IEEE Transactions on Cognitive Communications and Networking},
  vol.~3, no.~4, pp. 563--575, 2017.

\bibitem{TschenkRF}
W.~U. Khan \emph{et~al.}, \emph{{Opportunities for physical layer security in
  UAV communication enhanced with intelligent reflective surfaces}}, 2022.

\bibitem{Tarokh1998}
V.~Tarokh, N.~Seshadri, and A.~R. Calderbank, ``Space--time codes for high data
  rate wireless communication,'' \emph{IEEE Transactions on Information
  Theory}, vol.~44, no.~2, pp. 744--765, 1998.

\bibitem{LuJiangHu2022}
R.~Lu, H.~Jiang, and L.~Hu, ``Research on deep learning based end-to-end
  wireless communication system technology,'' in \emph{2022 IEEE 5th Advanced
  Information Management, Communicates, Electronic and Automation Control
  Conference (IMCEC)}, vol.~5, Dec 2022, pp. 180--185.

\bibitem{2channAEmag}
C.~Zou, F.~Yang, J.~Song, and Z.~Han, ``Channel autoencoder for wireless
  communication: State of the art, challenges, and trends,'' \emph{IEEE
  Communications Magazine}, vol.~59, no.~5, pp. 136--142, 2021.

\bibitem{11santos}
G.~Santos, ``When 5g meets deep learning: A systematic review,''
  \emph{Algorithms}, vol.~13, no. 208, 2020.

\bibitem{hybrid}
S.~Feng, X.~Lu, D.~Niyato, Y.~Wu, X.~Shen, and W.~Wang, ``System-level security
  solution for hybrid d2d communication in heterogeneous d2d-underlaid cellular
  network,'' \emph{IEEE Transactions on Wireless Communications}, vol.~23,
  no.~10, pp. 15\,054--15\,069, 2024.

\bibitem{sze2017}
V.~Sze, Y.-H. Chen, T.-J. Yang, and J.~S. Emer, ``Efficient processing of deep
  neural networks: A tutorial and survey,'' \emph{Proceedings of the IEEE},
  vol. 105, no.~12, pp. 2295--2329, Dec 2017.

\bibitem{4mobileDLsurvey}
C.~Zhang, P.~Patras, and H.~Haddadi, ``Deep learning in mobile and wireless
  networking: A survey,'' \emph{IEEE Communications Surveys \& Tutorials},
  vol.~21, no.~3, pp. 2224--2287, 2019.

\bibitem{minaee2022}
S.~Minaee, Y.~Boykov, F.~Porikli, A.~Plaza, N.~Kehtarnavaz, and D.~Terzopoulos,
  ``Image segmentation using deep learning: A survey,'' \emph{IEEE Transactions
  on Pattern Analysis and Machine Intelligence}, vol.~44, no.~7, pp.
  3523--3542, 2022.

\bibitem{deeprectifiers}
K.~He, X.~Zhang, S.~Ren, and J.~Sun, ``Delving deep into rectifiers: Surpassing
  human-level performance on imagenet classification,'' in \emph{2015 IEEE
  International Conference on Computer Vision (ICCV)}, 2015, pp. 1026--1034.

\bibitem{deepunfolding}
A.~Balatsoukas-Stimming and C.~Studer, ``Deep unfolding for communications
  systems: A survey and some new directions,'' in \emph{2019 IEEE International
  Workshop on Signal Processing Systems (SiPS)}, 2019, pp. 266--271.

\bibitem{sharmaUWB2021}
S.~Sharma, K.~Deka, and M.~Mandloi, ``Deep learning noncoherent uwb receiver
  design,'' \emph{IEEE Sensors Letters}, vol.~5, no.~6, pp. 1--4, June 2021.

\bibitem{68erpek}
\BIBentryALTinterwordspacing
T.~Erpek, T.~J. O'Shea, Y.~E. Sagduyu, Y.~Shi, and T.~C. Clancy, ``Deep
  learning for wireless communications,'' 2020. [Online]. Available:
  \url{https://arxiv.org/abs/2005.06068}
\BIBentrySTDinterwordspacing

\bibitem{8goodfellow}
I.~Goodfellow, Y.~Bengio, and A.~Courville, \emph{Deep Learning}.\hskip 1em
  plus 0.5em minus 0.4em\relax MIT Press, 2016,
  \url{http://www.deeplearningbook.org}.

\bibitem{DLbookch14}
\BIBentryALTinterwordspacing
I.~Goodfellow, Y.~Bengio, and A.~Courville, \emph{Autoencoders}.\hskip 1em plus
  0.5em minus 0.4em\relax MIT Press, 2016, pp. 499--523. [Online]. Available:
  \url{http://www.deeplearningbook.org}
\BIBentrySTDinterwordspacing

\bibitem{51lecun}
\BIBentryALTinterwordspacing
Y.~LeCun, Y.~Bengio, and G.~Hinton, ``Deep learning,'' \emph{Nature}, vol. 521,
  no. 7553, p. 436–444, May 2015. [Online]. Available:
  \url{http://dx.doi.org/10.1038/nature14539}
\BIBentrySTDinterwordspacing

\bibitem{53schmidhuber}
\BIBentryALTinterwordspacing
J.~Schmidhuber, ``Deep learning in neural networks: An overview,'' \emph{Neural
  Networks}, vol.~61, p. 85–117, Jan. 2015. [Online]. Available:
  \url{http://dx.doi.org/10.1016/j.neunet.2014.09.003}
\BIBentrySTDinterwordspacing

\bibitem{54yu}
L.~Deng and D.~Yu, \emph{Deep Learning: Methods and Applications}.\hskip 1em
  plus 0.5em minus 0.4em\relax Signal Processing, 2014, vol.~7, no. 3-4.

\bibitem{70backprop}
D.~E. Rumelhart, G.~E. Hinton, and R.~J. Williams, ``Learning representations
  by back-propagating errors,'' \emph{Nature}, vol. 323, no. 6088, pp.
  533--536, oct 1986.

\bibitem{55pouyanfar}
\BIBentryALTinterwordspacing
S.~Pouyanfar \emph{et~al.}, ``A survey on deep learning: Algorithms,
  techniques, and applications,'' \emph{ACM Comput. Surv.}, vol.~51, no.~5, sep
  2018. [Online]. Available: \url{https://doi.org/10.1145/3234150}
\BIBentrySTDinterwordspacing

\bibitem{52liu}
\BIBentryALTinterwordspacing
W.~Liu, Z.~Wang, X.~Liu, N.~Zeng, Y.~Liu, and F.~E. Alsaadi, ``A survey of deep
  neural network architectures and their applications,'' \emph{Neurocomputing},
  vol. 234, pp. 11--26, 2017. [Online]. Available:
  \url{https://www.sciencedirect.com/science/article/pii/S0925231216315533}
\BIBentrySTDinterwordspacing

\bibitem{3batchnorm}
\BIBentryALTinterwordspacing
S.~Ioffe and C.~Szegedy, ``Batch normalization: Accelerating deep network
  training by reducing internal covariate shift,'' in \emph{Proceedings of the
  32nd International Conference on Machine Learning}, ser. Proceedings of
  Machine Learning Research, F.~Bach and D.~Blei, Eds., vol.~37.\hskip 1em plus
  0.5em minus 0.4em\relax Lille, France: PMLR, 07--09 Jul 2015, pp. 448--456.
  [Online]. Available: \url{https://proceedings.mlr.press/v37/ioffe15.html}
\BIBentrySTDinterwordspacing

\bibitem{66berahmand}
\BIBentryALTinterwordspacing
K.~Berahmand, F.~Daneshfar, E.~S. Salehi, Y.~Li, and Y.~Xu, ``Autoencoders and
  their applications in machine learning: a survey,'' \emph{Artificial
  Intelligence Review}, vol.~57, no.~2, Feb. 2024. [Online]. Available:
  \url{http://dx.doi.org/10.1007/s10462-023-10662-6}
\BIBentrySTDinterwordspacing

\bibitem{67zimmermann}
H.~Zimmermann, ``Osi reference model - the iso model of architecture for open
  systems interconnection,'' \emph{IEEE Transactions on Communications},
  vol.~28, no.~4, pp. 425--432, 1980.

\bibitem{64jurdak}
\BIBentryALTinterwordspacing
R.~Jurdak, \emph{Physical Layer}.\hskip 1em plus 0.5em minus 0.4em\relax
  Boston, MA: Springer US, 2007, pp. 7--15. [Online]. Available:
  \url{https://doi.org/10.1007/978-0-387-39023-9_2}
\BIBentrySTDinterwordspacing

\bibitem{62rgheff}
\BIBentryALTinterwordspacing
M.~A. Abu-Rgheff, \emph{5G Physical Layer Technologies}.\hskip 1em plus 0.5em
  minus 0.4em\relax Wiley, Sep. 2019. [Online]. Available:
  \url{http://dx.doi.org/10.1002/9781119525547}
\BIBentrySTDinterwordspacing

\bibitem{road6G}
W.~Jiang, B.~Han, M.~A. Habibi, and H.~D. Schotten, ``The road towards {6G}: A
  comprehensive survey,'' \emph{IEEE Open Journal of the Communications
  Society}, vol.~2, pp. 334--366, 2021.

\bibitem{reconfig_intelligent_surfaces}
Y.~Liu \emph{et~al.}, ``Reconfigurable intelligent surfaces: Principles and
  opportunities,'' \emph{IEEE Communications Surveys \& Tutorials}, vol.~23,
  no.~3, pp. 1546--1577, 2021.

\bibitem{G6_wireless_chann_measure}
C.-X. Wang, J.~Huang, H.~Wang, X.~Gao, X.~You, and Y.~Hao, ``{6G} wireless
  channel measurements and models: Trends and challenges,'' \emph{IEEE
  Vehicular Technology Magazine}, vol.~15, no.~4, pp. 22--32, 2020.

\bibitem{mmWave_considerations}
I.~A. Hemadeh, K.~Satyanarayana, M.~El-Hajjar, and L.~Hanzo, ``Millimeter-wave
  communications: Physical channel models, design considerations, antenna
  constructions, and link-budget,'' \emph{IEEE Communications Surveys \&
  Tutorials}, vol.~20, no.~2, pp. 870--913, 2018.

\bibitem{spatialdesign}
P.~Yang, M.~Di~Renzo, Y.~Xiao, S.~Li, and L.~Hanzo, ``Design guidelines for
  spatial modulation,'' \emph{IEEE Communications Surveys \& Tutorials},
  vol.~17, no.~1, pp. 6--26, 2015.

\bibitem{57junejo}
\BIBentryALTinterwordspacing
N.~U.~R. Junejo \emph{et~al.}, ``A survey on physical layer techniques and
  challenges in underwater communication systems,'' \emph{Journal of Marine
  Science and Engineering}, vol.~11, no.~4, p. 885, Apr. 2023. [Online].
  Available: \url{http://dx.doi.org/10.3390/jmse11040885}
\BIBentrySTDinterwordspacing

\bibitem{NOMAstatus_makki}
B.~Makki, K.~Chitti, A.~Behravan, and M.-S. Alouini, ``A survey of noma:
  Current status and open research challenges,'' \emph{IEEE Open Journal of the
  Communications Society}, vol.~1, pp. 179--189, 2020.

\bibitem{28ozpoyraz}
B.~Ozpoyraz, A.~T. Dogukan, Y.~Gevez, U.~Altun, and E.~Basar, ``Deep
  learning-aided {6G} wireless networks: A comprehensive survey of
  revolutionary phy architectures,'' \emph{IEEE Open Journal of the
  Communications Society}, vol.~3, pp. 1749--1809, 2022.

\bibitem{21shi}
Y.~Shi \emph{et~al.}, ``Machine learning for large-scale optimization in {6G}
  wireless networks,'' \emph{IEEE Communications Surveys \& Tutorials},
  vol.~25, no.~4, pp. 2088--2132, 2023.

\bibitem{39lv}
C.~Lv and Z.~Luo, ``Deep learning for channel estimation in physical layer
  wireless communications: Fundamental, methods, and challenges,''
  \emph{Electronics}, vol.~12, p. 4965, 2023.

\bibitem{22szott}
S.~Szott \emph{et~al.}, ``Wi-fi meets ml: A survey on improving ieee 802.11
  performance with machine learning,'' \emph{IEEE Communications Surveys \&
  Tutorials}, vol.~24, no.~3, pp. 1843--1893, 2022.

\bibitem{42ye}
N.~Ye, S.~Miao, J.~Pan, Q.~Ouyang, X.~Li, and X.~Hou, ``Artificial intelligence
  for wireless physical-layer technologies (ai4phy): A comprehensive survey,''
  \emph{IEEE Transactions on Cognitive Communications and Networking}, vol.~10,
  no.~3, pp. 729--755, June 2024.

\bibitem{41wang}
T.~Wang, C.-K. Wen, H.~Wang, F.~Gao, T.~Jiang, and S.~Jin, ``Deep learning for
  wireless physical layer: Opportunities and challenges,'' \emph{China
  Communications}, vol.~14, no.~11, pp. 92--111, Nov 2017.

\bibitem{43qin}
Z.~Qin, H.~Ye, G.~Y. Li, and B.-H.~F. Juang, ``Deep learning in physical layer
  communications,'' \emph{IEEE Wireless Communications}, vol.~26, no.~2, pp.
  93--99, April 2019.

\bibitem{44huang}
H.~Huang \emph{et~al.}, ``Deep learning for physical-layer 5g wireless
  techniques: Opportunities, challenges and solutions,'' \emph{IEEE Wireless
  Communications}, vol.~27, no.~1, pp. 214--222, February 2020.

\bibitem{45akrout}
\BIBentryALTinterwordspacing
M.~Akrout, A.~Mezghani, E.~Hossain, F.~Bellili, and R.~W. Heath, ``From
  multilayer perceptron to gpt: A reflection on deep learning research for
  wireless physical layer,'' 2023. [Online]. Available:
  \url{https://arxiv.org/abs/2307.07359}
\BIBentrySTDinterwordspacing

\bibitem{47tanveer}
\BIBentryALTinterwordspacing
J.~Tanveer, A.~Haider, R.~Ali, and A.~Kim, ``Machine learning for physical
  layer in 5g and beyond wireless networks: A survey,'' \emph{Electronics},
  vol.~11, no.~1, 2022. [Online]. Available:
  \url{https://www.mdpi.com/2079-9292/11/1/121}
\BIBentrySTDinterwordspacing

\bibitem{50huynh}
\BIBentryALTinterwordspacing
N.~V. Huynh \emph{et~al.}, ``Generative ai for physical layer communications: A
  survey,'' 2023. [Online]. Available: \url{https://arxiv.org/abs/2312.05594}
\BIBentrySTDinterwordspacing

\bibitem{38islam}
\BIBentryALTinterwordspacing
N.~Islam and S.~Shin, ``Deep learning in physical layer: Review on data driven
  end-to-end communication systems and their enabling semantic applications,''
  2024. [Online]. Available: \url{https://arxiv.org/abs/2401.12800}
\BIBentrySTDinterwordspacing

\bibitem{9kaur}
J.~Kaur, M.~A. Khan, M.~Iftikhar, M.~Imran, and Q.~Emad Ul~Haq, ``Machine
  learning techniques for 5g and beyond,'' \emph{IEEE Access}, vol.~9, pp.
  23\,472--23\,488, 2021.

\bibitem{12akyildiz}
I.~F. Akyildiz, A.~Kak, and S.~Nie, ``6g and beyond: The future of wireless
  communications systems,'' \emph{IEEE Access}, vol.~8, pp. 133\,995--134\,030,
  2020.

\bibitem{16mao}
Q.~Mao, F.~Hu, and Q.~Hao, ``Deep learning for intelligent wireless networks: A
  comprehensive survey,'' \emph{IEEE Communications Surveys \& Tutorials},
  vol.~20, no.~4, pp. 2595--2621, 2018.

\bibitem{37nguyen}
D.~C. Nguyen \emph{et~al.}, ``Enabling ai in future wireless networks: A data
  life cycle perspective,'' \emph{IEEE Communications Surveys \& Tutorials},
  vol.~23, no.~1, pp. 553--595, 2021.

\bibitem{Wei2024Mem}
J.~Wei \emph{et~al.}, ``Memorization in deep learning: A survey,'' \emph{ACM
  Computing Surveys}, vol.~56, no.~12, pp. 1--38, 2024.

\bibitem{Gao2024EDL}
J.~Gao, M.~Chen, L.~Xiang, and C.~Xu, ``A comprehensive survey on evidential
  deep learning and its applications,'' \emph{IEEE Transactions on Pattern
  Analysis and Machine Intelligence}, vol.~PP, no.~99, pp. 1--20, 2024.

\bibitem{Gkarmpounis2024GNN}
G.~Gkarmpounis, C.~Vranis, N.~Vretos, and P.~Daras, ``Survey on graph neural
  networks,'' \emph{IEEE Access}, vol.~12, pp. 128\,816--128\,832, 2024.

\bibitem{pytorch}
\BIBentryALTinterwordspacing
A.~Paszke \emph{et~al.}, ``Pytorch: An imperative style, high-performance deep
  learning library,'' 2019. [Online]. Available:
  \url{https://arxiv.org/abs/1912.01703}
\BIBentrySTDinterwordspacing

\bibitem{tensorflow}
\BIBentryALTinterwordspacing
M.~Abadi \emph{et~al.}, ``Tensorflow: Large-scale machine learning on
  heterogeneous distributed systems,'' 2016. [Online]. Available:
  \url{https://arxiv.org/abs/1603.04467}
\BIBentrySTDinterwordspacing

\bibitem{69adam}
D.~P. Kingma and J.~Ba, ``Adam: A method for stochastic optimization,'' 2014.

\bibitem{empiricaloptimizers}
\BIBentryALTinterwordspacing
D.~Choi, C.~J. Shallue, Z.~Nado, J.~Lee, C.~J. Maddison, and G.~E. Dahl, ``On
  empirical comparisons of optimizers for deep learning,'' \emph{CoRR}, vol.
  abs/1910.05446, 2019. [Online]. Available:
  \url{http://arxiv.org/abs/1910.05446}
\BIBentrySTDinterwordspacing

\bibitem{JMLR:dropout}
\BIBentryALTinterwordspacing
N.~Srivastava, G.~Hinton, A.~Krizhevsky, I.~Sutskever, and R.~Salakhutdinov,
  ``Dropout: A simple way to prevent neural networks from overfitting,''
  \emph{Journal of Machine Learning Research}, vol.~15, no.~56, pp. 1929--1958,
  2014. [Online]. Available:
  \url{http://jmlr.org/papers/v15/srivastava14a.html}
\BIBentrySTDinterwordspacing

\bibitem{LayerNormalization}
J.~Ba, J.~Kiros, and G.~Hinton, ``Layer normalization,'' 07 2016.

\bibitem{baldiAE}
\BIBentryALTinterwordspacing
P.~Baldi, ``Autoencoders, unsupervised learning, and deep architectures,'' in
  \emph{Proceedings of ICML Workshop on Unsupervised and Transfer Learning},
  ser. Proceedings of Machine Learning Research, I.~Guyon, G.~Dror, V.~Lemaire,
  G.~Taylor, and D.~Silver, Eds., vol.~27.\hskip 1em plus 0.5em minus
  0.4em\relax Bellevue, Washington, USA: PMLR, 02 Jul 2012, pp. 37--49.
  [Online]. Available: \url{https://proceedings.mlr.press/v27/baldi12a.html}
\BIBentrySTDinterwordspacing

\bibitem{Guo2024FTLsurvey}
W.~Guo, X.~Yang, J.~Xu, and H.~Huang, ``A comprehensive survey of federated
  transfer learning: Methods, applications, and challenges,'' \emph{Frontiers
  of Computer Science}, vol.~18, no.~3, pp. 1--23, 2024.

\bibitem{Wazzeh2025DynamicSFL}
M.~Wazzeh, M.~M. Butt, H.~ElSawy, and M.~Alouini, ``Dynamic split federated
  learning for resource-constrained wireless networks,'' \emph{Computer
  Communications}, 2025, early Access.

\bibitem{Xin2024SemanticSurvey}
G.~Xin, X.~Liu, Y.~Tian, J.~Zhang, and H.~Wang, ``Semantic communication: A
  survey of its theoretical foundations and key techniques,'' \emph{Entropy},
  vol.~26, no.~2, p. 102, 2024.

\bibitem{Chaccour2025NextGenSemCom}
C.~Chaccour, M.~Debbah, and W.~Saad, ``Less data, more knowledge: Building
  next-generation semantic communications,'' \emph{IEEE Communications Surveys
  \& Tutorials}, 2025, early Access.

\bibitem{Ogenyi2025AINative6G}
F.~C. Ogenyi, A.~Al-Kababji, A.~Tassi, and M.~A. Imran, ``A comprehensive
  review of {AI}-native 6g: Integrating semantic communication, edge
  intelligence, and reconfigurable surfaces,'' \emph{Frontiers in
  Communications and Networks}, vol.~6, pp. 1--25, 2025.

\bibitem{caffe}
\BIBentryALTinterwordspacing
Y.~Jia \emph{et~al.}, ``Caffe: Convolutional architecture for fast feature
  embedding,'' 2014. [Online]. Available: \url{https://arxiv.org/abs/1408.5093}
\BIBentrySTDinterwordspacing

\bibitem{mxnet}
\BIBentryALTinterwordspacing
T.~Chen \emph{et~al.}, ``Mxnet: A flexible and efficient machine learning
  library for heterogeneous distributed systems,'' 2015. [Online]. Available:
  \url{https://arxiv.org/abs/1512.01274}
\BIBentrySTDinterwordspacing

\bibitem{liuStandard}
X.~Liu, Z.~Wei, A.~Pepe, Z.~Wang, and H.~Y. Fu, ``Autoencoder for optical
  wireless communication system in atmospheric turbulence,'' in \emph{2020
  Opto-Electronics and Communications Conference (OECC)}, 2020, pp. 1--3.

\bibitem{harshithaysStandard}
H.~Y~S, B.~T~S, G.~R. Sowkhya, R.~K. K~N, and V.~Karjigi, ``Wireless
  communication using autoencoder,'' in \emph{2023 International Conference on
  Smart Systems for applications in Electrical Sciences (ICSSES)}, 2023, pp.
  1--6.

\bibitem{baleviTurbo}
E.~Balevi and J.~G. Andrews, ``Autoencoder-based error correction coding for
  one-bit quantization,'' \emph{IEEE Transactions on Communications}, vol.~68,
  no.~6, pp. 3440--3451, 2020.

\bibitem{jiangTurbo}
Y.~Jiang, H.~Kim, H.~Asnani, S.~Kannan, S.~Oh, and P.~Viswanath, ``Joint
  channel coding and modulation via deep learning,'' in \emph{2020 IEEE 21st
  International Workshop on Signal Processing Advances in Wireless
  Communications (SPAWC)}, 2020, pp. 1--5.

\bibitem{wangTurbo}
L.~Wang, H.~Saber, H.~Hatami, M.~V. Jamali, and J.~H. Bae, ``Rate-matched turbo
  autoencoder: A deep learning based multi-rate channel autoencoder,'' in
  \emph{ICC 2023 - IEEE International Conference on Communications}, 2023, pp.
  6355--6360.

\bibitem{scikitlearn}
\BIBentryALTinterwordspacing
F.~Pedregosa \emph{et~al.}, ``Scikit-learn: Machine learning in python,'' 2018.
  [Online]. Available: \url{https://arxiv.org/abs/1201.0490}
\BIBentrySTDinterwordspacing

\bibitem{rstudio_dl_github}
\BIBentryALTinterwordspacing
S.~Varma and S.~Das, ``Deep learning,'' 2018, accessed: 2024-11-18. [Online].
  Available: \url{https://srdas.github.io/DLBook/DeepLearningWithR.html}
\BIBentrySTDinterwordspacing

\bibitem{nvidia_tesla_t4}
\BIBentryALTinterwordspacing
NVIDIA. (2021) Nvidia tesla t4 tensor core gpu. Accessed: 2024-07-29. [Online].
  Available:
  \url{https://www.nvidia.com/content/dam/en-zz/Solutions/Data-Center/tesla-t4/t4-tensor-core-product-brief.pdf}
\BIBentrySTDinterwordspacing

\bibitem{googleTPU}
\BIBentryALTinterwordspacing
N.~P. Jouppi \emph{et~al.}, ``In-datacenter performance analysis of a tensor
  processing unit,'' in \emph{44th International Symposium on Computer
  Architecture (ISCA)}, 2017. [Online]. Available:
  \url{https://arxiv.org/pdf/1704.04760.pdf}
\BIBentrySTDinterwordspacing

\bibitem{mode16}
S.~Dörner, S.~Cammerer, J.~Hoydis, and S.~t. Brink, ``Deep learning based
  communication over the air,'' \emph{IEEE Journal of Selected Topics in Signal
  Processing}, vol.~12, no.~1, pp. 132--143, 2018.

\bibitem{beyondonehot}
A.~Perotti, S.~Bertolotto, E.~Pastor, and A.~Panisson, ``Beyond
  one-hot-encoding: Injecting semantics to drive image classifiers,'' in
  \emph{Explainable Artificial Intelligence}, L.~Longo, Ed.\hskip 1em plus
  0.5em minus 0.4em\relax Cham: Springer Nature Switzerland, 2023, pp.
  525--548.

\bibitem{mode22}
A.~Elfikky and Z.~Rezki, ``Symbol detection and channel estimation for space
  optical communications using neural network and autoencoder,'' \emph{IEEE
  Transactions on Machine Learning in Communications and Networking}, vol.~2,
  pp. 110--128, 2024.

\bibitem{RNN_channpred}
W.~Jiang and H.~D. Schotten, ``Neural network-based fading channel prediction:
  A comprehensive overview,'' \emph{IEEE Access}, vol.~7, pp.
  118\,112--118\,124, 2019.

\bibitem{RNN_channelest_vehicular}
S.~A. Ul~Haq, B.~P. Dangwal, and S.~Darak, ``Rnn-based low complexity high
  speed channel estimation architectures for vehicular networks,'' in
  \emph{2024 IEEE 6th International Conference on AI Circuits and Systems
  (AICAS)}, 2024, pp. 51--55.

\bibitem{RNN_stella}
Y.~Lin, Y.~Tu, Z.~Dou, L.~Chen, and S.~Mao, ``Contour stella image and deep
  learning for signal recognition in the physical layer,'' \emph{IEEE
  Transactions on Cognitive Communications and Networking}, vol.~7, no.~1, pp.
  34--46, 2021.

\bibitem{LSTM_sigdetect}
S.~Wang, R.~Yao, T.~A. Tsiftsis, N.~I. Miridakis, and N.~Qi, ``Signal detection
  in uplink time-varying ofdm systems using rnn with bidirectional lstm,''
  \emph{IEEE Wireless Communications Letters}, vol.~9, no.~11, pp. 1947--1951,
  2020.

\bibitem{RNN_channcode}
R.~Sattiraju, A.~Weinand, and H.~D. Schotten, ``Performance analysis of deep
  learning based on recurrent neural networks for channel coding,'' in
  \emph{2018 IEEE International Conference on Advanced Networks and
  Telecommunications Systems (ANTS)}, 2018, pp. 1--6.

\bibitem{RNN_lindec}
A.~Bennatan, Y.~Choukroun, and P.~Kisilev, ``Deep learning for decoding of
  linear codes - a syndrome-based approach,'' in \emph{2018 IEEE International
  Symposium on Information Theory (ISIT)}, 2018, pp. 1595--1599.

\bibitem{RNN_commalgs}
\BIBentryALTinterwordspacing
H.~Kim, Y.~Jiang, R.~Rana, S.~Kannan, S.~Oh, and P.~Viswanath, ``Communication
  algorithms via deep learning,'' 2018. [Online]. Available:
  \url{https://arxiv.org/abs/1805.09317}
\BIBentrySTDinterwordspacing

\bibitem{LEARN_RNN}
Y.~Jiang, H.~Kim, H.~Asnani, S.~Kannan, S.~Oh, and P.~Viswanath, ``Learn codes:
  Inventing low-latency codes via recurrent neural networks,'' \emph{IEEE
  Journal on Selected Areas in Information Theory}, vol.~1, no.~1, pp.
  207--216, 2020.

\bibitem{mode55}
H.~Wu, Y.~Zhang, X.~Zhao, N.~Zhu, and M.~Coates, ``End-to-end physical layer
  communication using bi-directional grus for isi channels,'' in \emph{2020
  IEEE Globecom Workshops (GC Wkshps}, 2020, pp. 1--6.

\bibitem{mode50}
Y.~Zhang, H.~Wu, and M.~Coates, ``On the design of channel coding autoencoders
  with arbitrary rates for isi channels,'' \emph{IEEE Wireless Communications
  Letters}, vol.~11, no.~2, pp. 426--430, 2022.

\bibitem{attentionisallyouneed}
\BIBentryALTinterwordspacing
A.~Vaswani \emph{et~al.}, ``Attention is all you need,'' in \emph{Advances in
  Neural Information Processing Systems}, I.~Guyon \emph{et~al.}, Eds.,
  vol.~30.\hskip 1em plus 0.5em minus 0.4em\relax Curran Associates, Inc.,
  2017. [Online]. Available:
  \url{https://proceedings.neurips.cc/paper_files/paper/2017/file/3f5ee243547dee91fbd053c1c4a845aa-Paper.pdf}
\BIBentrySTDinterwordspacing

\bibitem{ogturbo}
C.~Berrou, A.~Glavieux, and P.~Thitimajshima, ``Near shannon limit
  error-correcting coding and decoding: Turbo-codes. 1,'' in \emph{Proceedings
  of ICC '93 - IEEE International Conference on Communications}, vol.~2, 1993,
  pp. 1064--1070 vol.2.

\bibitem{Jiang2019TurboAE}
\BIBentryALTinterwordspacing
Y.~Jiang, H.~Kim, H.~Asnani, S.~Kannan, S.~Oh, and P.~Viswanath, ``Turbo
  autoencoder: Deep learning based channel codes for point-to-point
  communication channels,'' \emph{ArXiv}, vol. abs/1911.03038, 2019. [Online].
  Available: \url{https://api.semanticscholar.org/CorpusID:202770026}
\BIBentrySTDinterwordspacing

\bibitem{ogturbo2}
C.~Berrou and A.~Glavieux, ``Near optimum error correcting coding and decoding:
  turbo-codes,'' \emph{IEEE Transactions on Communications}, vol.~44, no.~10,
  pp. 1261--1271, 1996.

\bibitem{turboprinciple}
J.~Hagenauer, ``The turbo principle: Tutorial introduction and state of the
  art,'' \emph{Proc. International Symposium on Turbo Codes and Related
  Topics}, 1997.

\bibitem{jointCNN}
B.~Zhu, J.~Wang, L.~He, and J.~Song, ``Joint transceiver optimization for
  wireless communication phy using neural network,'' \emph{IEEE Journal on
  Selected Areas in Communications}, vol.~37, no.~6, pp. 1364--1373, 2019.

\bibitem{jiang2019deepturbodeepturbodecoder}
\BIBentryALTinterwordspacing
Y.~Jiang, H.~Kim, H.~Asnani, S.~Kannan, S.~Oh, and P.~Viswanath, ``Deepturbo:
  Deep turbo decoder,'' 2019. [Online]. Available:
  \url{https://arxiv.org/abs/1903.02295}
\BIBentrySTDinterwordspacing

\bibitem{turbo7Clausius}
J.~Clausius, S.~Dörner, S.~Cammerer, and S.~t. Brink, ``Serial vs. parallel
  turbo-autoencoders and accelerated training for learned channel codes,'' in
  \emph{2021 11th International Symposium on Topics in Coding (ISTC)}, 2021,
  pp. 1--5.

\bibitem{xuTurbo}
C.~Xu \emph{et~al.}, ``Turbo detection aided autoencoder for multicarrier
  wireless systems: Integrating deep learning into channel coded systems,''
  \emph{IEEE Transactions on Cognitive Communications and Networking}, vol.~8,
  no.~2, pp. 600--614, 2022.

\bibitem{Ferdous2024CNNAE}
\BIBentryALTinterwordspacing
J.~Ferdous, M.~A. Mollah, and A.~Rahman, ``{CNN-Based End-to-End Deeper
  Autoencoders for Physical Layer of Wireless Communication System},'' in
  \emph{Proc. 2024 Int. Conf. on Advances in Computing, Communication,
  Electrical, and Smart Systems (iCACCESS)}, Dhaka, Bangladesh, Mar. 2024.
  [Online]. Available:
  \url{https://doi.org/10.1109/iCACCESS61735.2024.10499465}
\BIBentrySTDinterwordspacing

\bibitem{Sindal2024SparseAE}
\BIBentryALTinterwordspacing
S.~S. Sindal and Y.~N. Trivedi, ``Enhancing performance of end-to-end
  communication system using attention mechanism-based sparse autoencoder over
  {R}ayleigh fading channel,'' \emph{Physical Communication}, vol.~67, p.
  102534, Dec. 2024. [Online]. Available:
  \url{https://doi.org/10.1016/j.phycom.2024.102534}
\BIBentrySTDinterwordspacing

\bibitem{Zayat2024TMAE}
\BIBentryALTinterwordspacing
A.~Zayat, M.~A. Hasabelnaby, M.~Obeed, and A.~Chaaban, ``Transformer masked
  autoencoders for next-generation wireless communications: Architecture and
  opportunities,'' \emph{IEEE Communications Magazine}, vol.~62, no.~7, pp.
  88--94, Jul. 2024. [Online]. Available:
  \url{https://doi.org/10.1109/MCOM.002.2300257}
\BIBentrySTDinterwordspacing

\bibitem{Owfi2025OMLCAE}
\BIBentryALTinterwordspacing
A.~Owfi, J.~D. Ashdown, K.~A. Turck, and F.~Afghah, ``Online meta-learning
  channel autoencoder for dynamic end-to-end physical layer optimization,'' in
  \emph{Proc. IEEE Wireless Communications and Networking Conf. (WCNC)}, Mar.
  2025. [Online]. Available:
  \url{https://doi.org/10.1109/WCNC61545.2025.10978447}
\BIBentrySTDinterwordspacing

\bibitem{mode2}
J.~Ferdous, M.~Mollah, and A.~Rahman, ``Cnn-based end-to-end deeper
  autoencoders for physical layer of wireless communication system,'' in
  \emph{2024 International Conference on Advances in Computing, Communication,
  Electrical, and Smart Systems (iCACCESS)}, 2024, pp. 1--6.

\bibitem{mode3}
K.~Ramdani, H.~Shaiek, and D.~Roviras, ``Modified autoencoder structure for
  reducing ber performance,'' in \emph{2021 International Conference on
  Software, Telecommunications and Computer Networks (SoftCOM)}, 2021, pp.
  1--5.

\bibitem{mode43_JSCCImage}
E.~Bourtsoulatze, D.~Burth~Kurka, and D.~Gündüz, ``Deep joint source-channel
  coding for wireless image transmission,'' \emph{IEEE Transactions on
  Cognitive Communications and Networking}, vol.~5, no.~3, pp. 567--579, 2019.

\bibitem{mode18}
M.~Soltani, W.~Fatnassi, A.~Aboutaleb, Z.~Rezki, A.~Bhuyan, and P.~Titus,
  ``Autoencoder-based optical wireless communications systems,'' in \emph{2018
  IEEE Globecom Workshops (GC Wkshps)}, 2018, pp. 1--6.

\bibitem{mode19}
Z.-R. Zhu, J.~Zhang, R.-H. Chen, and H.-Y. Yu, ``Autoencoder-based transceiver
  design for owc systems in log-normal fading channel,'' \emph{IEEE Photonics
  Journal}, vol.~11, no.~5, pp. 1--12, 2019.

\bibitem{mode21}
A.~E.-R.~A. El-Fikky and Z.~Rezki, ``On the performance of autoencoder-based
  space optical communications,'' in \emph{GLOBECOM 2022 - 2022 IEEE Global
  Communications Conference}, 2022, pp. 1466--1471.

\bibitem{mode39}
J.~Guo, C.-K. Wen, S.~Jin, and G.~Y. Li, ``Overview of deep learning-based csi
  feedback in massive mimo systems,'' \emph{IEEE Transactions on
  Communications}, vol.~70, no.~12, pp. 8017--8045, 2022.

\bibitem{mode40}
J.~So and H.~Kwon, ``Universal auto-encoder framework for mimo csi feedback,''
  in \emph{GLOBECOM 2023 - 2023 IEEE Global Communications Conference}, 2023,
  pp. 01--07.

\bibitem{mode54}
C.-K. Wen, W.-T. Shih, and S.~Jin, ``Deep learning for massive mimo csi
  feedback,'' \emph{IEEE Wireless Communications Letters}, vol.~7, no.~5, pp.
  748--751, 2018.

\bibitem{mode24}
\BIBentryALTinterwordspacing
T.~J. O'Shea, T.~Erpek, and T.~C. Clancy, ``Deep learning based {MIMO}
  communications,'' \emph{CoRR}, vol. abs/1707.07980, 2017. [Online].
  Available: \url{http://arxiv.org/abs/1707.07980}
\BIBentrySTDinterwordspacing

\bibitem{mode30}
A.~G. Pathapati, N.~Chakradhar, P.~Havish, S.~A. Somayajula, and S.~Amuru,
  ``Supervised deep learning for mimo precoding,'' in \emph{2020 IEEE 3rd 5G
  World Forum (5GWF)}, 2020, pp. 418--423.

\bibitem{mode1}
P.~Ye, X.~Jia, X.~Yang, and H.~Hu, ``End-to-end physical layer optimization
  scheme based on deep learning autoencoder,'' in \emph{2019 IEEE 4th Advanced
  Information Technology, Electronic and Automation Control Conference
  (IAEAC)}, vol.~1, 2019, pp. 135--139.

\bibitem{mode23}
W.~Ye \emph{et~al.}, ``Autoencoder-based mimo communications with learnable
  adcs,'' in \emph{2021 IEEE 21st International Conference on Communication
  Technology (ICCT)}, 2021, pp. 525--530.

\bibitem{mode29}
J.~Song, C.~Häger, J.~Schröder, T.~J. O’Shea, E.~Agrell, and H.~Wymeersch,
  ``Benchmarking and interpreting end-to-end learning of mimo and multi-user
  communication,'' \emph{IEEE Transactions on Wireless Communications},
  vol.~21, no.~9, pp. 7287--7298, 2022.

\bibitem{mode37}
J.~Tao, J.~Chen, J.~Xing, S.~Fu, and J.~Xie, ``Autoencoder neural network based
  intelligent hybrid beamforming design for mmwave massive mimo systems,''
  \emph{IEEE Transactions on Cognitive Communications and Networking}, vol.~6,
  no.~3, pp. 1019--1030, 2020.

\bibitem{mode33}
S.~Shrestha \emph{et~al.}, ``Autoencoder-based spatial modulation for the next
  generation of wireless networks,'' \emph{IEEE Internet of Things Journal},
  pp. 1--1, 2024.

\bibitem{OFDM_AE_OG}
A.~Felix, S.~Cammerer, S.~Dörner, J.~Hoydis, and S.~Ten~Brink,
  ``Ofdm-autoencoder for end-to-end learning of communications systems,'' in
  \emph{2018 IEEE 19th International Workshop on Signal Processing Advances in
  Wireless Communications (SPAWC)}, 2018, pp. 1--5.

\bibitem{OFDM_AE_ensemble}
K.~M. Asif and A.~Trivedi, ``Ofdm ensemble autoencoder using cnn and spsa for
  end-to-end learning communication systems,'' in \emph{2020 IEEE 4th
  Conference on Information \& Communication Technology (CICT)}, 2020, pp.
  1--6.

\bibitem{OFDM_AEtrimmingfat}
F.~A. Aoudia and J.~Hoydis, ``Trimming the fat from ofdm: Pilot- and cp-less
  communication with end-to-end learning,'' in \emph{2021 IEEE International
  Conference on Communications Workshops (ICC Workshops)}, 2021, pp. 1--6.

\bibitem{PDNOMA_powerallocation}
M.~Hummert, N.~Bulk, C.~Bockelmann, D.~Wübben, and A.~Dekorsy, ``Improving
  noma performance by application of autoencoders and equidistant power
  allocation,'' in \emph{2024 27th International Workshop on Smart Antennas
  (WSA)}, 2024, pp. 1--6.

\bibitem{DLconstellation_downlinkNOMA}
F.~Alberge, ``Constellation design with deep learning for downlink
  non-orthogonal multiple access,'' in \emph{2018 IEEE 29th Annual
  International Symposium on Personal, Indoor and Mobile Radio Communications
  (PIMRC)}, 2018, pp. 1--5.

\bibitem{PDNOMA_weightedAE}
V.~Ninkovic, D.~Vukobratovic, A.~Pastore, and C.~Antón-Haro, ``A weighted
  autoencoder-based approach to downlink noma constellation design,'' in
  \emph{2023 IEEE 24th International Workshop on Signal Processing Advances in
  Wireless Communications (SPAWC)}, 2023, pp. 126--130.

\bibitem{CDAEbasedNOMA}
N.~Choubey, A.~Trivedi, and V.~S. Kushwah, ``Autoencoder for end-to-end
  learning communication system based on noma,'' in \emph{2022 IEEE 6th
  Conference on Information and Communication Technology (CICT)}, 2022, pp.
  1--5.

\bibitem{ogSCMA}
H.~Nikopour and H.~Baligh, ``Sparse code multiple access,'' in \emph{2013 IEEE
  24th Annual International Symposium on Personal, Indoor, and Mobile Radio
  Communications (PIMRC)}, 2013, pp. 332--336.

\bibitem{CDNOMA_DLaided_SCMA}
M.~Kim, N.-I. Kim, W.~Lee, and D.-H. Cho, ``Deep learning-aided scma,''
  \emph{IEEE Communications Letters}, vol.~22, no.~4, pp. 720--723, 2018.

\bibitem{CDNOMA_ElsevierAE_SCMA}
\BIBentryALTinterwordspacing
J.~Lin, S.~Feng, Y.~Zhang, Z.~Yang, and Y.~Zhang, ``A novel deep neural network
  based approach for sparse code multiple access,'' \emph{Neurocomputing}, vol.
  382, pp. 52--63, 2020. [Online]. Available:
  \url{https://www.sciencedirect.com/science/article/pii/S0925231219316686}
\BIBentrySTDinterwordspacing

\bibitem{CDNOMA_sparseAE_SCMA}
M.~Singh, D.~Mishra, and M.~Vanidevi, ``Sparse autoencoder for sparse code
  multiple access,'' in \emph{2021 International Conference on Artificial
  Intelligence in Information and Communication (ICAIIC)}, 2021, pp. 353--358.

\bibitem{CDNOMA_sergienko_SCMA}
A.~B. Sergienko, ``Direct optimization of codebooks for code-domain
  nonorthogonal multiple access with noncoherent reception,'' in \emph{2023
  XVIII International Symposium Problems of Redundancy in Information and
  Control Systems (REDUNDANCY)}, 2023, pp. 127--132.

\bibitem{E2E_SCMA}
Q.~Luo, Z.~Liu, G.~Chen, Y.~Ma, and P.~Xiao, ``A novel multitask learning
  empowered codebook design for downlink scma networks,'' \emph{IEEE Wireless
  Communications Letters}, vol.~11, no.~6, pp. 1268--1272, 2022.

\bibitem{CDNOMA_residualCNN_SCMA}
F.~Jiang, D.-W. Chang, S.~Ma, Y.-J. Hu, and Y.-H. Xu, ``A residual
  learning-aided convolutional autoencoder for scma,'' \emph{IEEE
  Communications Letters}, vol.~27, no.~5, pp. 1337--1341, 2023.

\bibitem{mac_Minsig_CDNOMA_2020}
M.~Han, H.~Seo, A.~T. Abebe, and C.~G. Kang, ``Deep learning-based multi-user
  multi-dimensional constellation design in code domain non-orthogonal multiple
  access,'' in \emph{2020 IEEE International Conference on Communications
  Workshops (ICC Workshops)}, 2020, pp. 1--6.

\bibitem{mac_Minsig_CDNOMA_2022}
M.~Han, H.~Seo, A.~T. Abebe, and C.~G. Kang, ``Deep learning-based codebook
  design for code-domain non-orthogonal multiple access: Approaching
  single-user bit-error rate performance,'' \emph{IEEE Transactions on
  Cognitive Communications and Networking}, vol.~8, no.~2, pp. 1159--1173,
  2022.

\bibitem{CDNOMA_6GWGAN_2023_SCMA}
L.~Miuccio, D.~Panno, and S.~Riolo, ``A flexible encoding/decoding procedure
  for {6G} scma wireless networks via adversarial machine learning
  techniques,'' \emph{IEEE Transactions on Vehicular Technology}, vol.~72,
  no.~3, pp. 3288--3303, 2023.

\bibitem{CDNOMA_MCMA}
Y.~Han, Z.~Wang, Q.~Guo, and W.~Xiang, ``Deep learning-based detection for
  moderate-density code multiple access in iot networks,'' \emph{IEEE
  Communications Letters}, vol.~24, no.~1, pp. 122--125, 2020.

\bibitem{MAC_SWIPT_RSMA}
M.~R. Camana, C.~E. Garcia, and I.~Koo, ``Deep learning-assisted power
  minimization in underlay miso-swipt systems based on rate-splitting multiple
  access,'' \emph{IEEE Access}, vol.~10, pp. 62\,137--62\,156, 2022.

\bibitem{mode20}
A.~Elfikky, M.~Soltani, and Z.~Rezki, ``Learning-based autoencoder for multiple
  access and interference channels in space optical communications,''
  \emph{IEEE Communications Letters}, vol.~27, no.~10, pp. 2662--2666, 2023.

\bibitem{mode25}
E.~Stauffer, A.~Wang, and N.~Jindal, ``Deep learning for the degraded broadcast
  channel,'' in \emph{2019 53rd Asilomar Conference on Signals, Systems, and
  Computers}, 2019, pp. 1760--1763.

\bibitem{mode28}
S.~Li, D.~Tuninetti, and N.~Devroye, ``Deep learning-aided coding for the
  fading broadcast channel with feedback,'' in \emph{ICC 2022 - IEEE
  International Conference on Communications}, 2022, pp. 3874--3879.

\bibitem{mode4}
J.~Park, D.~J. Ji, and D.-H. Cho, ``High-order modulation based on deep neural
  network for physical-layer network coding,'' \emph{IEEE Wireless
  Communications Letters}, vol.~10, no.~6, pp. 1173--1177, 2021.

\bibitem{mode8}
T.~Matsumine, T.~Koike-Akino, and Y.~Wang, ``Deep learning-based constellation
  optimization for physical network coding in two-way relay networks,'' in
  \emph{ICC 2019 - 2019 IEEE International Conference on Communications (ICC)},
  2019, pp. 1--6.

\bibitem{mode9}
A.~Gupta and M.~Sellathurai, ``End-to-end learning-based amplify-and-forward
  relay networks using autoencoders,'' in \emph{ICC 2020 - 2020 IEEE
  International Conference on Communications (ICC)}, 2020, pp. 1--6.

\bibitem{mode10}
A.~Gupta and M.~Sellathurai, ``End-to-end learning-based framework for
  amplify-and-forward relay networks,'' \emph{IEEE Access}, vol.~9, pp.
  81\,660--81\,677, 2021.

\bibitem{mode11}
A.~Gupta and M.~Sellathurai, ``End-to-end learning-based two-way af relay
  networks with i/q imbalance,'' in \emph{2021 IEEE 22nd International Workshop
  on Signal Processing Advances in Wireless Communications (SPAWC)}, 2021, pp.
  111--115.

\bibitem{mode12}
A.~Gupta and M.~Sellathurai, ``A novel average autoencoder-based
  amplify-and-forward relay networks with hardware impairments,'' \emph{IEEE
  Transactions on Cognitive Communications and Networking}, vol.~8, no.~2, pp.
  615--630, 2022.

\bibitem{mode13}
Y.~Lu, P.~Cheng, Z.~Chen, Y.~Li, W.~H. Mow, and B.~Vucetic, ``Deep autoencoder
  learning for relay-assisted cooperative communication systems,'' \emph{IEEE
  Transactions on Communications}, vol.~68, no.~9, pp. 5471--5488, 2020.

\bibitem{mode14}
A.~Gupta and M.~Sellathurai, ``A stacked-autoencoder based end-to-end learning
  framework for decode-and-forward relay networks,'' in \emph{ICASSP 2020 -
  2020 IEEE International Conference on Acoustics, Speech and Signal Processing
  (ICASSP)}, 2020, pp. 5245--5249.

\bibitem{mode15}
A.~Gupta, M.~Sellathurai, and R.~Tharmalingam, ``A stacked autoencoder-based
  decode-and-forward relay networks with i/q imbalance,'' \emph{CEUR Workshop
  Proceedings}, vol. 3189, no.~4, 2022.

\bibitem{mode7}
A.~Gupta, M.~Sellathurai, and T.~Ratnarajah, ``End-to-end learning-based
  full-duplex amplify-and-forward relay networks,'' \emph{IEEE Transactions on
  Communications}, vol.~71, no.~1, pp. 199--213, 2023.

\bibitem{mode45_semanticrelay}
X.~Luo, B.~Yin, Z.~Chen, B.~Xia, and J.~Wang, ``Autoencoder-based semantic
  communication systems with relay channels,'' in \emph{2022 IEEE International
  Conference on Communications Workshops (ICC Workshops)}, 2022, pp. 711--716.

\bibitem{mode53}
T.~Erpek, T.~J. O'Shea, and T.~C. Clancy, ``Learning a physical layer scheme
  for the mimo interference channel,'' in \emph{2018 IEEE International
  Conference on Communications (ICC)}, 2018, pp. 1--5.

\bibitem{mode31}
\BIBentryALTinterwordspacing
D.~Wu, M.~Nekovee, and Y.~Wang, ``An adaptive deep learning algorithm based
  autoencoder for interference channels,'' 2019. [Online]. Available:
  \url{https://arxiv.org/abs/1902.06841}
\BIBentrySTDinterwordspacing

\bibitem{mode6.5}
D.~Wu, M.~Nekovee, and Y.~Wang, ``Deep learning-based autoencoder for m-user
  wireless interference channel physical layer design,'' \emph{IEEE Access},
  vol.~8, pp. 174\,679--174\,691, 2020.

\bibitem{mode6}
L.~Pellatt, M.~Nekovee, and D.~Wu, ``A concurrent training method of
  deep-learning autoencoders in a multi-user interference channel,'' in
  \emph{2021 17th International Symposium on Wireless Communication Systems
  (ISWCS)}, 2021, pp. 1--6.

\bibitem{mode57}
X.~Zhang and M.~Vaezi, ``Deep autoencoder-based z-interference channels,'' in
  \emph{2023 IEEE Wireless Communications and Networking Conference (WCNC)},
  2023, pp. 1--6.

\bibitem{mode56}
X.~Zhang, M.~Vaezi, and L.~Zheng, ``Interference-aware constellation design for
  z-interference channels with imperfect csi,'' in \emph{ICC 2023 - IEEE
  International Conference on Communications}, 2023, pp. 6385--6390.

\bibitem{mode47}
X.~Zhang and M.~Vaezi, ``Deep autoencoder-based z-interference channels with
  perfect and imperfect csi,'' \emph{IEEE Transactions on Communications},
  vol.~72, no.~2, pp. 861--873, 2024.

\bibitem{mode48.5}
K.~Chahine, N.~Ye, and H.~Kim, ``Deepic: Coding for interference channels via
  deep learning,'' in \emph{2021 IEEE Global Communications Conference
  (GLOBECOM)}, 2021, pp. 01--06.

\bibitem{mode48}
K.~Chahine, Y.~Jiang, J.~Cho, and H.~Kim, ``Deepic+: Learning codes for
  interference channels,'' \emph{IEEE Transactions on Wireless Communications},
  vol.~23, no.~4, pp. 2740--2754, 2024.

\bibitem{mode51}
R.~K. Mishra, K.~Chahine, H.~Kim, S.~Jafar, and S.~Vishwanath, ``Distributed
  interference alignment for k-user interference channels via deep learning,''
  in \emph{2021 IEEE International Symposium on Information Theory (ISIT)},
  2021, pp. 2614--2619.

\bibitem{mode49}
T.~N.~T. Huynh, T.~K. Hoang, and T.-N. Dao, ``Performance analysis of an
  autoencoder-based communication system with numerous transmitter-receiver
  pairs,'' in \emph{2023 12th International Conference on Control, Automation
  and Information Sciences (ICCAIS)}, 2023, pp. 541--546.

\bibitem{mode52}
J.~Ji, Z.~Xiong, K.~Zhu, and T.~Quek, ``Deep learning-based multiuser physical
  layer communication without known channel,'' in \emph{2024 IEEE Wireless
  Communications and Networking Conference (WCNC)}, 2024, pp. 01--06.

\bibitem{communityAE}
\BIBentryALTinterwordspacing
O.~Tieleman, A.~Lazaridou, S.~Mourad, C.~Blundell, and D.~Precup, ``Shaping
  representations through communication: community size effect in artificial
  learning systems,'' 2019. [Online]. Available:
  \url{https://arxiv.org/abs/1912.06208}
\BIBentrySTDinterwordspacing

\bibitem{SICnet}
T.~Van~Luong, N.~Shlezinger, C.~Xu, T.~M. Hoang, Y.~C. Eldar, and L.~Hanzo,
  ``Deep learning based successive interference cancellation for the
  non-orthogonal downlink,'' \emph{IEEE Transactions on Vehicular Technology},
  vol.~71, no.~11, pp. 11\,876--11\,888, 2022.

\bibitem{mode27}
N.~Jindal, S.~Vishwanath, and A.~Goldsmith, ``On the duality of gaussian
  multiple-access and broadcast channels,'' \emph{IEEE Transactions on
  Information Theory}, vol.~50, no.~5, pp. 768--783, 2004.

\bibitem{aoudiaWithoutChannelModel}
F.~A. Aoudia and J.~Hoydis, ``End-to-end learning of communications systems
  without a channel model,'' in \emph{2018 52nd Asilomar Conference on Signals,
  Systems, and Computers}, 2018, pp. 298--303.

\bibitem{deep_deterministic_zhang_modelfree}
B.~Zhang and N.~Van~Huynh, ``Deep deterministic policy gradient for end-to-end
  communication systems without prior channel knowledge,'' in \emph{GLOBECOM
  2023 - 2023 IEEE Global Communications Conference}, 2023, pp. 5677--5682.

\bibitem{jovanovicModelfree}
O.~Jovanovic, M.~P. Yankov, F.~Da~Ros, and D.~Zibar, ``Gradient-free training
  of autoencoders for non-differentiable communication channels,''
  \emph{Journal of Lightwave Technology}, vol.~39, no.~20, pp. 6381--6391,
  2021.

\bibitem{modelfree_REINFORCEMENTLEARNING}
E.~Beck, C.~Bockelmann, and A.~Dekorsy, ``Model-free reinforcement learning of
  semantic communication by stochastic policy gradient,'' in \emph{2024 IEEE
  International Conference on Machine Learning for Communication and Networking
  (ICMLCN)}, 2024, pp. 367--373.

\bibitem{oshea2018_overtheair_GAN}
\BIBentryALTinterwordspacing
T.~J. O'Shea, T.~Roy, N.~West, and B.~C. Hilburn, ``Physical layer
  communications system design over-the-air using adversarial networks,'' 2018.
  [Online]. Available: \url{https://arxiv.org/abs/1803.03145}
\BIBentrySTDinterwordspacing

\bibitem{approximatingthevoid}
T.~J. O’Shea, T.~Roy, and N.~West, ``Approximating the void: Learning
  stochastic channel models from observation with variational generative
  adversarial networks,'' in \emph{2019 International Conference on Computing,
  Networking and Communications (ICNC)}, 2019, pp. 681--686.

\bibitem{metalearning_modelfree}
S.~Park, O.~Simeone, and J.~Kang, ``End-to-end fast training of communication
  links without a channel model via online meta-learning,'' in \emph{2020 IEEE
  21st International Workshop on Signal Processing Advances in Wireless
  Communications (SPAWC)}, 2020, pp. 1--5.

\bibitem{aoudiaModelfree}
F.~A. Aoudia and J.~Hoydis, ``Model-free training of end-to-end communication
  systems,'' \emph{IEEE Journal on Selected Areas in Communications}, vol.~37,
  no.~11, pp. 2503--2516, 2019.

\bibitem{poissongradient}
A.~Elfikky, M.~Soltani, and Z.~Rezki, ``End-to-end learning framework for space
  optical communications in non-differentiable poisson channel,'' \emph{IEEE
  Wireless Communications Letters}, vol.~13, no.~8, pp. 2090--2094, 2024.

\bibitem{agrell_database}
\BIBentryALTinterwordspacing
E.~Agrell. Database of sphere packing. Accessed: Nov. 10, 2024. [Online].
  Available: \url{https://codes.se/packings/8.htm}
\BIBentrySTDinterwordspacing

\bibitem{continuouscontroldeepreinforcement}
\BIBentryALTinterwordspacing
T.~P. Lillicrap \emph{et~al.}, ``Continuous control with deep reinforcement
  learning,'' 2019. [Online]. Available: \url{https://arxiv.org/abs/1509.02971}
\BIBentrySTDinterwordspacing

\bibitem{modelaware_semantic}
\BIBentryALTinterwordspacing
E.~Beck, C.~Bockelmann, and A.~Dekorsy, ``Semantic information recovery in
  wireless networks,'' \emph{Sensors}, vol.~23, no.~14, 2023. [Online].
  Available: \url{https://www.mdpi.com/1424-8220/23/14/6347}
\BIBentrySTDinterwordspacing

\bibitem{causal_policy_gradient_modelfree}
S.~Shirodkar and S.~Banerjee, ``Causal policy gradient for end-to-end
  communication systems,'' in \emph{2024 16th International Conference on
  COMmunication Systems \& NETworkS (COMSNETS)}, 2024, pp. 572--576.

\bibitem{towardCRL}
\BIBentryALTinterwordspacing
H.~Sun and T.~Wang, ``Toward causal-aware rl: State-wise action-refined
  temporal difference,'' 2022. [Online]. Available:
  \url{https://arxiv.org/abs/2201.00354}
\BIBentrySTDinterwordspacing

\bibitem{MAML}
C.~Finn, P.~Abbeel, and S.~Levine, ``Model-agnostic meta-learning for fast
  adaptation of deep networks,'' in \emph{Proceedings of the 34th International
  Conference on Machine Learning - Volume 70}, ser. ICML'17.\hskip 1em plus
  0.5em minus 0.4em\relax JMLR.org, 2017, p. 1126–1135.

\bibitem{residual_aids_jiang_modelfree}
H.~Jiang, S.~Bi, L.~Dai, H.~Wang, and J.~Zhang, ``Residual-aided end-to-end
  learning of communication system without known channel,'' \emph{IEEE
  Transactions on Cognitive Communications and Networking}, vol.~8, no.~2, pp.
  631--641, 2022.

\bibitem{DeepMIMO_dataset}
\BIBentryALTinterwordspacing
A.~Alkhateeb, ``Deepmimo: A generic deep learning dataset for millimeter wave
  and massive mimo applications,'' 2019. [Online]. Available:
  \url{https://arxiv.org/abs/1902.06435}
\BIBentrySTDinterwordspacing

\bibitem{Hao_channelagnostic}
H.~Ye, G.~Y. Li, B.-H.~F. Juang, and K.~Sivanesan, ``Channel agnostic
  end-to-end learning based communication systems with conditional gan,'' in
  \emph{2018 IEEE Globecom Workshops (GC Wkshps)}, 2018, pp. 1--5.

\bibitem{Hao_conditionalGANS}
H.~Ye, L.~Liang, G.~Y. Li, and B.-H. Juang, ``Deep learning-based end-to-end
  wireless communication systems with conditional gans as unknown channels,''
  \emph{IEEE Transactions on Wireless Communications}, vol.~19, no.~5, pp.
  3133--3143, 2020.

\bibitem{Hao_bilinear}
H.~Ye, G.~Ye~Li, and B.-H.~F. Juang, ``Bilinear convolutional auto-encoder
  based pilot-free end-to-end communication systems,'' in \emph{ICC 2020 - 2020
  IEEE International Conference on Communications (ICC)}, 2020, pp. 1--6.

\bibitem{GAIforPHY_Survey2024}
N.~Van~Huynh \emph{et~al.}, ``Generative ai for physical layer communications:
  A survey,'' \emph{IEEE Transactions on Cognitive Communications and
  Networking}, vol.~10, no.~3, pp. 706--728, 2024.

\bibitem{poissondnn}
L.-H. Si-Ma, Z.-R. Zhu, and H.-Y. Yu, ``Model-aware end-to-end learning for
  siso optical wireless communication over poisson channel,'' \emph{IEEE
  Photonics Journal}, vol.~12, no.~6, pp. 1--15, 2020.

\bibitem{owc5fiber}
B.~Karanov \emph{et~al.}, ``End-to-end deep learning of optical fiber
  communications,'' \emph{Journal of Lightwave Technology}, vol.~36, no.~20,
  pp. 4843--4855, 2018.

\bibitem{owc6fiber}
W.~Jiang \emph{et~al.}, ``End-to-end learning of constellation shaping for
  optical fiber communication systems,'' \emph{IEEE Photonics Journal},
  vol.~15, no.~6, pp. 1--7, 2023.

\bibitem{owc7fiber}
Y.~Li \emph{et~al.}, ``A cgan-aided autoencoder supporting joint geometric
  probabilistic shaping for optical fiber communication system,'' in \emph{2023
  Asia Communications and Photonics Conference/2023 International Photonics and
  Optoelectronics Meetings (ACP/POEM)}, 2023, pp. 1--3.

\bibitem{owc9fiber}
M.~Li, D.~Wang, Q.~Cui, Z.~Zhang, L.~Deng, and M.~Zhang, ``End-to-end learning
  for optical fiber communication with data-driven channel model,'' in
  \emph{2020 Opto-Electronics and Communications Conference (OECC)}, 2020, pp.
  1--3.

\bibitem{owc15fiber}
O.~Jovanovic, R.~T. Jones, S.~Gaiarin, M.~P. Yankov, F.~Da~Ros, and D.~Zibar,
  ``Optimization of fiber optics communication systems via end-to-end
  learning,'' in \emph{2020 22nd International Conference on Transparent
  Optical Networks (ICTON)}, 2020, pp. 1--3.

\bibitem{owc18fiber}
Z.~Liu, X.~Liu, S.~Xiao, W.~Yang, and W.~Hu, ``Bi-gru enhanced cost-effective
  memory-aware end-to-end learning for geometric constellation shaping in
  optical coherent communications,'' \emph{IEEE Photonics Journal}, vol.~16,
  no.~1, pp. 1--10, 2024.

\bibitem{owc3}
F.~Tian, Z.~Liu, J.~Qi, M.~Leeson, G.~Zheng, and T.~Xu, ``Gmi optimisation for
  end-to-end learning-based probabilistic constellation shaping in
  400-gbits/s/$\lambda$ dp-64qam optical communication system,'' in \emph{2024
  24th International Conference on Transparent Optical Networks (ICTON)}, 2024,
  pp. 1--4.

\bibitem{owc8decoding}
M.~S. Neves, P.~A. Loureiro, T.~Slavov, and P.~Georgieva, ``End-to-end learning
  system for symbol decoding in optical communication,'' in \emph{2024 IEEE
  12th International Conference on Intelligent Systems (IS)}, 2024, pp. 1--6.

\bibitem{owc10OFDM}
W.~Jiang \emph{et~al.}, ``End-to-end learning based bit-wise autoencoder for
  optical ofdm communication system,'' in \emph{2021 Asia Communications and
  Photonics Conference (ACP)}, 2021, pp. 1--3.

\bibitem{owc16wasserstein}
Z.~Cheng, R.~Gao, Q.~Xu, F.~Wang, Y.~Cui, and X.~Xin, ``Wasserstein autoencoder
  based end-to-end learning strategy of geometric shaping for an oam mode
  division multiplexing im/dd transmission,'' in \emph{2023 Asia Communications
  and Photonics Conference/2023 International Photonics and Optoelectronics
  Meetings (ACP/POEM)}, 2023, pp. 1--3.

\bibitem{colorVLC}
\BIBentryALTinterwordspacing
H.~Lee, I.~Lee, and S.~H. Lee, ``Deep learning based transceiver design for
  multi-colored vlc systems,'' \emph{Opt. Express}, vol.~26, no.~5, pp.
  6222--6238, Mar 2018. [Online]. Available:
  \url{https://opg.optica.org/oe/abstract.cfm?URI=oe-26-5-6222}
\BIBentrySTDinterwordspacing

\bibitem{soltaniOWC}
M.~Soltani, W.~Fatnassi, A.~Aboutaleb, Z.~Rezki, A.~Bhuyan, and P.~Titus,
  ``Autoencoder-based optical wireless communications systems,'' in \emph{2018
  IEEE Globecom Workshops (GC Wkshps)}, 2018, pp. 1--6.

\bibitem{zhuOWC}
Z.-R. Zhu, J.~Zhang, R.-H. Chen, and H.-Y. Yu, ``Autoencoder-based transceiver
  design for owc systems in log-normal fading channel,'' \emph{IEEE Photonics
  Journal}, vol.~11, no.~5, pp. 1--12, 2019.

\bibitem{owc2}
Q.~Zhang \emph{et~al.}, ``An autoencoder-based transceiver for uav-to-ground
  free space optical communication,'' in \emph{2023 Opto-Electronics and
  Communications Conference (OECC)}, 2023, pp. 1--3.

\bibitem{owc4}
M.~Cao, R.~Wang, Y.~Zhang, H.~Deng, L.~Zhou, and H.~Wang, ``An end-to-end
  autoencoder for fso system under unknown csi scenarios,'' in \emph{2023 21st
  International Conference on Optical Communications and Networks (ICOCN)},
  2023, pp. 1--3.

\bibitem{owc12}
X.~Liu, Z.~Wei, A.~Pepe, Z.~Wang, and H.~Y. Fu, ``Autoencoder for optical
  wireless communication system in atmospheric turbulence,'' in \emph{2020
  Opto-Electronics and Communications Conference (OECC)}, 2020, pp. 1--3.

\bibitem{owc13underwater}
C.~Zou, F.~Yang, J.~Song, and Z.~Han, ``Underwater wireless optical
  communication with one-bit quantization: A hybrid autoencoder and generative
  adversarial network approach,'' \emph{IEEE Transactions on Wireless
  Communications}, vol.~22, no.~10, pp. 6432--6444, 2023.

\bibitem{uwocchannelGAN}
C.~Zou, F.~Yang, J.~Song, and Z.~Han, ``Underwater optical channel generator: A
  generative adversarial network based approach,'' \emph{IEEE Transactions on
  Wireless Communications}, vol.~21, no.~11, pp. 9394--9403, 2022.

\bibitem{owc14biLSTM}
T.~Zhang \emph{et~al.}, ``End-to-end learning for free space optical
  communication with bilstm-based channel model,'' in \emph{2023 IEEE
  International Symposium on Broadband Multimedia Systems and Broadcasting
  (BMSB)}, 2023, pp. 1--4.

\bibitem{binaryVLC}
\BIBentryALTinterwordspacing
H.~Lee, T.~Q.~S. Quek, and S.~H. Lee, ``A deep learning approach to universal
  binary visible light communication transceiver,'' 2019. [Online]. Available:
  \url{https://arxiv.org/abs/1910.12048}
\BIBentrySTDinterwordspacing

\bibitem{binaryVLCbaseline}
S.~Zhao, ``A serial concatenation-based coding scheme for dimmable visible
  light communication systems,'' \emph{IEEE Communications Letters}, vol.~20,
  no.~10, pp. 1951--1954, 2016.

\bibitem{owc1}
H.~Safi, I.~Tavakkolnia, and H.~Haas, ``Deep learning based end-to-end optical
  wireless communication systems with autoencoders,'' \emph{IEEE Communications
  Letters}, vol.~28, no.~6, pp. 1342--1346, 2024.

\bibitem{soc2}
A.~Elfikky, M.~Soltani, and Z.~Rezki, ``Learning-based autoencoder for multiple
  access and interference channels in space optical communications,''
  \emph{IEEE Communications Letters}, vol.~27, no.~10, pp. 2662--2666, 2023.

\bibitem{soc3}
A.~Elfikky and Z.~Rezki, ``Symbol detection and channel estimation for space
  optical communications using neural network and autoencoder,'' \emph{IEEE
  Transactions on Machine Learning in Communications and Networking}, vol.~2,
  pp. 110--128, 2024.

\bibitem{soc1}
A.~E.-R.~A. El-Fikky and Z.~Rezki, ``On the performance of autoencoder-based
  space optical communications,'' in \emph{GLOBECOM 2022 - 2022 IEEE Global
  Communications Conference}, 2022, pp. 1466--1471.

\bibitem{transferlearning}
S.~J. Pan and Q.~Yang, ``A survey on transfer learning,'' \emph{IEEE
  Transactions on Knowledge and Data Engineering}, vol.~22, no.~10, pp.
  1345--1359, 2010.

\bibitem{Cammerer2023NeuralRx}
\BIBentryALTinterwordspacing
S.~Cammerer \emph{et~al.}, ``A neural receiver for 5g nr multi-user mimo,''
  \emph{arXiv preprint arXiv:2312.02601}, 2023. [Online]. Available:
  \url{https://arxiv.org/abs/2312.02601}
\BIBentrySTDinterwordspacing

\bibitem{Bonati2024ORANML}
L.~Bonati, M.~Polese, S.~Basagni, and T.~Melodia, ``Machine learning for o-ran:
  Design principles, challenges, and future directions,'' \emph{IEEE
  Communications Magazine}, vol.~62, no.~5, pp. 78--85, 2024.

\bibitem{3gpp_tr_38843}
\BIBentryALTinterwordspacing
{3GPP}, ``{Study on Artificial Intelligence (AI)/Machine Learning (ML) for NR
  air interface},'' 3rd Generation Partnership Project (3GPP), Technical Report
  TR 38.843, 2023, release 18, v0.1.0. [Online]. Available:
  \url{https://www.3gpp.org/technologies/ai-ml-nr}
\BIBentrySTDinterwordspacing

\bibitem{3gpp_tr_28908}
\BIBentryALTinterwordspacing
{3GPP}, ``{Study on Artificial Intelligence/Machine Learning (AI/ML)
  Management},'' 3rd Generation Partnership Project (3GPP), Technical Report TR
  28.908, 2024, release 18, v18.0.0. [Online]. Available:
  \url{https://www.3gpp.org/technologies/ai-ml-management}
\BIBentrySTDinterwordspacing

\bibitem{lin_ai_5g_adv}
\BIBentryALTinterwordspacing
X.~Lin, ``{Artificial Intelligence in 3GPP 5G-Advanced: A Survey},''
  \emph{arXiv preprint arXiv:2305.05092}, 2023. [Online]. Available:
  \url{https://arxiv.org/abs/2305.05092}
\BIBentrySTDinterwordspacing

\bibitem{bonati_intelligence_oran}
L.~Bonati, S.~D'Oro, M.~Polese, S.~Basagni, and T.~Melodia, ``{Intelligence and
  Learning in O-RAN for Data-Driven NextG Cellular Networks},'' \emph{IEEE
  Communications Magazine}, vol.~59, no.~10, pp. 21--27, 2021.

\bibitem{yungaicela2024misconfig}
N.~M. Yungaicela-Naula, V.~Sharma, and S.~Scott-Hayward, ``{Misconfiguration in
  O-RAN: Analysis of the Impact of AI/ML},'' \emph{Computer Networks}, vol.
  240, p. 110455, 2024.

\bibitem{itu_y3181}
\BIBentryALTinterwordspacing
{ITU-T}, ``{Recommendation ITU-T Y.3181: Architectural Framework for Machine
  Learning Sandbox in Future Networks Including IMT-2020},'' International
  Telecommunication Union, Telecommunication Standardization Sector (ITU-T),
  2022, version 07/22. [Online]. Available:
  \url{https://www.itu.int/rec/T-REC-Y.3181}
\BIBentrySTDinterwordspacing

\bibitem{cammerer2023neural}
\BIBentryALTinterwordspacing
S.~Cammerer \emph{et~al.}, ``{A Neural Receiver for 5G NR Multi-User MIMO},''
  \emph{arXiv preprint arXiv:2312.02601}, 2023. [Online]. Available:
  \url{https://arxiv.org/abs/2312.02601}
\BIBentrySTDinterwordspacing

\bibitem{rohde2023neural}
\BIBentryALTinterwordspacing
{Rohde \& Schwarz and NVIDIA}, ``{Towards 6G: Rohde \& Schwarz Showcases
  AI/ML-Based Neural Receiver with NVIDIA at MWC Barcelona},'' Press release,
  Feb. 2023. [Online]. Available: \url{https://www.rohde-schwarz.com}
\BIBentrySTDinterwordspacing

\bibitem{Chen2022FLIoT}
H.~Chen, S.~Huang, D.~Zhang, M.~Xiao, M.~Skoglund, and H.~V. Poor, ``Federated
  learning over wireless {IoT} networks with optimized communication and
  resources,'' \emph{IEEE Internet of Things Journal}, vol.~9, no.~17, pp.
  16\,592--16\,605, 2022.

\bibitem{Wu2023SplitSL}
W.~Wu \emph{et~al.}, ``Split learning over wireless networks: Parallel design
  and resource management,'' \emph{IEEE Journal on Selected Areas in
  Communications}, vol.~41, no.~4, pp. 1051--1066, 2023.

\bibitem{Zhang2024SemCom6G}
Q.~Zhang, Y.~Liu \emph{et~al.}, ``Semantic communication empowered {6G}
  networks: Techniques, applications, and challenges,'' \emph{IEEE Access},
  2024, early Access.

\bibitem{Ma2023TOESemCom}
S.~Ma \emph{et~al.}, ``Task-oriented explainable semantic communications,''
  \emph{IEEE Transactions on Wireless Communications}, vol.~22, no.~12, pp.
  9248--9262, 2023.

\bibitem{transformermaskedAE}
\BIBentryALTinterwordspacing
A.~Zayat, M.~A. Hasabelnaby, M.~Obeed, and A.~Chaaban, ``Transformer masked
  autoencoders for next-generation wireless communications: Architecture and
  opportunities,'' 2024. [Online]. Available:
  \url{https://arxiv.org/abs/2401.06274}
\BIBentrySTDinterwordspacing

\bibitem{frombitstosemantics}
Z.~Qin \emph{et~al.}, ``Ai empowered wireless communications: From bits to
  semantics,'' \emph{Proceedings of the IEEE}, vol. 112, no.~7, pp. 621--652,
  2024.

\end{thebibliography}

\end{document}